\documentclass[11pt]{article}
\pdfoutput=1

\usepackage{jcappub}
\usepackage{caption}
\usepackage{subcaption}
\usepackage{graphicx}
\usepackage{mathtools}
\usepackage{physics}
\usepackage{siunitx}
\usepackage{tablefootnote}
\usepackage{soul}
\usepackage{verbatim}
\usepackage{float}
\usepackage{bm}
\usepackage[normalem]{ulem}
\usepackage[dvipsnames]{xcolor}
\usepackage{orcidlink}
\usepackage[nameinlink, capitalise]{cleveref}
\usepackage{braket}
\usepackage{dirtytalk}
\usepackage{mathtools}
\usepackage{underscore}
\usepackage{slashed}

\usepackage{color,appendix,latexsym,amsmath,amssymb,graphicx,booktabs,epsfig,hyperref,url}

\numberwithin{equation}{section}
\usepackage[english]{babel}

\usepackage{letltxmacro}
\LetLtxMacro{\originaleqref}{\eqref}
\usepackage[sorting=none]{biblatex} 
\addto\extrasenglish{%
}

\definecolor{MyBlue}{rgb}{0.15,0.15,0.70}
\hypersetup{
colorlinks=true,
citecolor=MyBlue,
linkcolor=MyBlue,
urlcolor=MyBlue
}

\definecolor{orange}{rgb}{0.98, 0.6, 0.01}
\definecolor{darkolivegreen}{rgb}{0.33, 0.42, 0.18}
\definecolor{tealblue}{rgb}{0.21, 0.46, 0.53}

\usepackage{listings}
\definecolor{codegreen}{rgb}{0,0.6,0}
\definecolor{codegray}{rgb}{0.5,0.5,0.5}
\definecolor{codepurple}{rgb}{0.58,0,0.82}
\definecolor{backcolour}{rgb}{0.95,0.95,0.92}
\graphicspath{figures/} 

\begin{document}

\title{Beyond Standard Model equation of state and primordial black holes}

\author[1]{Xavier Pritchard\,\orcidlink{0009-0007-6543-8563},}
\author[1]{Matthew Starbuck\,\orcidlink{0009-0003-2306-8083},}
\author[1]{Wingfung Leung\,\orcidlink{0009-0006-8077-8497}}
\emailAdd{X.Pritchard@sussex.ac.uk}
\emailAdd{M.Starbuck@sussex.ac.uk}
\emailAdd{WF.Leung@sussex.ac.uk}
\affiliation[1]{Department of Physics and Astronomy, University of Sussex, Brighton BN1 9QH, UK\\}

\abstract{The Standard Model of particle physics successfully describes all known fundamental particles and their interactions; however, it leaves several unanswered questions. Theories beyond the Standard Model typically introduce new particles and symmetries to address these issues. In the early universe, when such particles become non-relativistic, or the symmetries are broken, there are associated reductions in the equation of state of the primordial plasma. These reductions lead to an exponential enhancement in the formation rate of primordial black holes. In this paper, we calculate the equation of state for several supersymmetric and composite Higgs models, which naturally predict a large number of additional degrees of freedom. Using these equations of state, we compute some example primordial black hole abundances, which we find can be enhanced by up to 20 orders of magnitude.}

\maketitle

\section{Introduction}

The Standard Model of particle physics (SM) has so far enjoyed remarkable phenomenological success, demonstrating a competency for describing physics at all energy scales yet probed by collider experiments \cite{LEP, 20251, AtlasDrel}. Despite this, it is widely believed that the SM is an effective field theory, having only a finite energy range within which its physical predictions remain accurate \cite{Isidori_2024, gouttenoire2022}. Furthermore, it is conceivable (and hoped) that the energies probed at the Large Hadron Collider lie in close proximity to the end of the SM domain of validity, thus, potentially revealing new degrees of freedom at the TeV scale \cite{Belfatto_2020, craig2022}. 

The existence of physics beyond the SM (BSM) is further motivated by several persistent, theoretical issues. These include, but are not limited to, neutrino oscillations \cite{Maki:1962mu,Super-Kamiokande:1998kpq}, the strong CP problem \cite{Peccei:1977hh,Weinberg:1977ma,Wilczek:1977pj} and the flavour hierarchy problem \cite{Kobayashi:1973fv,Froggatt:1978nt}. However, perhaps the most compelling indication of BSM physics comes from naturalness arguments \cite{PhysRevD.14.1667, PhysRevD.20.2619, peskin2025}. The Higgs mass is radiatively unstable against quantum corrections, exhibiting a quadratic sensitivity to the energy scale at which the SM is UV completed \cite{Schwartz_2013}. From quantum gravity arguments alone, this scale could be as high as the Planck scale ($10^{19}$GeV) \cite{WITTEN1981513}, yet experimentally the Higgs mass lies comfortably at around 125GeV \cite{Chatrchyan_2012, Aad_2012}. This large gap in scales has a tendency to imply a hugely fine-tuned cancellation between the parameters of the SM and its UV extension, such that the correct physical Higgs mass is recovered.

Two of the most contemplated remedies to this problem are the existence of a TeV-scale supersymmetry (SUSY) \cite{CS_KI_1996, Martin:1997ns, krippendorf2010}, and models in which the Higgs is realised as the lightest bound state of some new strongly interacting dynamics, i.e. composite Higgs models (CHMs) \cite{DIMOPOULOS1979237, contino2010tasi, Panico_2016}. Both of these solutions, whilst qualitatively very different, share a common feature, namely the introduction of a large number of new degrees of freedom with a mass spectrum characterised by a particular energy scale. For CHMs, this is the scale of confinement for the new strong dynamics, with the new degrees of freedom being the microscopic constituents from which the Higgs and other bound states are comprised \cite{Giudice_2007}. For SUSY, the degrees of freedom are the superpartners to those of the SM, as well as an additional Higgs doublet, whose masses are expected to surround the scale of SUSY breaking \cite{Martin:1997ns}.


Both models involve new symmetries that are manifest at high temperatures. As the universe cools, the symmetries are spontaneously broken, leading to a phase transition (PT) that produces the low-temperature effective theory resembling the SM. This has parallels to the electroweak PT, whereby the Higgs mechanism generates masses for gauge bosons \cite{Englert:1964et,Higgs:1964pj,Weinberg:1967tq}. Naturalness arguments favour the BSM PT to take place at temperatures as low as phenomenologically allowed \cite{Panico_2013}, typically a few TeV \cite{Glioti_2025}. However, if these concerns are neglected, then the scales may in principle be much higher \cite{Barnard:2014tla, Ellis:2015jwa}. 

The PTs associated with SUSY and composite Higgs models are not only of interest from a particle theory perspective, but also have cosmological significance \cite{Kirzhnits:1972iw,Weinberg:1974hy,Linde:1978px}. In this work, we are particularly interested in the corresponding drops in the radiation equation of state, which depend on the order of the PT and the number of degrees of freedom that promptly become non-relativistic. For first-order PTs, thermodynamic quantities, and thus the equation of state, sharply change at the critical temperature \cite{1981MNRAS.195..467S}. On the other hand, in the case of a crossover or second-order transition, at the critical temperature, thermodynamic quantities vary more smoothly. This generally results in a smaller, more gradual reduction in the equation of state.


Regardless of the order, the drop is expected to be proportional to the number of the degrees of freedom becoming non-relativistic, making our considered BSM models distinctively relevant. There are several cosmological calculations which are sensitive to the equation of state parameter \cite{Byrnes:2018clq,Saikawa:2018rcs,Saikawa:2020swg}. In this work, we focus on primordial black holes (PBHs) \cite{Zeldovich}, which are typically thought to form from the collapse of density fluctuations \cite{Carr1,Hawking}. Black hole formation is sensitive to both the inwards gravitational force, and thus energy density, as well as outwards pressure gradients. Therefore, if such objects were to form in the early universe, they are naturally expected to be sensitive to the equation of state. In fact, the abundance of PBHs is exponentially sensitive to this parameter. Therefore, in this work we focus on the BSM models predicting the largest drops in equation of state, and calculate the corresponding enhancements in PBH abundances. Work has been done to this end \cite{Lu:2022yuc,Escriva:2023nzn}, however we expand on this by focusing on specific BSM models.

Recently, there has been renewed interest in PBHs, both as a potential dark matter candidate \cite{Chapline, Esser:2025pnt} and as sources of gravitational waves (GWs) \cite{Sasaki:2018dmp,Byrnes:2018clq,Franciolini,Escriva1}, which links BSM PTs directly to observable phenomena in cosmology. With a large proportion of these studies being relevant to lower-mass PBHs ($M\lesssim10^{-6}M_\odot$), they are thus relevant to these higher-energy particle physics PTs.

The rest of the paper is organised as follows: 
In \cref{section2}, using the SM as an example, we demonstrate the methodologies used later in the determination of the evolution of the equation of state, and exemplify the generic sensitivities to BSM degrees of freedom using a toy model. In \cref{SUSYsection}, we detail the mass spectrum and degrees of freedom for the Minimal Supersymmetric SM, and use this to calculate the equation of state. In \cref{CHsection} we apply the methodology developed in \cref{section2} to a UV realisation of composite Higgs models, and study the equation of state during the PT. The enhanced spectrum of PBHs resulting from the equations of state produced in \cref{SUSYsection} and \cref{CHsection} are presented in \cref{PBHsection}, which are the main results of this work. In \cref{Conclusion}, we summarise our conclusions, and give our outlook for future work on this subject. Finally, we give some more details as to our calculations in the appendices.

\section{From Standard Model to beyond} \label{section2}

Before discussing BSM physics, we begin by describing the thermodynamics of the SM itself in some detail. Our purpose for doing this is twofold. Firstly, it is a good opportunity to, within a familiar context, develop the methodologies that we later employ. Secondly, this provides further background information for those less familiar with the SM and the QCD PT. Those more comfortable with SM thermodynamics may wish to skip to \cref{Toy model}, where we use a perturbative toy model to describe the equation of state in a general BSM scenario.

\subsection{Relativistic degrees of freedom} \label{Rel dofs}

After reheating, the early universe is filled with a hot, dense plasma of relativistic particles \cite{Abbott2,Traschen} (see \cite{Bassett,Allahverdi,Jedamzik:2024wtq} for reviews). As the universe expands, the density and pressure of the plasma scale as radiation. However, when the temperature drops below the mass of a given particle species, it becomes non-relativistic, leaving a slight offset in the scaling laws. The effective numbers of relativistic degrees of freedom are given by
\begin{equation}\label{gstarEq}
    g_{*\rho}(T)=\frac{30\rho(T)}{\pi^2T^4},\hspace{15mm}g_{*s}(T)=\frac{45s(T)}{2\pi^2T^3},
\end{equation}
where $\rho(T)$ is the energy density and $s(T)$ is the entropy density. We note that fitting functions for these quantities have been provided in \cite{Saikawa:2018rcs}. The equation of state parameter, which tracks the ratio of the pressure to the energy density, is a pivotal quantity in cosmology. Using $p(T)=s(T)T+\rho(T)$, we may write 
\begin{equation}\label{EoS}
\omega(T)\equiv\frac{p(T)}{\rho(T)}=\frac{4g_{*\rho}(T)}{3g_{*s}(T)}-1.
\end{equation}
When in thermal equilibrium, the energy density and pressure of an ideal gas is determined from statistical mechanics
\begin{align}\label{lowtempEoS}
    \rho_i&=\frac{g_i}{2\pi^2}\int_{m_i}^\infty \frac{\sqrt{E^2-m_i^2}}{\text{exp}(E/T)\pm1}E^2\text{d}E,\\
     p_i&=\frac{g_i}{6\pi^2}\int_{m_i}^\infty\frac{(E^2-m_i^2)^{3/2}}{\text{exp}(E/T)\pm1}\text{d}E,
\end{align}
where $m_i$ is the particle’s mass, $g_i$ is the particle's degrees of freedom, and the plus (minus) sign in the denominator applies to fermions (bosons). These formula may be used to rewrite the equation of state
\begin{equation} \label{omegasum}
    \omega(T)=\frac{1}{3}-\sum _i\delta\omega_i(T),
\end{equation}
where the departure of the plasma from pure radiation due to a given particle species is described by
\begin{equation} \label{eq:deltaomega}
\delta\omega_i(T)=\frac{5}{\pi^4}\frac{g_i}{g_{*\rho}(T)}\left(\frac{m_i}{T}\right)^2\int_{m_i/T}^\infty\frac{\sqrt{u^2-(m_i/T)^2}}{e^u\pm1}\text{d}u.
\end{equation}
Within the SM, the first drop in the equation of state occurs at the electroweak (EW) scale. The top quark, Z, W, and Higgs boson, contributing a total of 20.5 effective degrees of freedom, each become non-relativistic during this period. Due to the observed Higgs mass \cite{ATLAS:2015yey}, this PT is thought to be a crossover \cite{Kajantie:1996mn}. However, for the models considered in this paper, a first-order electroweak PT may be realised \cite{Cline:1998hy,Bruggisser:2018mrt}. In this case, GW signals are expected from bubble collisions \cite{Hindmarsh:2013xza}, which could be detectable in future experiments. Following the EW crossover, the bottom and charm quarks, as well as the $\tau$ leptons annihilate, accounting for a total of 24.5 degrees of freedom. Subsequently, there is the QCD PT, which has also been found to be a crossover, and is the topic of the next subsection. Shortly after this, a further 6.5 degrees of freedom decouple, corresponding to the muons, $\mu$, and pions, $\pi^\pm,\pi^0$. Lastly, $e^-e^+$ annihilation and neutrino decoupling occur, leaving $g_{*\rho}\approx3.36$ relativistic degrees of freedom remaining \cite{Husdal:2016haj}.

\subsection{The QCD phase transition} \label{QCD phase transition}

QCD is a gauge theory based on a local SU(3) symmetry group. In addition to 16 gluonic degrees of freedom, there are 6 species of quarks that possess colour charge. Of these 6 quarks, from a QCD perspective, the up and down are effectively massless, with the strange quarks being light \cite{Carrasco_2014}, leading to an approximate chiral flavour symmetry $G_{\text{F}} = \text{SU(3)}_{\text{L}}\times \text{SU(3)}_{\text{R}}$. The flavour symmetry is spontaneously broken into its vectorial subgroup $H_{\text{F}} = \text{SU(3)}_{\text{V}}$ by the QCD vacuum, producing 8 pseudo-Nambu-Goldstone bosons \cite{Weinberg_1996}. At the QCD confinement scale, there are 31.5 relativistic fermionic degrees of freedom coming from the up, down and strange quarks.

An understanding of the QCD PT in the context of cosmology is highly motivated \cite{Byrnes:2018clq,Saikawa:2018rcs,Saikawa:2020swg}. Moreover, numerous theories beyond the SM are strongly interacting, such as composite Higgs models. Despite occurring at higher energy scales, it is expected that the composite Higgs PT will share some qualitative features with that of QCD. Therefore, in the remainder of the subsection we describe the SM QCD PT. By demonstrating these techniques in the context of QCD, we will subsequently apply them to our study of composite Higgs models in \cref{CHsection}. 


\subsubsection{The M.I.T. bag model}\label{HadronicAndMIT}


The quantity responsible for defining the scale of the QCD PT is the critical temperature, $T_c$. This value is typically extracted from lattice simulations \cite{Aoki:2006br,Borsanyi:2010bp,Borsanyi:2020fev}, where it has been found to be around 155MeV. For $T\lesssim T_c$, it is expected that the plasma behaves like an ideal gas in thermal equilibrium. Whilst the most precise determination of the critical temperature is extracted from the lattice, analytical techniques still provide a reasonable estimate. These techniques offer a simpler alternative calculation, without resorting to numerical methods. 


One such example is the M.I.T. bag model \cite{PhysRevD.10.2599}, whereby hadrons are approximated as bubbles of perturbative vacuum, within which quarks freely propagate \cite{Le_Bellac_1996}. Qualitatively, the PT occurs once neighboring bubbles overlap and fill the entire volume of the plasma. Thus, following \cite{Ishii:2002ys}, we define the ratio of the volume occupied by bubbles to the total volume as
\begin{equation} \label{bagratio}
r_{V}\left(T\right) = \sum_{i}\frac{4\pi g_{i}}{3}R^{3}_{i}T^{3} \int \frac{d^{3}k}{\left(2\pi\right)^{3}}\frac{1}{e^{\sqrt{k^{2} +m^{2}_{i}}/T}\pm1}
\end{equation}
where $R_{i}$ is the radius of the hadron. A precise determination of the radius must come from experiment, however, since we are only interested in an estimate of $T_{c}$, we use dimensional analysis $R_{i} \sim \frac{1}{m_{i}}$. The critical temperature is then defined through solving $r_{V} \left(T_{C}\right) = 1$. Applying this to the spectrum of QCD resonances, we find the critical temperature to be $\approx133\text{MeV}$, which is within 15$\%$ of the best fit value. Thus, we expect to obtain a critical temperature of the correct order of magnitude when applied to composite Higgs models.

\subsubsection{Full QCD equation of state} 

For $T\gg T_{c}$, we are able to make use of finite-temperature QCD. The pressure of the quark-gluon plasma is typically written as an expansion in the strong coupling constant, $g_3$. Currently, this is known up to $\mathcal{O}(g_3^6\ln(1/g_3))$ \cite{Kajantie:2002wa}, however major progress has been made towards calculating the full $g_3^6$ pressure \cite{Navarrete:2024ruu}, which is now within reach if applied in the context of \cite{Braaten:1995jr}. For completeness, we describe the full finite-temperature QCD pressure calculation in \cref{alphabeta}. Furthermore, a large amount of work has been done providing alternative perturbative techniques for calculating the thermodynamics of hot QCD \cite{Braaten:1995ju,Andersen:2010wu,Kneur:2015moa,Kneur:2021feo,Ihssen:2024miv}. 

\begin{figure}[ht!]
    \centering
\includegraphics[width=\linewidth]{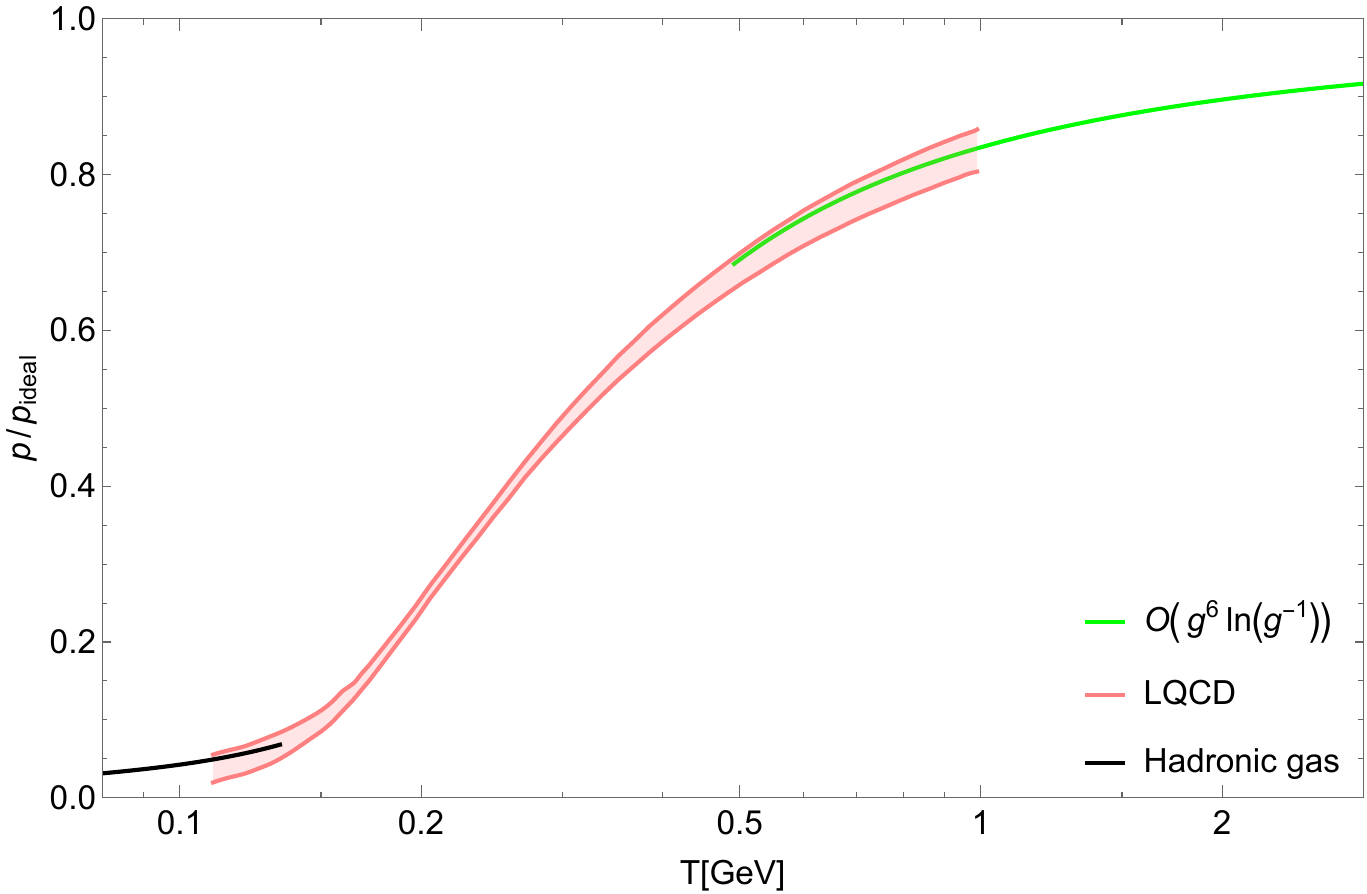}
    \caption{The pressure during the SM QCD PT, normalised to the Stefan-Boltzmann limit. At lower temperatures, we use the hadronic gas model. At high temperatures we employ finite-temperature QCD results, up to $\mathcal{O}(g_3^6\ln (1/g_3))$ in the strong coupling constant. During the non-perturbative phase we use lattice data from \cite{Borsanyi:2016ksw}.}\label{QCDpressure}
\end{figure}

Having described the perturbative aspects of the QCD plasma at both high and low temperatures, we are left with the non-perturbative phase. The most robust description we have of this regime is from the lattice, which is then extrapolated to the continuum limit. This has been the topic of extensive work, with full results given for $N_f=2+1$ flavors \cite{Borsanyi:2013bia, Bazavov:2017dsy, Bresciani:2025vxw}, as well as $N_f=2+1+1$ \cite{Borsanyi:2016ksw}. We take the results of the latter to plot the full QCD pressure in \cref{QCDpressure}. Furthermore, we normalize to the ideal Stefan-Boltzmann gas, defined in a general $\text{SU}(N_c)$ gauge theory with $N_f$ fermions
\begin{equation} \label{SBlimit}
    p_\text{ideal}(T)=\frac{T^4\pi^2}{90}\left[(2N_c^2-2)+(4N_cN_f\frac{7}{8})\right],
\end{equation}
where the term in the square brackets simply counts the degrees of freedom. For the SM QCD PT, we set $N_c=3$ and $N_f=4$.

For the hadronic gas model, we include each of the QCD bound state degrees of freedom of the SM with masses up to $\mathcal{O}(T_c)$. This information can be found in \cite{Husdal:2016haj}, for example.  We plot these results up to the critical temperature found in \cref{HadronicAndMIT} using the M.I.T. bag model. 
We see a good agreement between the hadronic gas and lattice results up to this point, which has already been pointed out, for example, in \cite{Borsanyi:2013bia, Bellwied:2015lba, Saikawa:2018rcs}. We likewise see excellent agreement between the lattice results and those of the finite-temperature QCD, down to temperatures as low as $0.5\text{GeV}$. Here, we use \cref{hotqcdpressure}, and set the renormalisation scale $\mu=2\pi T$. 

Using the pressure, we define the trace anomaly
\begin{equation}\label{Trace}
    \Delta_\text{QCD}(T)=T\frac{\text{d}}{\text{d}T}\left\{\frac{p_\text{QCD}(T)}{T^4}\right\}.
\end{equation}
Finally, we can write $\rho(T)$ and $s(T)$, appearing in \cref{gstarEq}, as functions of $p(T)$ and $\Delta(T)$. These can then be estimated as
\begin{equation}
 \begin{split}
    &\rho_\text{QCD}(T)=T^4\left[\Delta_\text{QCD}(T)+\frac{3p_\text{QCD}(T)}{T^4}\right],\\
    &s_{\text{QCD}}(T)=T^3\left[\Delta_\text{QCD}(T)+\frac{4p_\text{QCD}(T)}{T^4}\right].
 \end{split}   
\end{equation}
In the early universe, the QCD PT does not occur in isolation, but in a thermal bath of colourless particles, comprising 17.25 relativistic degrees of freedom. At temperatures greater than $T_{c}$, the energy density of the background plasma is expected to be related to its pressure through $\rho_{\gamma} = 3p_{\gamma}$. For our purposes, it is the total equation of state for the thermal plasma, including both QCD and the remanent relativistic degrees of freedom that is of interest. For $T \sim T_{c}$ and greater, this is
\begin{equation} \label{EOSQCD}
    \omega(T) =  \frac{1}{3 + \frac{T^4\tilde{\Delta}_\text{QCD}(T)}{r_{\gamma}+ \tilde{p}_{\text{QCD}}\left(T\right)}}.
\end{equation}
where the tilde notation denotes a quantity that is normalised to the SB limit of $p_{\text{QCD}}$ in \cref{SBlimit}. The temperature dependence of $\tilde{\Delta}_{\text{QCD}}$ and $\tilde{p}_{\text{QCD}}$ have been measured on the lattice \cite{Borsanyi:2016ksw}, and $r_{\gamma} = g_{*,\gamma}/g_{*,\text{QCD}}$ is the ratio of degrees of freedom within the background plasma to those of quarks and gluons in the UV.



\subsection{A Toy Model} \label{Toy model}

\begin{figure}[ht!]
    \centering
\includegraphics[width=\linewidth]{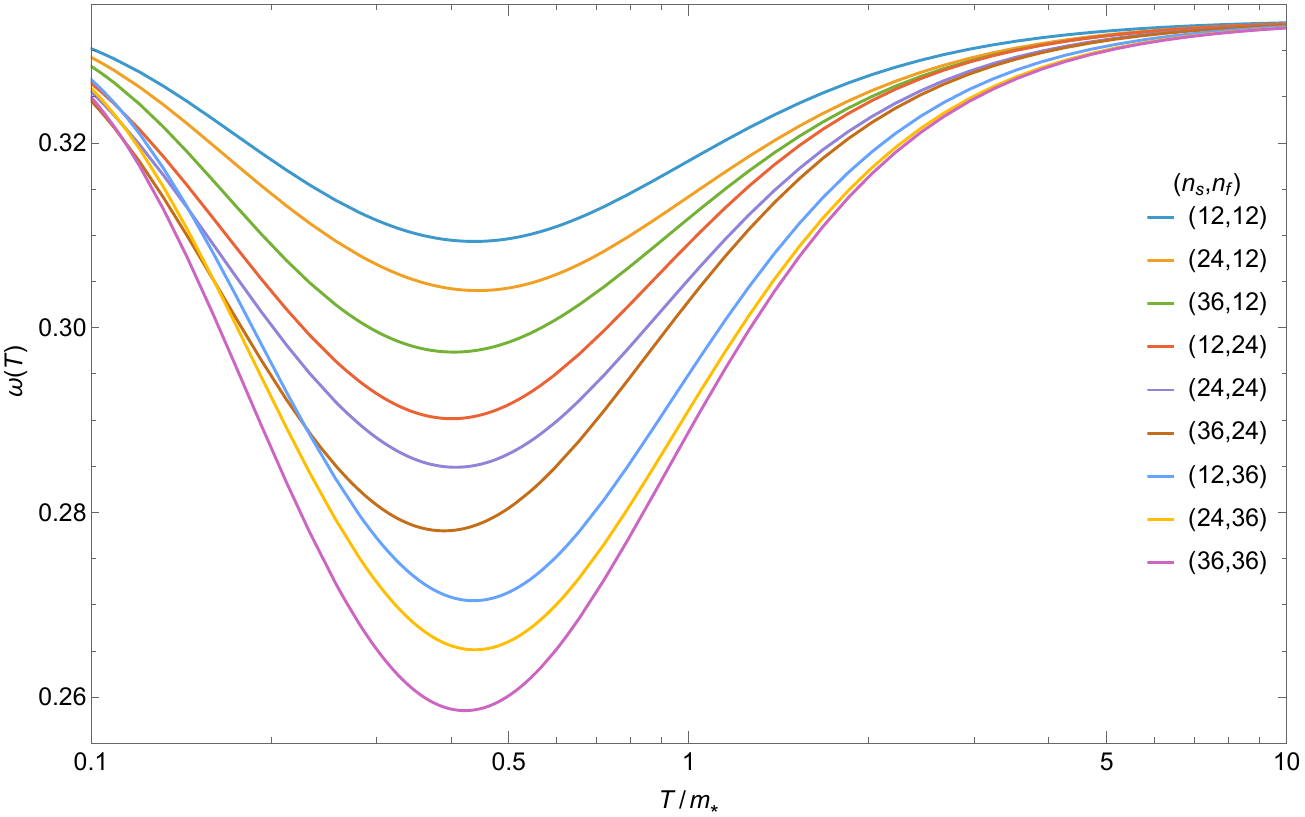}
    \caption{The equation of state for our toy model, as a function of $T/m_*$, where $m_*$ is some input energy scale. We vary the number of scalars and fermions within the model, showing explicitly the sensitivity of the equation of state to the number of degrees of freedom becoming non-relativistic.}\label{ToyLargeG}
\end{figure}

Having established the basics of SM thermodynamics, we now study a toy model for a BSM sector. This will set up our later discussions, in which we investigate motivated BSM models. Specifically, we exemplify some generic features that can be expected from drops in relativistic degrees of freedom, and the imprints left on the equation of state.  Consider the following toy model 
\begin{equation}
    \mathcal{L}_{\text{Toy}}= \mathcal{L}_{\text{SM}}+\mathcal{L}_{\text{BSM}} + \mathcal{L}_{\text{int}}
\end{equation}
where 
\begin{equation}
    \mathcal{L}_{\text{BSM}} = \frac{1}{2}\partial_{\mu}\phi^{a}\partial^{\mu}\phi^{a} - \frac{1}{2}m_{*}\mathcal{O}^{ab}_{S}\phi^{a}\phi^{b} + i\Psi^{i} \slashed{\partial}\Psi^{i} - m_{*}\mathcal{O}^{ij}_{F}\bar{\Psi}^{i}\Psi^{j}
\end{equation}
In addition to the SM, there is a BSM sector characterized by a common mass scale $m_{*}$, with $n_{s}$ real scalar fields and $n_{f}$ Dirac fields. The eigenvalues of the matrices $\mathcal{O}^{ab}_{S}$ and $\mathcal{O}^{ij}_{F}$ determine the physical mass spectrum of the model. The BSM sector is coupled to itself, and to the SM, through the unspecified $\mathcal{L}_{\text{int}}$, keeping it in thermal equilibrium with the plasma of the early universe. For the present discussion, it will be assumed that the couplings within $\mathcal{L}_{\text{int}}$ are perturbative. At $T\sim m_{*}$, the BSM degrees of freedom become non-relativistic, and the equation of state is obtained using \cref{eq:deltaomega}.
Through varying $n_{s}$ and $n_{f}$, with the eigenvalues of $\mathcal{O}^{ab}_{S}$ and $\mathcal{O}^{ij}_{F}$ randomly distributed around 1, we explore the response of the equation of state in \cref{ToyLargeG}. 

From the figure, two interesting features arise.
Firstly, as expected, increasing $n_s$ and $n_f$ leads to larger drops in the equation of state. Notably, we find that $(n_s,n_f)=(24,24)$ gives roughly the same number of degrees of freedom as in the SM. However, as can be seen, the resulting drop in $\omega(T/m_*)$ is only $\approx0.285$. Even with $(n_s,n_f)=(36,36)$, corresponding to 162 effective degrees of freedom, we see that we are unable to perturbatively obtain a drop as large as that of the QCD PT, in which $\omega_\text{min}\approx0.23$. This attests to the effectiveness of non-perturbative strongly interacting PTs in softening the equation of state. 

Less trivially, for large $n_{f}$ and $n_{s}$, the drop becomes less sensitive to the exact distribution of the masses. The argument here follows the central limit theorem \cite{Hald2007,STIGLER2005329}. That is, given a large enough number of particles randomly distributed about some scale, the total distribution will approximate a normal distribution centred at that scale. If we considered a larger window to randomly distribute our masses, we would expect more degrees of freedom to be needed in order to recover this result. On the other hand, if this window was smaller, we would expect a larger drop in $\omega(T)$, for the same number of decoupling degrees of freedom.

\section{Supersymmetry}\label{SUSYsection}

Since its first introduction as a formal workaround to the Coleman-Mandula theorem \cite{Coleman:1967ad,Fayet:1977yc,Farrar:1978xj}, supersymmetry (SUSY) has become the single most studied extension to the SM. Of greatest phenomenological relevance is the application of $\mathcal{N}=1$ SUSY to the SM, known as the Minimal Supersymmetric Standard Model (MSSM) \cite{Haber:1993wf,Csaki:1996ks}. 
In this framework, the SM degrees of freedom are embedded within supermultiplets which are either chiral or gauge, with the spin of the SM particles and their superpartners differing by 1/2. 


Of course, SUSY must not be exact at the current physics scales we can probe, as no sparticles have been detected so far in experiments. Thus, it is a key requirement within the MSSM that SUSY must be broken at some higher scale, conventionally called $M_\text{SUSY}$. The SM is then a low-energy effective theory of a MSSM, obtained by integrating out the heavy supersymmetric degrees of freedom. Consequentially, the running couplings of the SM are, to a good approximation, the same up to $M_\text{SUSY}$. The MSSM has the nice property that, for certain values of $M_\text{SUSY}$, the gauge couplings of the SM run into an intersection point, and thus may possibly unify into a single fundamental interaction at some higher-energy scale, called the grand unified scale, or $M_\text{GUT}$. We discuss this feature of the theory further in \cref{gaugeuniapp}. Unfortunately, the lack of detection of any sparticles in colliders has pushed the experimental bound on $M_\text{SUSY}$ such that a precise intersection is no longer viable. Nonetheless, in the remainder of the section, and following the excellent work of \cite{Martin:1997ns}, we focus on the mass spectrum of the MSSM for the purpose of obtaining the equation of state.

To make this model theoretically viable, there is required to be an extra Higgs doublet. The mass eigenstates associated to these are denoted as $h^0$, and $H^{0}$, $A^{0}$ and $H^{\pm}$, contributing an extra 4 bosonic degrees of freedom. Secondly, the superpartners of the quarks, known as squarks have mass eigenstates denoted by $\tilde{u}_{L}$, $\tilde{c}_{L}$, $\tilde{t}_{1}$, $\tilde{u}_{R}$, $\tilde{c}_{R}$, $\tilde{t}_{2}$, $\tilde{d}_{L}$, $\tilde{s}_{L}$, $\tilde{b}_{1}$, $\tilde{d}_{R}$, $\tilde{s}_{R}$ and $\tilde{b}_{2}$. As each of these come as a particle-antiparticle pair, as well as a triplet of colour, they contribute a total of 72 bosonic degrees of freedom. For the sleptons we have $\tilde{e}_{L}$, $\tilde{\mu}_{L}$, $\tilde{\tau}_{1}$, $\tilde{e}_{R}$, $\tilde{\mu}_{R}$, $\tilde{\tau}_{2}$, $\tilde{\nu}_{e}$, $\tilde{\nu}_{\mu}$ and $\tilde{\nu}_{\tau}$. The degrees of freedom associated to them are the same as the squarks, with the obvious exception of not being in colour triplets. Thus, they contribute 18 bosonic degrees of freedom. The total extra bosonic degrees of freedom in the MSSM is therefore 94.

For the new fermionic degrees of freedom, we get contributions from the Higgs and SM gauge boson superpartners. The mass eigenstates are denoted as neutralinos $\tilde{N}_{1}$ $\tilde{N}_{2}$ $\tilde{N}_{3}$ $\tilde{N}_{4}$  and charginos $\tilde{C}^{\pm}_{1}$ $\tilde{C}^{\pm}_{2}$. These are all chiral fermions with 2 degrees of freedom, and thus contribute a total of 16 additional fermionic degrees of freedom. Finally we have the gluions $\tilde{g}$, of which there are 8 new chiral fermions, thus giving a total contribution of 16 extra fermionic degrees of freedom. Therefore, the total number of new fermionic degrees of freedom is 32. To convert this to effective degrees of freedom there is the $7/8$ conversion factor, which gives 28. Overall, the total number of new effective degrees of freedom beyond the SM in the MSSM is $94+28 =122$.

\subsection{SUSY masses}

Having calculated the degrees of freedom for each of the sparticles in the MSSM, we move on to the more technical task of calculating the corresponding masses. Once completed, we will be in a position to calculate the equation of state for each of our SUSY breaking scenarios. We take the majority of our mass formulae from \cite{Martin:1997ns}, where the derivations are explained in more detail. The majority of these masses depend on quantities which run with renormalisation scale. Hence, in order to determine the masses of each sparticle, we must solve a number of RGEs. For completeness, we include each of the SUSY 1-loop RGEs that we solve in \cref{SUSYRGEs}. 

\subsubsection{Higgs bosons}

We start with the masses of the Higgs bosons, which are given by the following
\begin{equation}
 \begin{split}
 &m_{A^0}^2=\frac{2b}{\text{sin}2\beta},\\
 &m_{h^0,H^0}^2=\frac{1}{2}\left(m_{A^0}^2+m_Z^2\mp\sqrt{(m_{A^0}^2-m_Z^2)^2+4m_Z^2m_{A^0}^2\text{sin}2\beta}\right),\\
 &m_{H^\pm}^2=m_{A^0}^2+m_W^2,
 \end{split}
\end{equation}
where we take $m_Z=91.188\text{GeV}$, $m_W=80.377\text{GeV}$ \cite{10.1093/ptep/ptaa104} and $b$ is a real and positive constant appearing in the scalar Higgs potential. Specifically, we have
\begin{equation}
    b=\frac{\text{sin}2\beta}{2}(2|\mu|^2+m_{H_u}^2+m_{H_d}^2),
\end{equation}
where $H_u=(H_u^+,H_u^0)$ and $H_d=(H_d^0,H_d^-)$ define the two complex Higgs doublets and $\mu$ is the Higgsino mass parameter. Setting $H_u^+=H_d^-=0$, we arrive at one of the free parameters of the MSSM, $\tan\beta$, which is given by the ratio of the two vacuum expectation values
\begin{equation}
    \tan\beta\equiv\frac{\langle H_u^0\rangle}{\langle H_d^0 \rangle}.
\end{equation}
A particularly interesting feature of the MSSM Higgs sector is that the parameter $\beta$ which, as well as the $Z$-boson mass, sets a tree-level bound on the mass of $h_0$ \cite{Inoue:1982pi,Flores:1982pr}
\begin{equation}
    m_{h_0}<m_Z|\cos(2\beta)|\leq m_Z.
\end{equation}
The tree-level Higgs mass receives large radiative corrections, which are required to recover the observed value \cite{Draper:2016pys}. The remainder of the Higgs $A^0,H^0,H^\pm$ have very similar masses to each other in each of our scenarios. The running of which depends largely on the running of $H_d$.

\subsubsection{Charginos and neutralinos}

The six mass eignestates of the neutralinos and charginos are obtained due to mixing between the Higgs bosons and gauginos. Four of these eigenstates have neutral charge, named neutralinos $\tilde{N}_i$($i$=1,2,3,4). These sparticles are the combination of neutral Higgsinos ($\tilde{H}_u^0$,  $\tilde{H}_d^0$) and neutral gauginos ($\tilde{B}, \tilde{W}^0$). The remaining two carry $\pm1$ charge, named charginos $\tilde{C}^\pm_i(i=1,2)$. These come from the combination of charged Higgsinos ($\tilde{H}_u^+$ and  $\tilde{H}_d^-$) and winos ($\tilde{W}^+$ and $\tilde{W}^-$).

Diagonalizing the corresponding mass matrices, we obtain the mass formulae for the neutralinos and charginos as follows
\begin{equation}
    \begin{split}
        &m_{\tilde{N}_1}= M_1-\frac{m_Z^2s_W^2(M_1+\mu\text{sin}2\beta)}{\mu^2-M_1^2}+...,\\
        &m_{\tilde{N}_2}= M_2-\frac{m_W^2(M_2+\mu\text{sin}2\beta)}{\mu^2-M_2^2}+...,\\
        &m_{\tilde{N}_3}= |\mu|+\frac{m_Z^2(I-\text{sin}2\beta)(\mu+M_1c_W^2+M_2s_W^2)}{2(\mu+M_1)(\mu+M_2)}+...,\\
        &m_{\tilde{N}_4}= |\mu|+\frac{m_Z^2(I+\text{sin}2\beta)(\mu-M_1c_W^2-M_2s_W^2)}{2(\mu-M_1)(\mu-M_2)}+...,\\
        &m_{\tilde{C}^\pm_1}= M_2-\frac{m_W^2(M_2+\mu\text{sin}2\beta)}{\mu^2-M_2^2}+...,\\
        &m_{\tilde{C}^\pm_2}=|\mu|+\frac{Im_W^2(\mu+M_2\text{sin}2\beta)}{\mu^2-M_2^2}+...,
    \end{split}
\end{equation}
where $M_1$ and $M_2$ are the U(1) and SU(2) electroweak gaugino mass parameters respectively and $I$ is the sign of $\mu$. 

We find that the runnings of each of these masses are dominated by the leading-order term. Thus, there are roughly 3 masses to be considered in this sector, governed by the gaugino parameters $M_1,M_2$ and $\mu$. Furthermore, we find that $\mu$ hardly runs, and thus it can be well approximated by it's value given at the GUT scale. Then, it is interesting to note that in the case of GUT-scale SUSY breaking, the boundary conditions we use set the scale of $\mu$ to be $10^{-3}$ times that of the scalars. This will be discussed further in the following subsection.

\subsubsection{Squarks and sleptons}
 
For the first-family squark and slepton masses, we use the following
\begin{equation}\label{squarkslepton}
    \begin{split}
        &m_{\tilde{d}_L}^2=m_0^2+K_3+K_2+\frac{1}{36}K_1+\Delta_{\tilde{d}_L},\\
        &m_{\tilde{u}_L}^2=m_0^2+K_3+K_2+\frac{1}{36}K_1+\Delta_{\tilde{u}_L} ,\\
        &m_{\tilde{u}_R}^2=m_0^2+K_3 \ \, \ \ \ \ \ \ +\frac{4}{9}K_1+\Delta_{\tilde{u}_R},\\
        &m_{\tilde{d}_R}^2=m_0^2+K_3 \ \, \ \ \ \ \ \ +\frac{4}{9}K_1+\Delta_{\tilde{d}_R},\\
        &m_{\tilde{e}_L}^2=m_0^2 \ \; \ \ \ \ \ \ +K_2+\frac{1}{4}K_1+\Delta_{\tilde{e}_L},\\ 
        &m_{\tilde{\nu}_e}^2=m_0^2 \ \, \; \ \ \ \ \ \ +K_2+\frac{1}{4}K_1+\Delta_{\tilde{\nu}_e},\\
        &m_{\tilde{e}_R}^2=m_0^2 \ \, \, \, \ \ \ \ \ \ \ \ \ \ \ \ \ +K_1+\Delta_{\tilde{e}_R},
    \end{split}
\end{equation}
where $K_a$ terms arise from RG running proportional to the gaugino masses, given at one-loop by
\begin{equation}
    K_a(Q)= \begin{Bmatrix}
              3/5  \\
              3/4  \\
              4/3 
              \end{Bmatrix}\times\frac{2}{\pi}\int_{\text{ln}Q}^{\text{ln}Q_0}\text{d}t \alpha_a(t)\vert M_a(t)\vert^2 \quad \quad (a=1,2,3),
\end{equation}
where $\alpha_a$ are the gauge, and $M_a$ the gaugino runnings described in \cref{gaugeuniapp}. There is also a contribution to each squark and slepton mass arising from D-term quartic interactions, $\Delta_\phi$. These are given by
\begin{equation}\label{ScalarSquarkInteraction}
    \Delta_{\phi}=(T_{3\phi}-Q_{\phi}\text{sin}^2\theta_{W})\text{cos}(2\beta)m_{Z}^2,
\end{equation}
where $T_{3\phi}$ and $Q_\phi$ are the third component of weak isospin and the electric charge of the left-handed chiral supermultiplet to which $\phi$ belongs. We show each of these contributions in \cref{DtermTable}.

\begin{table}[h!]
\centering
 \begin{tabular}{||c c c c||} 
 \hline
 $\phi$ & $T_{3\phi}$ & $Q_\phi$ & $\Delta_\phi$ \\ [0.5ex] 
 \hline
 \rule{0pt}{2.5ex} $\tilde{d}_L$ & -$\frac{1}{2}$ & -$\frac{1}{3}$ & $(-\frac{1}{2}+\frac{1}{3}S_w^2)C_{\beta z}m_Z^2$ \\ 
\rule{0pt}{2.5ex} $\tilde{u}_L$ & $\frac{1}{2}$ & $\frac{1}{3}$ & $(\frac{1}{2}-\frac{1}{3}S_w^2)C_{\beta z}m_Z^2$ \\
\rule{0pt}{2.5ex} $\tilde{u}_R$ & 0 & $\frac{2}{3}$ & $-\frac{2}{3}S_w^2C_{\beta z}m_Z^2$ \\
\rule{0pt}{2.5ex}$\tilde{d}_R$ & 0 & -$\frac{1}{3}$ & $\frac{1}{3}S_w^2C_{\beta z}m_Z^2$ \\
\rule{0pt}{2.5ex} $\tilde{e}_L$ & -$\frac{1}{2}$ & -1 & $(-\frac{1}{2}+
S_w^2)C_{\beta z}m_Z^2$ \\
\rule{0pt}{2.5ex} $\tilde{\nu}_e$ & $\frac{1}{2}$ & 0 & $\frac{1}{2}S_w^2C_{\beta z}m_Z^2$ \\
\rule{0pt}{2.5ex} $\tilde{e}_R$ & 0 & -1 & $S_w^2C_{\beta z}m_Z^2$ \\[1ex] 
 \hline
 \end{tabular}
 \caption{D-term quartic interaction contributions to squark and slepton masses. Here, $C_{\beta z}=\text{cos}(2\beta)$ and $S_w=\text{sin}(\theta_w)$ is the Weinberg angle. }
 \label{DtermTable}
\end{table}

The Yukawa couplings of each of the first- and second-family squarks and sleptons are negligible, which leads to 7 almost degenerate unmixed pairs. Thus, we set: $m_{\tilde{e}_R}=m_{\tilde{\mu}_R}$, $m_{\tilde{\nu}_e}=m_{\tilde{\nu}_\mu}$, $m_{\tilde{e}_L}=m_{\tilde{\mu}_L}$, $m_{\tilde{u}_R}=m_{\tilde{c}_R}$, $m_{\tilde{d}_R}=m_{\tilde{s}_R}$, $m_{\tilde{u}_L}=m_{\tilde{c}_L}$ and $m_{\tilde{d}_L}=m_{\tilde{s}_L}$. On the other hand, there can be significant mixing between 
stop, sbottom and stau sparticles. The mass matrices are given as follows
\begin{equation}
    \begin{split}
      & m_{\tilde{t}}^2=  \begin{pmatrix}
m_{Q_3}^2+m_t^2+\Delta_{\tilde{u}_L} & v(a_t^*\sin\beta-\mu y_t\cos\beta) \\
 v(a_t\sin\beta-\mu^*y_t\cos\beta) & m_{\bar{u}_3}^2+m_t^2+\Delta_{\bar{u}_R}  
\end{pmatrix},\\
& m_{\tilde{b}}^2=  \begin{pmatrix}
m_{Q_3}^2+\Delta_{\tilde{d}_L} & v(a_b^*\cos\beta-\mu y_b\sin\beta) \\
v(a_b\cos\beta-\mu^*y_b\sin\beta) & m_{\bar{d}_3}^2+\Delta_{\tilde{d}_R}  
\end{pmatrix},\\
&m_{\tilde{\tau}}^2=  \begin{pmatrix}
m_{L_3}^2+\Delta_{\tilde{e}_L} & v(a_\tau^*\cos\beta-\mu y_\tau\sin\beta)\\
v(a_\tau\cos\beta-\mu^*y_\tau\sin\beta) & m_{\bar{e}_3}^2+\Delta_{\tilde{e}_R}  
\end{pmatrix},
    \end{split}
\end{equation}
where $v=\langle H\rangle$ is the vacuum expectation value of the SM Higgs ($\approx174$GeV), $m_t$ is the top quark mass ($\approx173$GeV) and we define the third-family squark and slepton masses ($m_{Q_3}, m_{\bar{u}_3},m_{\bar{d}_3},m_{L_3},m_{\bar{e}_3}$), the soft terms ($a_{t,b,\tau}$) and yukawa couplings ($y_{t,b,\tau}$) in \cref{SUSYRGEs}. We then obtain the mass eigenstates of the stop, sbottom and stau particles by diagonalising these mass matrices.

For the squarks and sleptons, it is relatively straightforward to classify the running masses. Beginning with the sleptons, we find that there are mainly two masses in each SUSY breaking scenario. These masses may be distinguished as masses proportional to $K_1$ and masses proportional to $K_2$. Because we typically expect $K_3\gg K_2\gg K_1$, the slepton running masses proportional to $K_{1/2}$ closely resemble the running of $K_{1/2}$. However, as you increase the SUSY breaking scale, the size of the gauge couplings at the scales of interest lead to deviations from this inequality.

Likewise, for the squark masses, we find a largely degenerate mass distribution. This is due to the fact that each of the running masses are proportional to $K_3$, meaning they are largely insensitive to the remainder of the terms in \cref{squarkslepton}. This turns out to be important in later discussions, due to the large number of squark degrees of freedom.

\begin{figure}[ht] 

  \begin{minipage}[b]{0.5\linewidth}
    \centering
    \includegraphics[width=0.95\linewidth]{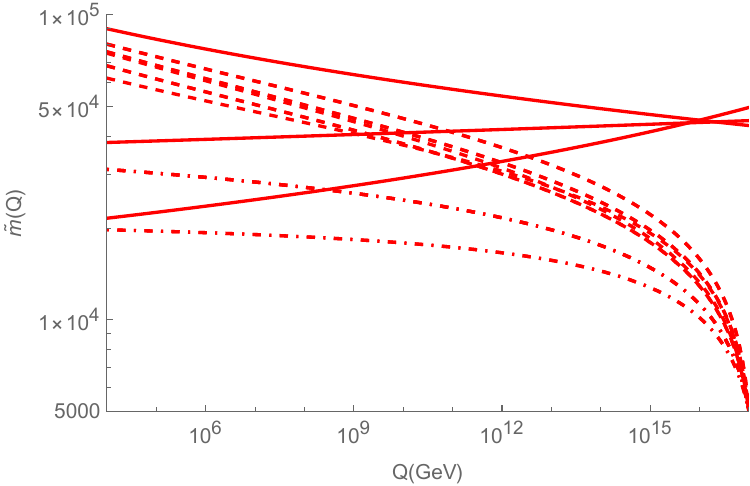} 
  \end{minipage}
  \begin{minipage}[b]{0.5\linewidth}
    \centering
    \includegraphics[width=0.95\linewidth]{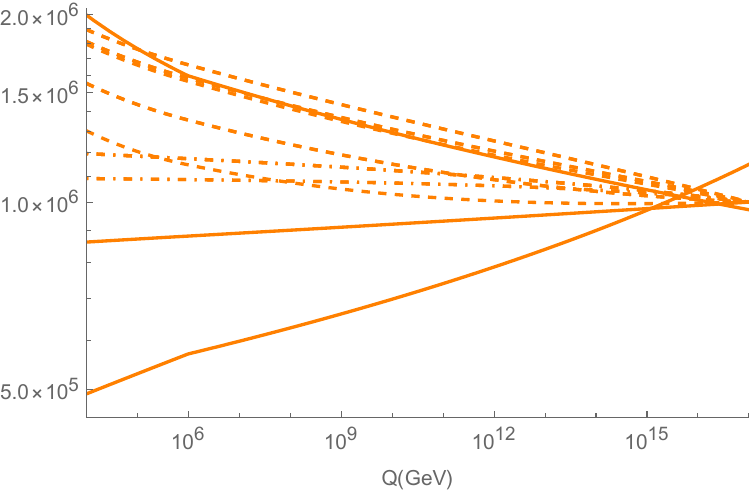} 
  \end{minipage} 
  \begin{minipage}[b]{0.5\linewidth}
    \centering
    \includegraphics[width=0.95\linewidth]{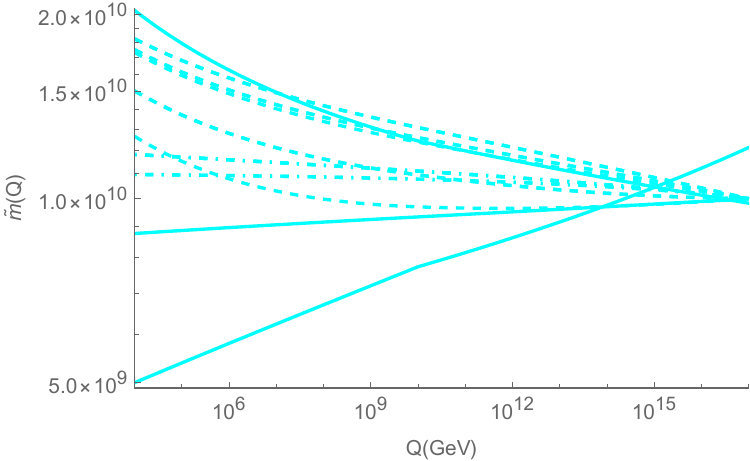} 
  \end{minipage}
  \begin{minipage}[b]{0.5\linewidth}
    \centering
    \includegraphics[width=0.95\linewidth]{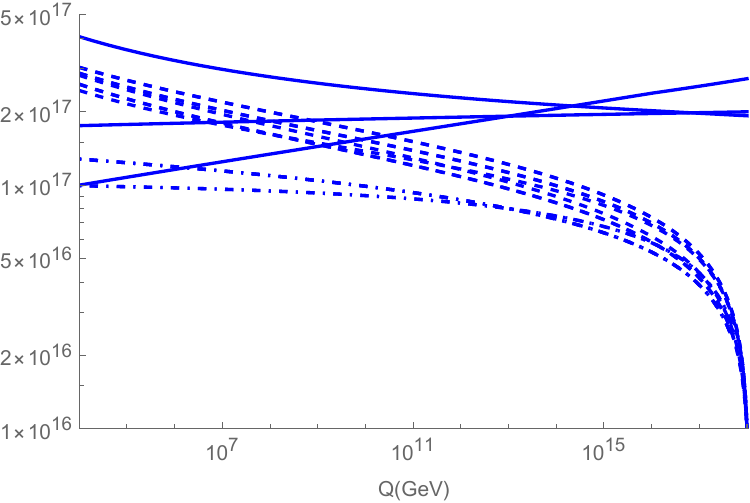} 
  \end{minipage} 
  \caption{We show the runnings of several sparticle masses as functions of renormalisation scale, for each of the SUSY breaking scenarios we consider: $Top\hspace{1mm}Left:$ TeV-scale, $Top\hspace{1mm}Right:$ PeV-scale, $Bottom\hspace{1mm}Left:$ Intermediate-scale and $Bottom\hspace{1mm}right:$ GUT-scale. We plot the running masses of the gauginos in solid, squarks in dashed and sleptons in dotdashed. As can be seen from the figure, in each of the scenarios we consider, there is not significant running.} \label{RunningMinipage} 
\end{figure}

The squark masses, as well as the gluino mass, are discussed in the following subsection. For the remainder of the MSSM sparticles, we take the tree-level pole masses. This is equivalent to solving the following equation
\begin{equation}\label{PoleMassEq}
    \tilde{m}(Q)=Q,
\end{equation}
where $\tilde{m}(Q)$ is the running mass. In \cref{RunningMinipage} we show the runnings of each of the gaugino parameters, which roughly correspond to several chargino and neutralino running masses, as well as the gluino. Furthermore, we plot each of the squark and slepton running masses, including the stop, sbottom and stau. Throughout this work, motivated by the boundary conditions provided in \cite{Ellis:2017erg}, we consider several SUSY breaking scales: $5\times10^3$GeV (TeV scale), $10^6$GeV (PeV scale), $10^{10}$GeV (Intermediate scale) and $10^{16}$GeV (GUT scale).



\subsection{Mass corrections}\label{MassCorrectionsSection}

As discussed in the previous section, in order to obtain the pole masses of the majority of the sparticles, it is usually enough to solve \cref{PoleMassEq}. Whilst approximate, it is not expected for the corrections to be large for non-strongly interacting sparticles. Then, for the scope of this work we believe it is justified to not include these corrections. On the other hand, due to the fact that the gluino, as well as the squarks, interact strongly, they receive the largest mass corrections. Thankfully, these are the simplest of the mass renormalisations. For the gluino one-loop corrected mass we use the results of \cite{Pierce:1996zz}
\begin{equation}
    m_{\tilde{q}}=M_3(Q)\left[1-\left(\frac{\Delta M_3}{M_3}\right)^{g\tilde{g}}-\left(\frac{\Delta M_3}{M_3}\right)^{q\tilde{q}}\right]^{-1},
\end{equation}
where
\begin{equation}
    \left(\frac{\Delta M_3}{M_3}\right)^{g\tilde{g}}=\frac{3\alpha_3}{4\pi}\left[6\ln \left(\frac{Q}{M_3}\right)+5\right],
\end{equation}
and
\begin{equation*}
    \left(\frac{\Delta M_3}{M_3}\right)^{q\tilde{q}}=-\frac{3\alpha_3}{\pi}\left(\ln\left(\frac{Q}{M}\right)+1-\frac{1}{2x}\left[1+\frac{(x-1)^2}{x}\ln |x-1|\right]+\frac{1}{2}\theta(x-1)\ln x\right),
\end{equation*}
where $M=\max (M_3^2,M_{Q_1}^2)$ and $x=M_3^2/M_{Q_1}^2$. 

For the squark one-loop corrected masses we again follow \cite{Pierce:1996zz}. Focussing on the dominant QCD corrections, it is found
\begin{equation}
    m_{\tilde{q}}^2=\hat{m}_{\tilde{q}}^2(Q)\left[1+\left(\frac{\Delta m_{\tilde{q}}^2}{m_{\tilde{q}}^2}\right)\right],
\end{equation}
where
\begin{equation}
    \left(\frac{\Delta m_{\tilde{q}}^2}{m_{\tilde{q}}^2}\right)=\frac{2\alpha_3}{3\pi}\left[1+3x+(x-1)^2\ln|x-1|-x^2\ln x+4x\ln \left(\frac{Q}{m_{\tilde{q}}}\right) \right],
\end{equation}
where $x=m_{\tilde{g}}^2/m_{\tilde{q}}^2$.

In scenarios where there is a hierarchy among superpartner masses, these mass corrections can become large \cite{Martin:2006ub}. We find the ratio between the average squark mass and gluino mass to be: 0.799, 0.929 and 0.961 for TeV, PeV and intermediate scale SUSY breaking respectively. Thus, we expect relatively small 1-loop corrections to the gluino mass. Specifically, we find the mass corrections to be: $2.89\%$, $3.57\%$ and $2.37\%$, which is clearly in good agreement with expectations. For GUT-scale SUSY breaking we find the biggest difference between gluino and squark masses. This appears to be a result of the boundary conditions described in \cref{SUSYRGEs}. However, we find the corrections to the gluino mass to only be $3.46\%$. Then, we take the 1-loop masses to produce our final mass spectra.


\subsection{SUSY equation of state}

Upon calculating a given SUSY mass spectrum, we are able to calculate the corresponding equation of state. Using the degrees of freedom outlined in the introduction to the section, this can be done purely statistically, in a similar fashion to \cref{omegasum}. For each of the SUSY breaking scenarios considered, we find a nearly degenerate mass spectrum. That is, each of the sparticles have a similar mass. As alluded to in \cref{Toy model}, this has significant implications; if a given number of degrees of freedom decouple at the same scale, this will lead to a bigger reduction in the equation of state. 

\begin{figure}[ht!]
    \centering
\includegraphics[width=0.95\linewidth]{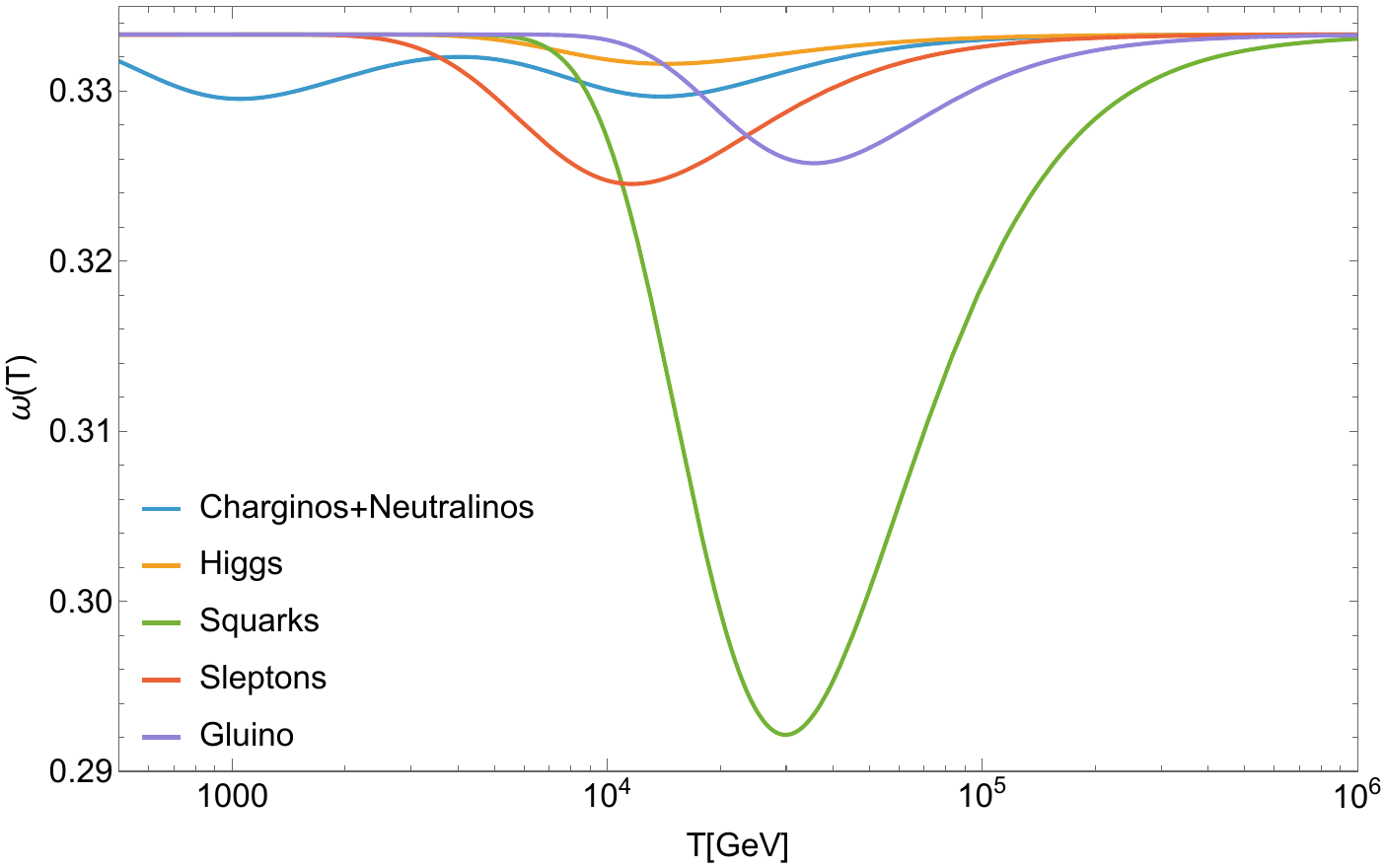}
    \caption{The equation of state for TeV-scale SUSY breaking. We split this into each sparticles contribution, with the largest contribution coming from the squarks.}\label{TeVEoSPlot}
\end{figure}

We show the contributions of each sparticle species to the equation of state in \cref{TeVEoSPlot}, for the TeV-scale SUSY breaking scenario. From the figure, it is clear to see that the squarks contribute most significantly to the drop in equation of state. This is to be expected, as the they comprise over half of the MSSM degrees of freedom. To obtain the overall equation of state, we simply sum over each of the different sparticles. Therefore, we can see from the figure that it is still important to include each of the species when considering the overall change in $\omega(T)$.

\begin{figure}[ht!]
    \centering
\includegraphics[width=\linewidth]{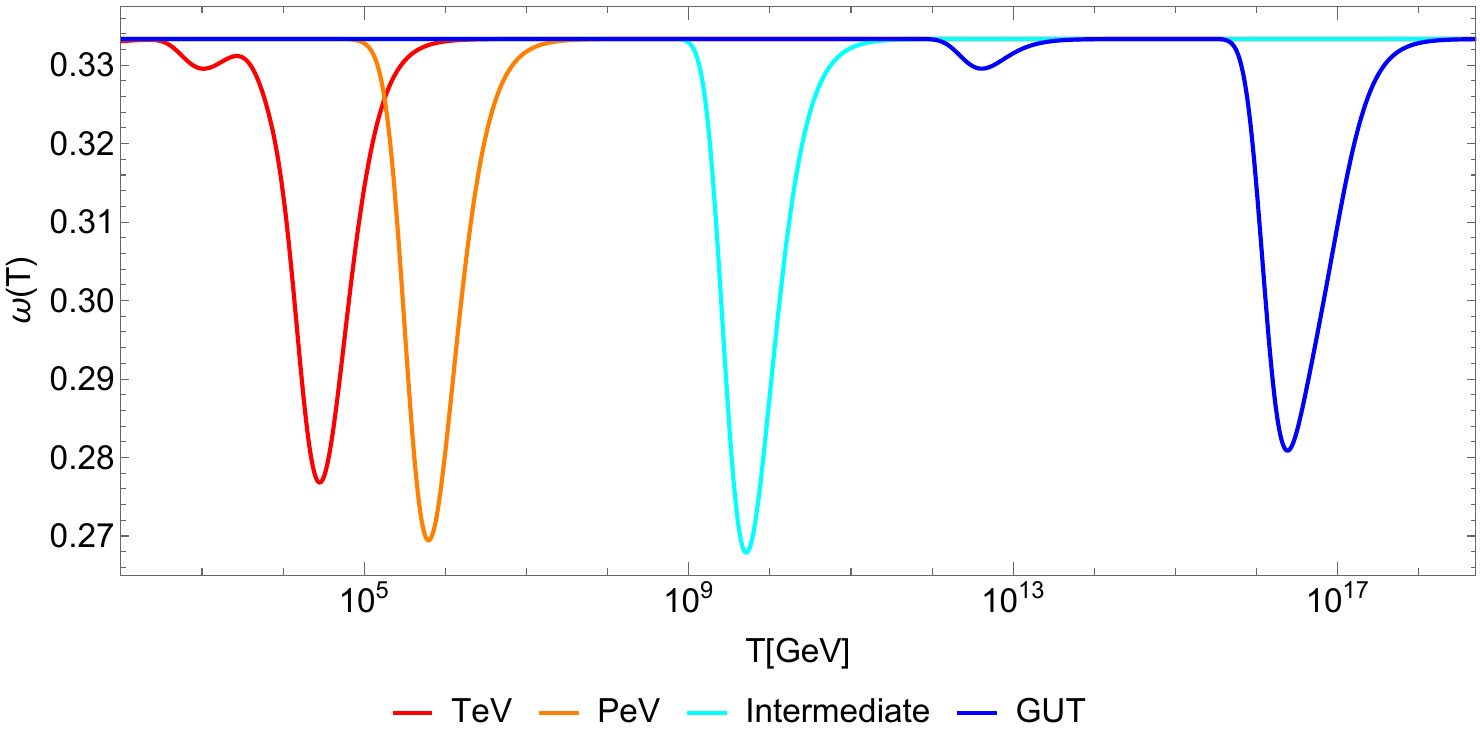}
    \caption{The equation of state for each of the SUSY breaking scenarios we considered. We found most of the sparticle mass spectra to be close to degenerate, leading to sharper drops.}\label{JointSUSYPlot}
\end{figure}

We show the overall equation of state for the four different SUSY breaking scenarios in \cref{JointSUSYPlot}. Here, we see some interesting features. Firstly, in each scenario, the largest drop in the equation of state occurs close to the corresponding SUSY breaking scale. This comes as no surprise; the boundary conditions of our scenarios \cite{Ellis:2017erg} are each close to that of the SUSY breaking. Then, unless a given quantity runs by more than several orders of magnitude, we should be able to safely approximate $\tilde{m}(Q)=Q\approx M_{\text{SUSY}}$. Assuming the MSSM, this has the exciting implication that, if cosmological evidence of softening in the equation of state at some scale was found, we could associate this scale to $M_\text{SUSY}$.

Secondly, for both the TeV and GUT scale SUSY breaking $\mu$ is given as $0.5m_0$ and $10^{-3}m_0$ respectively, where $m_0$ is the mass of the scalars. Then, we can see the smaller drop corresponding to the charginos and neutralinos, with masses proportional to $\mu$, decoupling on these slightly different scales. Of course, each of these models has the same number of degrees of freedom, therefore the largest drop in $\omega(T)$ occurs in the other two scenarios, where the mass spectra are more degenerate. 

Finally, we remind the reader that this work has focused on the \say{minimal} model for simplicity. However, we are motivated by maximising the additional degrees of freedom, so it is exciting that we are able to obtain such large drops in the equation of state in, more or less, the worst-case scenario. We expect any extension to the MSSM predicting additional degrees of freedom to produce even softer equations of state. Furthermore, we have only considered the case of one superfield, $\mathcal{N}=1$. This does not have to be the case at all \cite{Tachikawa:2013kta}, and with more (super)fields, there would be more field content, so one should expect more degrees of freedom. We leave this for future study.

\section{Composite Higgs}\label{CHsection}

Composite Higgs models were originally introduced as a compelling solution to protecting the higgs mass from extremely large radiative corrections \cite{KAPLAN1984183, KAPLAN1984187}. In this scenario, rather than being elementary, the Higgs is assumed to be the lightest bound state of a new strongly interacting sector. Additionally, if this sector possesses an approximate global symmetry that is spontaneously broken, then composite pseudo-Nambu Goldstone bosons (pNGBs) emerge, making some of the bound states parametrically lighter than others \cite{Panico_2016}. In modern incarnations, the symmetry breaking pattern is chosen such that the Higgs can be identified as a subset of the pNGBs, helping to explain why it is significantly lighter than the other bound states of this new composite sector, which are yet to be discovered \cite{Agashe_2005, Gripaios_2009}.

Microscopically, CHMs are expected to originate from gauge theories involving fermions \cite{Ferretti_2014, Cacciapaglia_2014, Agugliaro_2017, Dong_2021}, and therefore large drops could be expected at around the confinement scale. Firstly, we demonstrate the capacity of CHMs to realise large drops in relativistic degrees of freedom explicitly. Secondly, we investigate the equation of state for a particular class of UV models in greater detail. By applying the methodologies detailed in \cref{QCD phase transition}, we will obtain an estimate of both the critical temperature, along with the evolution of the equation of state with respect to temperature. To reiterate, our aim is to give a qualitative description of $\omega(T)$ within these models. This will help to better understand the cases in which CHMs are interesting in the context of cosmology sensitive to the equation of state. Additionally, our hope is to further motivate lattice simulations of the thermodynamics of these particular models, such that a more precise understanding may be obtained.  

\subsection{UV degrees of freedom}

At energies significantly higher than the confinement scale $m_{\rho}$, the relevant degrees of freedom of the composite sector are assumed to be fundamental fermions $\Psi$, transforming in irreducible representations of the confining hypercolour gauge group $\text{G}_{HC}$, along with the associated gauge fields $A_{\mu}$ \cite{Ferretti_2014, Cacciapaglia_2014, Agugliaro_2017, Dong_2021}. The total number of degrees of freedom will depend both upon the gauge group, the number of $\Psi$ denoted $N_{f}$, and their associated representations under $\text{G}_{HC}$. Upon confinement, these will no longer be the relevant degrees of freedom. Instead, composite resonance states with masses $\sim m_{\rho}$ will emerge, along with parametrically lighter pNGB states of $\text{G}_{F}/\text{H}_{F}$ flavour symmetry breaking. As the temperature of the plasma drops further, these states will promptly become non-relativistic, which eventually, if the composite sector was in isolation, would evolve the equation of state towards that of non-relativistic matter.

The behaviour of the equation of state within an isolated CHM is therefore expected to be very similar to that of QCD in isolation. The distinguishing feature between the two scenarios comes from the relativistic degrees of freedom external to these strong sectors within the plasma. As shown in \cref{EOSQCD}, the large drop in the QCD equation of state may be mainly traced to the ratio $r_{\gamma}$ being small. In other words, the QCD degrees of freedom dominate over the relativistic degrees of freedom leftover in the plasma. For CHM, the left over degrees of freedom are the entire SM. Consequentially, for a composite Higgs model to modify the equation of state in a significant way, there must be a larger number of fundamental degrees of freedom associated to it.

\begin{figure}[ht!]
    \centering
\includegraphics[width=0.85\linewidth]{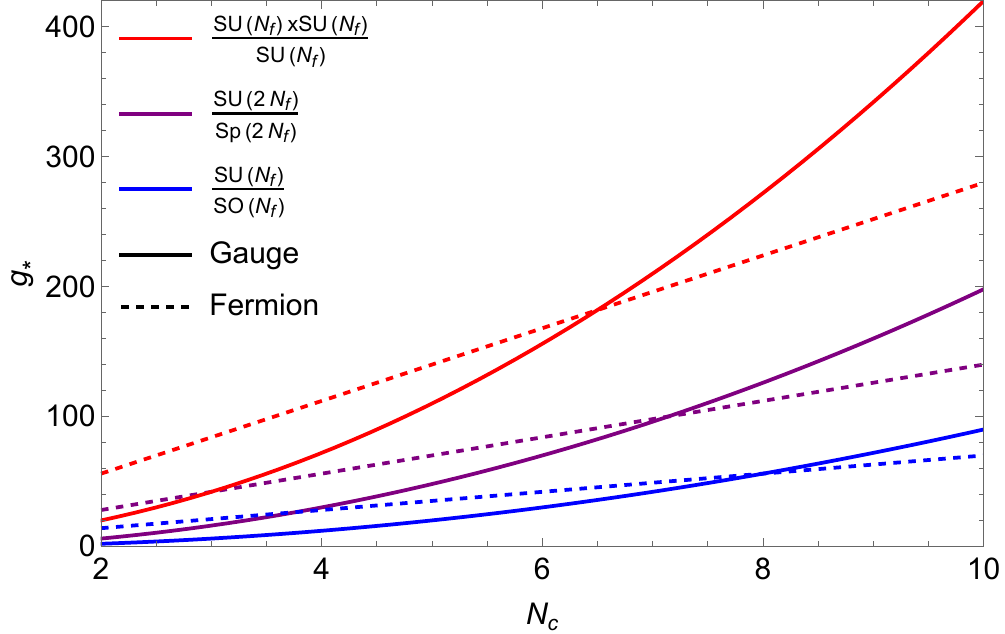}
    \caption{We show the degrees of freedom for each of the minimal composite Higgs symmetry breaking patterns, as a function of $N_c$. We split these degrees of freedom into gauge (solid) and fermion (dashed).}\label{gstar(Nc)}
\end{figure}

By using the minimal UV CHMs given in \cite{Agugliaro_2017} we exemplify the capacity for CHMs to readily contain enough degrees of freedom to dominate over those of the SM. A summary of each model is given in \cref{CH table}. All models are quadratically sensitive to the number of colours $N_{c}$, due to the presence of gauge fields, along with a linear sensitivity from fermions. Even moderate values of $N_{c}$ are capable of containing very substantial numbers of fundamental degrees of freedom, which we show in \cref{gstar(Nc)}. For example, in the $\text{SU}(N_{c})$ model, for $N_{c}>5$, $r_\text{SM}<1$. As $N_{c}$ is essentially a free parameter for our purposes, there are a wide class of models that will modify the equation of state in an interesting way for the present discussion.


\begin{table}[h]
    \centering
    \begin{tabular}{|p{3.5cm}|c|c|c|}
        \hline
        $\text{G}_{\text{HC}}$ & $\text{SU}(N_c)$ & $\text{Sp}(2N_c)$ & $\text{SO}(N_c)$ \\ \hline
        Rep. of fermions & $(\mathbf{N_c},\mathbf{N_f})$, Dirac & $(\mathbf{2N_c},\mathbf{2N_f})$, Weyl & $(\mathbf{N_c},\mathbf{N_f})$, Weyl \\ \hline
        $\text{G}_{F}/\text{H}_{F}$ breaking & $\frac{\text{SU}(N_f) \times \text{SU}(N_f)}{\text{SU}(N_f)}$ & $\frac{\text{SU}(2N_f)}{\text{Sp}(2N_f)}$ & $\frac{\text{SU}(N_f)}{\text{SO}(N_f)}$ \\ \hline
        Minimal $N_{f}$ & 4 &  2 & 5 \\ \hline
        Gauge field d.o.f. & $2(N_c^2-1)$ & $2N_c(2N_c+1)$ & $N_c (N_c-1)$ \\ \hline

        Fermion d.o.f. & $\frac{7}{8} \times 4 N_c N_f$ & $\frac{7}{8} \times 8 N_c N_f$ & $\frac{7}{8} \times 2 N_c N_f$  \\ \hline
        
        pNGB d.o.f. & $N_f^2-1$ & $2N_f^2 - N_f - 1$ & $ \frac{N_f^2}{2}+ \frac{N_f}{2} - 1$ \\ \hline
        
    \end{tabular}
    \caption{We count, for several different symmetry breaking patterns, the effective number of degrees of freedom. As can be seen, this depends on both $N_c$ and $N_f$.}
     \label{CH table}
\end{table}

\subsection{Critical temperature}

The critical temperature for CHMs will be model dependent, since it relies on the precise mass spectrum of the associated bound states. For the present purposes, we are interested in an estimate of the critical temperature, obtained by applying the methodology developed in \cref{HadronicAndMIT}. Similar to QCD, the full spectrum of resonances that emerge from a UV composite Higgs model is determined from group theoretic considerations. Furthermore, the masses of such states are expected to cluster around the scale $m_{\rho}$. It is common to make the one-scale-one-coupling assumption (1S1C) \cite{Panico_2016,Liu_2016, Glioti_2025}, in which these resonances interact with each other through a single coupling, $g_{\text{CS}}$. From large-$N_{c}$ considerations, first proposed in \cite{tHooft:1973alw,Witten:1979kh}, the coupling is expected to be related to $N_{c}$ as
\begin{equation}
    g_{\text{CS}} \sim \frac{4\pi}{\sqrt{N_{c}}}.
\end{equation}
From \cref{bagratio} the contributions from more massive states are exponentially suppressed. Therefore, to calculate the critical temperature, it is sufficient to consider only the lightest resonances within the spectrum, namely the pNGBs. For the minimal model with gauge group $\text{SU}(N_{c})$, besides the Higgs, 11 pNGB's exist. Given the 1S1C assumption, from dimensional analysis the decay constant for the pNGBs will be
\begin{equation}
    f_{\text{CS}} \sim \frac{m_{\rho}}{g_{\text{CS}}}.
\end{equation}
In order to generate a Higgs mass of the correct size, a certain amount of fine-tuning within the Higgs potential is required \cite{Panico_2013, Banerjee_2017, Murnane_2019}, the size of which is quantified by the parameter $\xi = \frac{v^{2}}{f_{\text{CS}}^{2}}$, and thus models with a lower $f_{\text{CS}}$ are deemed more compatible with naturalness expectations. Provided that no additional fine-tuning is present, it is expected that the masses of the remaining pNGBs should cluster around the scale $m_{\Pi}$, which we estimate to be
\begin{equation}
     m_{\Pi} \sim \frac{f_{\text{CS}}}{v}m_{h}
\end{equation}
where $m_{h}$ is the Higgs mass. Thus, once $N_{c}$ and $m_{\rho}$ are specified, we simulate the spectrum of pNGB masses by taking a random distribution of values, tightly centred around this scale. This schematically resembles the mass spectrum of pNGBs extracted from lattice data \cite{Appelquist_2021}. Again, to estimate the radius of the composite states, we use dimensional analysis to obtain $R_{i} \sim \frac{1}{m_{i}}$. Our conclusions are that the estimated critical temperature is consistently slightly lower than the typical pNGB mass $m_{\Pi}$, and thus the two scales are closely related. 

\subsection{Composite Higgs equation of state}

Our aim is to now extrapolate available lattice data to classify the possible evolution of the equation of state for CHMs based on a minimal SU($N_{c}$) gauge theory with $N_{f}=4$. Analogously to \cref{EOSQCD}, the temperature dependence of the equation of state for the universe is
 \begin{equation}\label{CHeos}
    \omega(T) = \frac{1}{3+ \frac{T^4\tilde{\Delta}_\text{CS}(T)}{r_{\text{SM}}+ \tilde{p}_{\text{CS}}\left(T\right)}},
\end{equation}
where $r_{\text{SM}}=g_{*,SM}/g_{*,CS}$. 

At present, lattice data for $\tilde{\Delta}_\text{CS}(T)$ and $\tilde{p}_{\text{CS}}\left(T\right)$ is not available for $N_{c}>3$, however, we consider two possibilities for extrapolation to larger $N_{c}$ values. One possibility is that, as $N_{c}$ increases, all the thermodynamic quantities become qualitatively similar to that of the pure gauge theory (with $N_f=0$). The argument here is based on the gauge degrees of freedom dominating over the fermionic ones for larger values of $N_c$, thus leaving the fermions with less of a thermodynamic impact \cite{fujikura2023}. Lattice data is available in the pure case for various values of $N_{c}$ \cite{Panero}. Specifically, it has been shown that these quantities are essentially universal, showing little sensitivity to $N_{c}$ up to degree of freedom rescaling. This would imply that, for the larger $N_{c}$ values of interest, the pure gauge theory data may be used as an approximation of the $N_{f}=4$ case.
 
Alternatively, since the critical properties of a confining gauge theory are dependent only upon their spectrum, which is itself independent of $N_{c}$, it might be expected that the universality observed within the pure gauge case extends to cases where $N_{f} \ne 0$ \cite{DeGrand:2021zjw}. Lattice studies for $N_{c} =3,4,5$ with $N_{f} =2$ were performed in \cite{DeGrand:2021zjw} for the temperature dependent chiral condensate and screening masses, and find preliminary evidence to support this conclusion. If correct, this would imply that lattice data of $\tilde{\Delta}_\text{CS}(T)$ and $\tilde{p}_{\text{CS}}\left(T\right)$ for an SU(3) with $N_{f}=4$ simulation would be  representative of all $N_{c}$ cases. Such a simulation, which will be used to describe this possibility, was performed for two different distributions of fermion masses in \cite{Engels_1997}. This study, despite being relatively old, is consistent with that of the pure gauge case if we rescale by the ratio of gluons to total degrees of freedom. We believe it would be of interest to follow up this work with more modern lattice QCD techniques, and hope to motivate this.

Interestingly, this rescaling can also be applied to the cases of $N_f=2+1$ and $N_f=2+1+1$ lattice QCD \cite{Borsanyi:2013bia,Borsanyi:2016ksw}, if we consider them as $N_f=2$ studies. It can be seen by comparing these studies that the peak in the trace anomaly reaches roughly the same value, with the distinction occurring at higher temperatures. This itself can be attested to the difference in fermionic degrees of freedom decoupling at these temperatures. Then, the peak in the trace anomaly, and thus the drop in the equation of state, is closely related to the total number of degrees of freedom becoming non relativistic, despite the highly non-perturbative physics of a QCD PT. 

In practice, the main qualitative difference between the two scenarios is the order of the PT. For the pure gauge theory, the transition is necessarily first order, whilst in the fermionic case we may realise a smooth crossover, as well as a first/second order PT. This discussion is the subject of much ongoing study \cite{Cuteri:2018wci,Cuteri:2021ikv,Klinger:2025xxb}, and depends on several factors. Firstly, the specific values of ($N_c,N_f$) set the phase boundaries of the Columbia plot. Then, the order of the PT is determined by the mass distribution of the fermions. As the CH constituent fermions have not yet been observed, the mass spectrum is not known, therefore we must remain somewhat agnostic about this specific aspect. 

In the case of a first-order PT, since the universe is no longer in thermal equilibrium at temperatures below $T_{c}$ \cite{Morgante_2023}, we are unable to reliably predict the equation of state in this regime. Instead, focussing on $T\gtrsim T_c$, we are able to give a description of the equilibrium thermodynamics. In doing so, we hope to encapsulate the main qualitative behaviour, and thus describe the equation of state regardless of the detailed PT thermodynamics.

\begin{figure}[ht!]
    \centering
\includegraphics[width=0.9\linewidth]{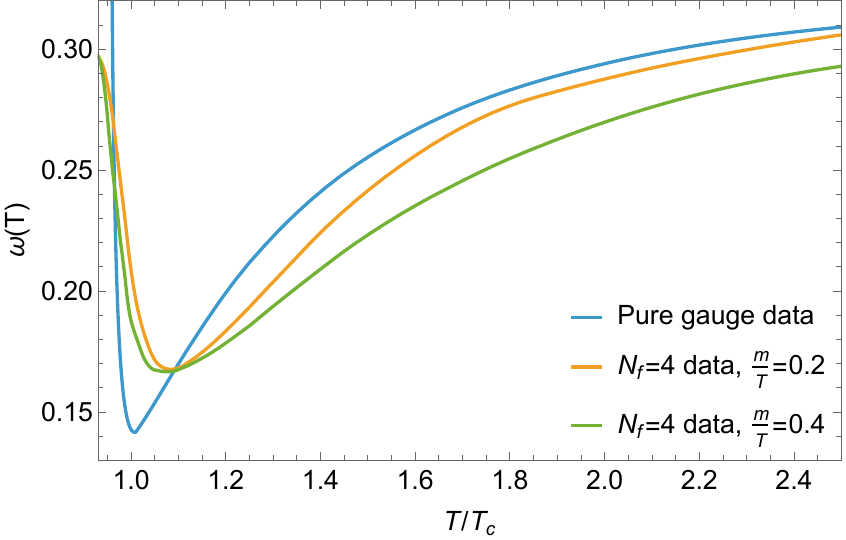}
    \caption{We show the composite Higgs equation of state for several different scenarios. Namely, the first-order pure gauge case, as well as the crossover case, with varying fermion masses. We rescale by the degrees of freedom associated with a $N_c=5$, $N_f=4$ gauge theory.}\label{QCDorder}
\end{figure}

In \cref{QCDorder}, we show the estimation of the equation of state for the two possible scenarios considered. The blue curve represents the scenario where the thermodynamics approach the form of the pure gauge case. Thus, we have used lattice data taken from pure gauge simulations in \cite{Panero}, and rescaled with both the gauge and fermionic degrees of freedom. The yellow and green curve represent the second scenario, where the thermodynamics is the same for a fixed $N_{f}$, and is independent of $N_c$. The only remanent dependence is the masses of the fermions, of which we consider 2 cases: $m/T=0.2$ and $m/T=0.4$, as provided in \cite{Engels_1997}. As previously discussed, we rescale the data appropriately with the degrees of freedom associated to an $N_{f}=4$, $N_{c}=5$ gauge theory. 

\section{Primordial black holes} \label{PBHsection}

Over 50 years ago it was proposed that we may observe radiation accreted from compact objects formed in the early universe \cite{Zeldovich}. Shortly after, a mechanism for generating such objects was developed \cite{Carr1, Hawking}, based on the collapse of large density fluctuations created during inflation. These objects, now labelled as primordial black holes (PBHs), are the topic of a large amount of ongoing research. One reason for such popularity is the possibility that they may account for the totality, or at least a large fraction, of the dark matter abundance \cite{Chapline, Green, Carr2, Esser:2025pnt}. Furthermore, there have been several recent advancements in GW science, such as LIGO, Virgo and KAGRA \cite{Abbott,Abbott1,KAGRA:2013rdx}, as well as pulsar timing array collaborations \cite{NANOGrav, EPTA, PPTA, CPTA}. These advancements have led to new perspectives on PBHs \cite{Bird,Carr:2019kxo,Iovino:2024tyg,Pritchard:2024vix,Kumar:2025jfi}, and attest to the versatility of the field.

\subsection{Standard scenario}

In order to calculate the abundance of PBHs, we require two key pieces of information. Firstly, the equation of state of the early universe, which we hope to have sufficiently described in the previous sections. Using the equation of state, following the method of \cite{Musco2,Byrnes:2018clq}, we are able to define the threshold for collapse, $\delta_c$. The threshold for collapse determines the minimum value of a perturbation amplitude such that a PBH is formed upon horizon re-entry. 

Secondly, we are interested in the primordial curvature power spectrum, $\mathcal{P}_\zeta$. Assuming perturbations follow a Gaussian distribution, the curvature power spectrum gives an exact description of them. Importantly, in order for any PBHs to form, the amplitude of the power spectrum needs to be enhanced by many orders of magnitude from the measured value on the large scales ($10^{-4}\text{Mpc}^{-1}\lesssim k\lesssim1\text{Mpc}^{-1}$) of the cosmic microwave background \cite{Planck:2018jri}. Achieving such an enhancement of power without severe fine-tuning remains a difficulty within single-field inflationary models \cite{37}, however multi-field models offer more promising results \cite{Qin,Stamou}.

Throughout the remainder of the work, we assume that there is such an enhancement of power, and parametrise it in two different ways. Firstly, we consider a cut-power-law spectrum, given as
\begin{equation}\label{CutP}
    \mathcal{P}_\zeta(k)=A\left(\frac{k}{k_\text{min}}\right)^{n_s-1}\Theta(k-k_\text{min})\Theta(k_\text{max}-k),
\end{equation}
where $k_\text{min}$ and $k_\text{max}$ define cut-off scales, and $n_s$ is the tilt of the power spectrum. 

Secondly, we consider a lognormal distribution, given as
\begin{equation}\label{lognorm}
    \mathcal{P}_\zeta(k) = A\frac{1}{\sqrt{2\pi}\Delta}\text{exp}\left(-\frac{\text{ln}^2(k/k_p)}{2\Delta^2}\right),
\end{equation}
where \(\Delta\) is the width of the power spectrum, \(k_p\) is the peak scale of the power spectrum.

Once we have chosen a curvature power spectrum we can define the radiation domination smoothed density power spectrum as
\begin{equation}\label{PdeltaR}
    \mathcal{P}_{\delta_R} (k) = \frac{16}{81}(kR)^4W^2(k,R)\mathcal{P}_\zeta(k),
\end{equation}
where \(W(k,R)\) is a window function used to relate $\mathcal{P}_\zeta(k)$ in Fourier space to the probability distribution function of $\delta$ in real space \cite{Ando,Kalaja:2019uju}. In this work, we consider a Gaussian window function modified by a factor of 2 in the exponent \cite{Young}, defined as
\begin{equation}\label{GaussianWindow}
    W(k,R)=\text{exp}\left(-\frac{(kR)^2}{4}\right).
\end{equation}
We can then use the smoothed density power spectrum to calculate the variance of the distribution, given as
\begin{equation}\label{Variance}
    \sigma^2(R)=\int_{0}^{\infty} \frac{\text{d}k}{k} \mathcal{P}_{\delta_R}.
\end{equation}
Finally, we may calculate the fraction of the universe which collapses to form PBHs, at the time of formation. We calculate this quantity, typically labelled $\beta_\text{PBH}$, using Press--Schechter theory \cite{Press} (Although see also \cite{Bardeen,Kushwaha:2025zpz}). In this formalism, and assuming that PBHs form with exactly the horizon mass of the universe, we can approximate the abundance as
\begin{equation}
    \beta_\text{PBH}(M_H)=\text{erfc}\left( \frac{\delta_c(M_H)}{\sqrt{2\sigma^2(M_H)}} \right), \label{AbundanceBeta}
\end{equation}
where erfc is the complementary error function. We have also related the horizon mass and wavelength via the following relation \cite{Nakama:2016gzw}
\begin{equation}
    M_H\simeq17\left(\frac{g_{*\rho}(k)}{10.75}\right)^{-1/6}\left(\frac{k}{10^6\text{Mpc}^{-1}}\right)^{-2}M_\odot.
\end{equation}
Alternatively, we can describe the abundance of PBHs in terms of the fraction of cold dark matter, given as
\begin{equation}\label{fPBH}
    f(M_H)=\frac{1}{\Omega_{\text{CDM}}}\frac{\text{d}\Omega_{\text{PBH}}}{\text{dln}M_H},
\end{equation}
where $\Omega_{\text{PBH}}$ is the total abundance of PBHs relative to the critical density and $\Omega_{\text{CDM}}=0.245$ \cite{aghanim}. This may be written as a function of $\beta$
\begin{equation}\label{fPBH1}
    f(M_H) = \frac{2.4}{f_{\text{PBH}}} \beta_\text{PBH}(M_H) \left( \frac{M_H}{M_{\text{eq}}}\right) ^{-1/2},
\end{equation}
where the factor 2.4 comes from 2(1+$\Omega_\text{b}/\Omega_{\text{CDM}})$, with $\Omega_{\text{b}}=0.0456$ \cite{aghanim} and $M_{\text{eq}}$ is the horizon mass at the time of matter-radiation equality, $M_{\text{eq}} \approx2.8\times10^{17}$ $M_\odot$. Furthermore, we have assumed that $\beta$ is constant when $\delta_c$ is constant.

\subsection{Beyond standard scenario}

\begin{figure}[ht!]
    \centering
\includegraphics[width=\linewidth]{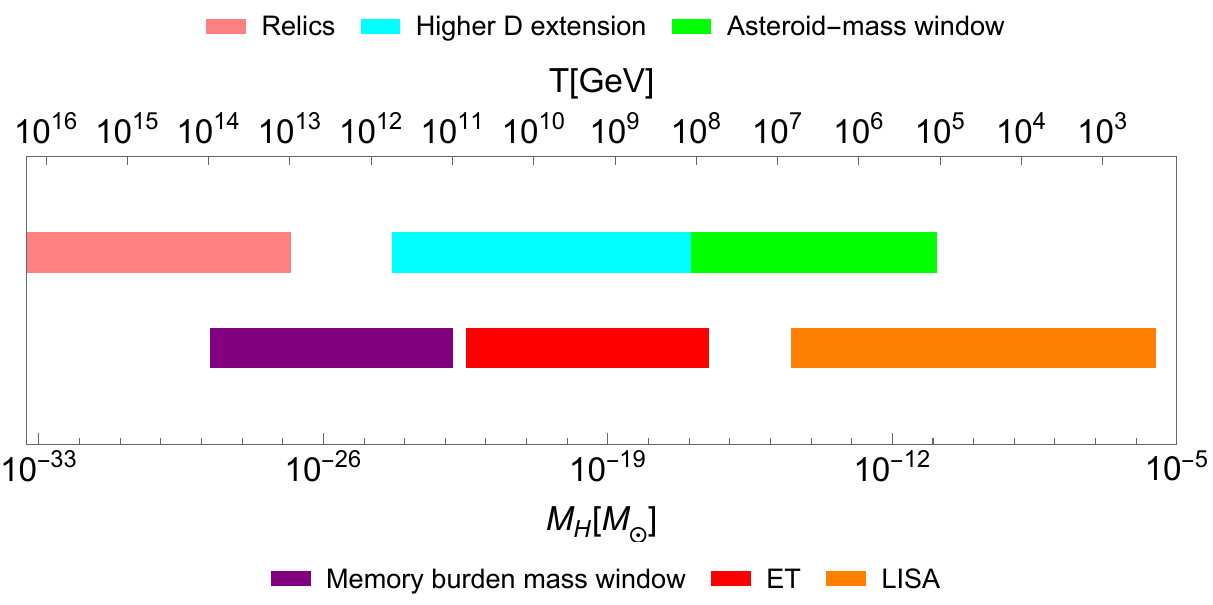}
    \caption{We highlight several interesting high-energy mass scales for PBHs. This includes several future GW experiments, as well as windows in which PBHs could constitute the totality of dark matter.}\label{InterestingMassPlot}
\end{figure}

The horizon mass of the universe, an important observable in the study of PBHs, will naturally be low during the high-scale SUSY breaking and composite Higgs models we have considered. This can be seen from the relation between horizon mass and temperature \cite{Nakama:2016gzw}
\begin{equation}\label{horizonmass}
 M_H \simeq 0.15\left(\frac{g_{\ast\rho}(T)}{10.75}\right)^{-1/2}\left(\frac{T}{\text{1GeV}}\right)^{-2}M_\odot.
\end{equation}
Therefore, drops in the equation of state at high temperatures can plausibly enhance the abundance of PBHs at several interesting mass scales, which are unconstrained by current observations \cite{Gow:2020bzo}. We show some examples of interesting mass ranges, along with the corresponding temperature of the universe in \cref{InterestingMassPlot}. From the figure, we see several future planned GW experiments, and the corresponding scales which they will probe. These are of interest to us as, if there were the large enhancement in power required for an abundance of PBHs, there will also be detectable scalar-induced GWs (see \cite{Domenech:2021ztg} for a review). 

We also see scales relevant to PBHs as a dark matter candidate. Firstly, we have the asteroid-mass window \cite{Gorton:2024cdm,Esser:2025pnt}, along with the extended range if one considers higher dimensions \cite{Guedens:2002km,Aldecoa-Tamayo:2025dqe}. Secondly, we have the memory burden mass window \cite{Dvali:2020wft,Thoss:2024hsr} (however see \cite{Montefalcone:2025akm}).  This region has been made even more interesting by the recent proposal that a memory burden effect through black hole mergers may be observed \cite{Dvali:2025sog}. If PBHs were to form within either of these mass ranges, it is possible that they constitute the totality of dark matter. Lastly, we see the window in which PBHs may have stopped evaporating before today, and left stable relics which may account for dark matter \cite{Carr:1994ar,Chen:2004ft,Aljazaeri:2025ftv}.

Considering the SUSY-breaking scenarios, we see that LISA \cite{LISA:2017pwj} will probe scales relevant for both the TeV and PeV scales (see also \cite{Escriva:2024ivo}). On the other hand, the Einstein Telescope \cite{Punturo:2010zz} will probe Intermediate-scale SUSY. We may also see that PeV and intermediate scales will be within the asteroid mass windows. Lastly, GUT-scale SUSY offers an enhancement in PBH abundance at scales relevant for relics. For the composite Higgs models, each of these windows are relevant, as the drop in equation of state will occur on a similar scale to that of confinement, which is not bounded if we neglect the naturalness argument.

\begin{figure}[ht!]
    \centering
\includegraphics[width=\linewidth]{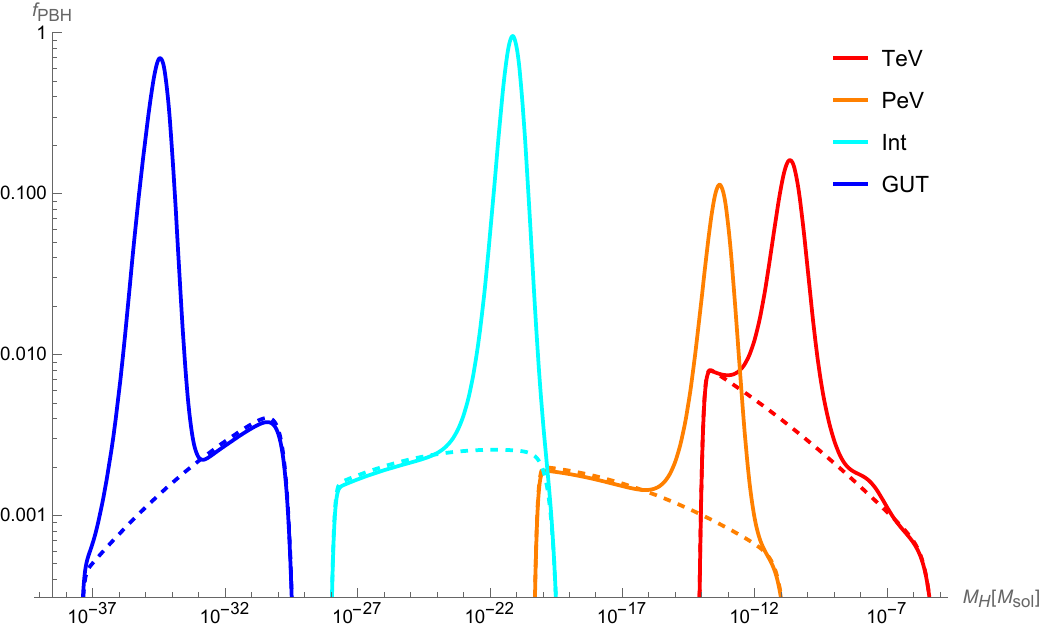}
    \caption{The PBH mass distributions corresponding to each of our SUSY equations of state. At lower horizon masses we see a larger enhancement from a similar drop, due to $\beta_\text{PBH}$ naturally being lower. We see from the previous figure that each of these enhancement peak at exciting cosmological scales.}\label{SUSYf}
\end{figure}

Using the SUSY equations of state obtained in \cref{SUSYsection}, we show some example primordial black hole mass spectra in \cref{SUSYf}. Specifically, we use \cref{CutP} and avoid having overly broad spectra by cutting off irrelevant scales. As with the SM PTs, we still need our power spectrum to peak at similar scales to our soft equation of state, if we wish to realise an enhancement. Interestingly, \say{isocurvature pumping} \cite{Germani:2017bcs} can occur naturally in a supergravity framework \cite{Khlopov:1985fch,Ketov:2021fww,Aldabergenov:2022rfc} via scalar field induced gravitational instabilities. It would be of interest to explore these models with our predicted equations of state.

In each of the SUSY breaking scenarios we find an enhancement in the PBH abundance of at least an order of magnitude, which is roughly the enhancement obtained over the SM QCD PT. We note, however, that the minimum of the SUSY equation of state we obtain is $\omega_\text{min}\approx0.27$, as opposed to the QCD dropping to $\omega_\text{min}\approx0.23$. Furthermore, we see from \cref{SUSYf} that the GUT-scale SUSY breaking leads to an increase in PBH abundance of almost 3 orders of magnitude, despite having the smallest drop in equation of state. Both of these facts arise because we are dealing with higher energy theories. As alluded to at the start of the subsection, the horizon mass of the universe is naturally smaller in these scenarios. This means that $\beta_\text{PBH}$ will naturally be smaller. Then, because we are deeper in the tail of the probability distribution function of the curvature perturbation, there is a higher sensitivity to changes in the threshold for collapse, $\delta_c$. 

We next consider PBH mass spectra in the context of composite Higgs. Here, we consider the minimal SU($N_c$) model with 4 fermion flavours of mass $m/T=0.4$, varying $N_c$ from 5 to 11. We take $f_{\text{CS}}=3\text{TeV}$, which is inspired by naturalness, however in theory this could take higher values \cite{Barnard:2014tla}. Intriguingly, is has been shown that such models may naturally realise ultra-slow-roll inflation \cite{Merchand:2025bzt}, which can provide the additional power necessary for PBH formation \cite{Motohashi:2017kbs,Garcia-Bellido:2017mdw}. It is also of interest that, in these theories, if PBHs were to form well before the confinement-deconfinement phase they could have significant colour charge \cite{Alonso-Monsalve:2023brx}.

\begin{figure}[ht!]
    \centering
\includegraphics[width=\linewidth]{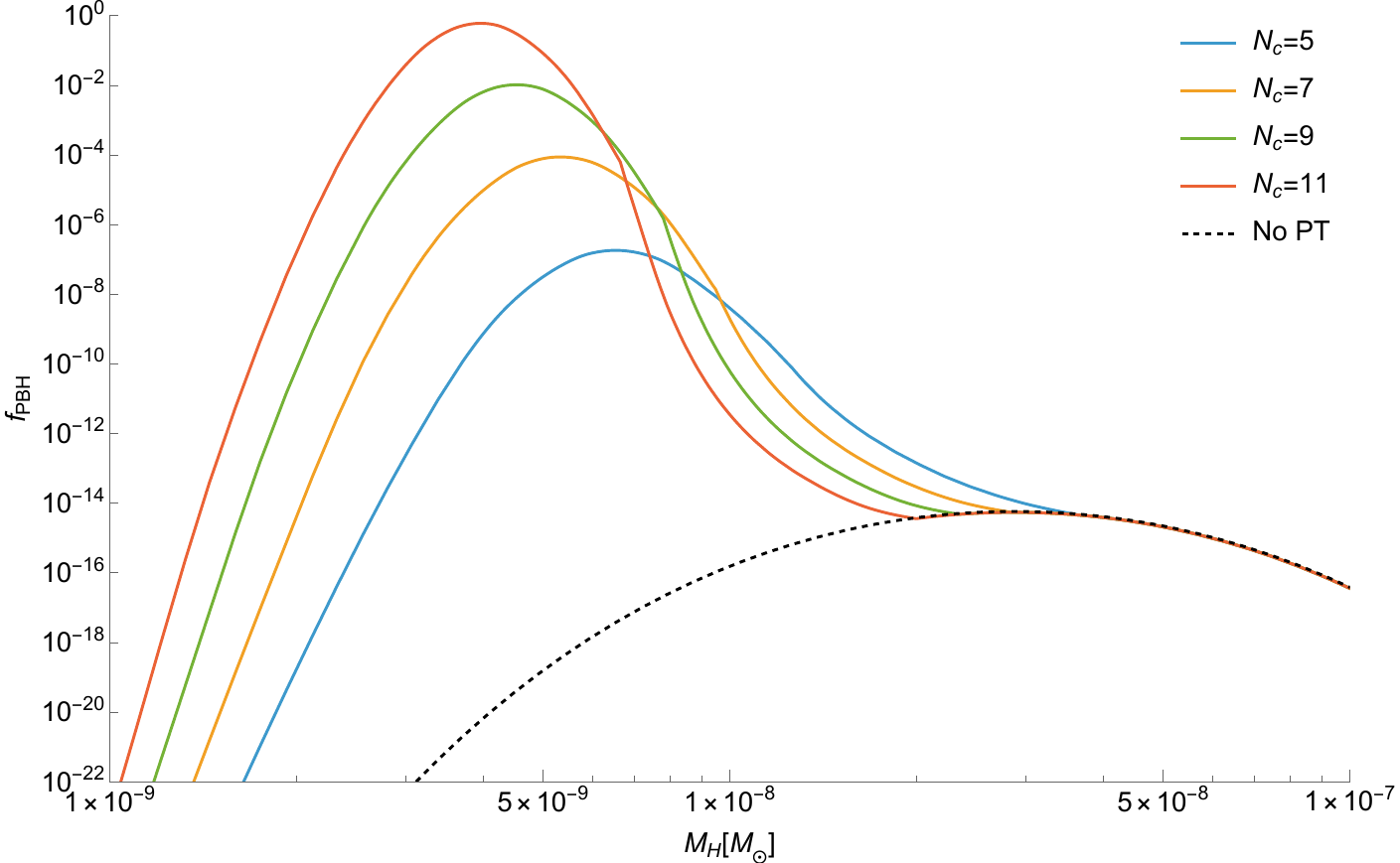}
    \caption{PBH mass spectra for various composite Higgs scenarios. Using a lognormal power spectrum with $\Delta=1.5$, we vary $N_c$ and see the impact this has on the drop in equation of state and thus the enhancement of PBHs. Notably, for $N_c=11$, we find that the PBH abundance is increased by 20 orders of magnitude. For a more conservative value of $N_c=5$, we find an enhancement of 10 orders of magnitude. However, we reiterate that $N_c$ is a free parameter.}\label{CHf}
\end{figure}

Upon choosing $f_{\text{CS}}$, using the M.I.T. bag model described in \cref{HadronicAndMIT}, we find the critical temperature to be $\approx2\text{TeV}$. In this theory there are 11 pNGB degrees of freedom, which is more than around the pion scale in the SM. However, we find that because of the suppression of the relativistic SM degrees of freedom, the drop due to decoupling pNGBs is negligible. Thus, we neglect these particles, only using them for finding the critical temperature. We show our results in \cref{CHf}, where we consider a lognormal $\mathcal{P}_\zeta$ defined in \cref{lognorm}, with $\Delta=1.5$ and $k_p\approx4\times10^{10}\text{Mpc}^{-1}$. 

From \cref{CHf}, we can see the capability that strongly interacting theories have in softening the equation of state, and thus enhancing the abundance of PBHs. Due to $N_c$ being a free parameter, as well as the degrees of freedom scaling like $N_c^2$, these models quickly surpass any SM enhancement. Specifically, in these examples we find that the minimum in the equation of state varies between $0.1\lesssim\omega_\text{min}\lesssim0.17$. Taking the case of $N_c=11$, we find a significant enhancement of almost 20 orders of magnitude compared to the case of pure radiation. We expect this to be true regardless of the order of the PT, as previously discussed. However, in the case of a first order PT, we wouldn't be able to trust our thermodynamic description for $T\lesssim T_c$, due to a large departure from thermal equilibrium.

Taking the limit that $r_{\text{SM}}\rightarrow0$, which is equivalent to taking $N_c\rightarrow\infty$, we obtain that the minimum in equation of state also tends to 0. This is because the pressure of pure Yang-Mills is exponentially Boltzmann suppressed for $T<T_c$ \cite{Panero}, therefore we obtain a $T^4\tilde{\Delta}_\text{CS}(T)/0$ term in the denominator of \cref{CHeos}. There is a subtlety here; in theory, each independent thermodynamic quantity will be exponentially Boltzmann suppressed for $T\ll m$, meaning the trace $\tilde{\Delta}_\text{CS}$ will also tend to 0. However, because of the derivative factor appearing in \cref{Trace}, the pressure reaches 0 before the trace. This can be seen by taking the non-relativistic (Boltzmann) limit of the hadronic gas thermodynamic quantities \cref{lowtempEoS}
\begin{equation}
    \begin{split}
        &\rho_\text{B}(T)\approx g\left(m+\frac{3T}{2}\right)\left(\frac{mT}{2\pi}\right)^{3/2}e^{-m/T},\\
        &p_\text{B}(T)\approx gT\left(\frac{mT}{2\pi}\right)^{3/2}e^{-m/T},
    \end{split}
\end{equation}
such that
\begin{equation}
    \begin{split}
        &\Delta_B(T)=\rho_B(T)-3p_B(T)\approx g\left(m-\frac{3T}{2}\right)\left(\frac{mT}{2\pi}\right)^{3/2}e^{-m/T},\\
        &\hspace{15.3mm}\frac{\rho_B(T)-3p_B(T)}{p_B(T)}\approx\frac{m}{T} \hspace{5mm} (m\gg T).
    \end{split}
\end{equation}
So, although both quantities tend to 0, the trace is parametrically larger than the pressure. One might hope that we can use this to arbitrarily enhance a given abundance of PBHs. We note, however, that if $\omega$ becomes too small, one has to start worrying about non-spherical effects of density perturbation collapse \cite{Bardeen:1985tr,Germani:2025fkh,Ebrahimian:2025syf,Ye:2025wif}. We thus decide to stop our studies once the minimum value of $\omega(T)$ reaches 0.1. However, with more tolerance of these effects, we would be able to enhance the PBH abundance further still. 

\section{Conclusions} \label{Conclusion}

In this paper we have investigated the impact of several motivated beyond Standard Model theories on the abundance of primordial black holes. We began by summarising the physics of each Standard Model phase transition, with a particular focus on the strongly interacting QCD phase transition. This focus is a result of the non-perturbative nature of the transition, as well as it offering the largest known drop in equation of state. In itself, this has led to much study within the primordial black hole literature \cite{Jedamzik:1996mr,Byrnes:2018clq,Musco:2023dak,Pritchard:2024vix,Bhaumik:2025vlb}, as well as in other cosmological contexts \cite{Saikawa:2018rcs,Saikawa:2020swg,Franciolini:2023wjm}. The interest here arises due to the fact that the primordial black hole abundance is exponentially sensitive to changes in the threshold for collapse, which in turn depends on the equation of state. 

When investigating beyond Standard Model theories, we chose to focus on the Minimal Supersymmetric Standard Model, as well as composite Higgs models. Both predict a large number of additional degrees of freedom, therefore it is expected that large drops in equation of state may be realised. This is exactly what we find; starting with the Minimal Supersymmetric Standard Model, we first gave some motivations for the theory before calculating the mass spectra of several models \cite{Ellis:2015jwa,Ellis:2017erg}. The distinction between the models is the scale at which SUSY breaking occurs, which in turn leads to different horizon mass scales in which one could expect a softer equation of state. Due to the perturbative nature of the Minimal Supersymmetric Standard Model, we were then able to calculate the equation of states using the ideal gas approximation,

Following this, we studied composite Higgs models. We first calculated the degrees of freedom predicted by several UV realisations of these models, before focusing on the SU($N_c$) gauge theory. We develop a similar description of phase transitions within this theory to that of the Standard Model QCD. Namely, we find that the drop in equation of state, upon rescaling each of our quantities to the Stefan-Boltzmann limit, is dictated by the ratio of relativistic Standard Model degrees of freedom to the number of fundamental degrees of freedom predicted by the new composite sector. As we show, this depends quadratically on the number of colours of the theory, $N_c$, which for our purposes is a free parameter. This gives us the power to quickly produce large drops in our composite Higgs equation of state. We end this section with a discussion on the order of the phase transition, highlighting the main qualitative differences expected. Importantly, we expect the drop to remain a similar value regardless of the order. 

Having calculated examples of equations of state in both theories, we provide a mini-review of primordial black hole formation in the standard picture. We then arrive at the impact our beyond Standard Model equations of state have on the abundance of primordial black holes. We find that in the case of the Minimal Supersymmetric Standard Model, we are able to achieve an enhancement of 3 orders of magnitude from that of pure radiation. Notably, this is achieved perturbatively. Then, likely our most exciting result is regarding the composite Higgs models. We find that non-perturbatively we are able to generate an enhancement in primordial black hole abundance of $\sim\mathcal{O}(10^{20})$. This is achieved by introducing 11 colours, leading to the minimum in the equation of state being around 0.1. We emphasise that, although we focussed on a scale motivated by naturalness within a composite Higgs theory, one could in principle have this enhancement at any scale, set by $f_\text{CS}$. We show that, regardless of the scale these phase transitions occur, they will necessarily be of some cosmological significance. 

Lastly, we wish to emphasise an important point; it is unthinkable that colliders will be able to probe physics at, say, $10^{10}\text{GeV}$ any time soon. However, in the theories we have studied, evidence for a soft equation of state could be found from future gravitational wave experiments, bubble wall collisions, or even from a population of primordial black holes, for example. This could lead to precious information as to the scales of physics beyond that of the Standard Model, therefore offering one of our most promising avenues for progress.

{\bf Acknowledgements}

The authors greatly thank Manuel Reichert for comments on the draft. The authors further thank Szabolcs Borsanyi, Chris Byrnes, Charlie Cresswell-Hogg, Stephen P. Martin, Marco Panero, Ennio Salvioni and Andrea Tesi for fruitful discussions. XP and MS are supported by an STFC studentship, and WL is supported by a Croucher Scholarship.

\appendix

\section{Finite-temperature quantum chromodynamics}\label{alphabeta}

In this appendix we provide some more details on finite-temperature QCD calculations. The beta function of the theory is written in a loop expansion of the QCD running coupling, $g_3$, which runs with the renormalisation scale, $\mu$. Re-expressing the coupling as $a_\mu=g_3^2(\mu)/16\pi^2$, the renormalisation group equation (RGE) reads 
\begin{equation}\label{QCDbeta}
    \mu^2\frac{\text{d}a_\mu}{\text{d}\mu^2}=\beta(a_\mu)=-a_\mu^2\sum^\infty_{i=0}\beta_ia_\mu^i
\end{equation}
where $\beta_i$ are coefficients computed from $i+1$ loop integrals. Throughout our analysis, we use the 5-loop beta functions provided in  \cite{Herzog:2017ohr}. Following \cite{Wu:2019mky}, we integrate \cref{QCDbeta}, which leads to 
\begin{equation}
    \text{ln}\mu_0^2-\frac{1}{\beta_0a_{\mu_0}}-\frac{\beta_1}{\beta_0^2}\text{ln}a_{\mu_0}-\int^{a_{\mu_0}}_0\frac{\text{d}a}{\tilde{\beta}(a)}= \text{ln}\mu^2-\frac{1}{\beta_0a_{\mu}}-\frac{\beta_1}{\beta_0^2}\text{ln}a_{\mu}-\int^{a_{\mu}}_0\frac{\text{d}a}{\tilde{\beta}(a)},
\end{equation}
This integral, up to 5-loop order, can be written as the following power series in $a_\mu$,
\begin{equation}
 \begin{split}\int^{a_{\mu}}_0\frac{\text{d}a}{\tilde{\beta}(a)}=& \left( \frac{\beta_2}{\beta_0^2}-\frac{\beta_1^2}{\beta_0^3}\right) a_\mu + \left( \frac{\beta_3}{2\beta_0^2}-\frac{\beta_2\beta_1}{\beta_0^3}+\frac{\beta_1^3}{2\beta_0^4}\right)a_\mu^2 \\
    &+\left(\frac{\beta_4}{3\beta_0^2}-\frac{\beta_2^2}{3\beta_0^3}-\frac{2\beta_3\beta_1}{3\beta_0^3}+\frac{\beta_2\beta_1^2}{\beta_0^4}-\frac{\beta_1^4}{3\beta_0^5}\right)a_\mu^3+\mathcal{O}(a_\mu^4),
 \end{split}
\end{equation}
which can then be solved iteratively \cite{Kniehl:2006bg}, to arrive at
\begin{equation}\label{aSeries}
    \begin{split}
        a_\mu=& \frac{1}{\beta_0L}-\frac{b_1\text{ln}L}{(\beta_0L)^2}+\frac{1}{(\beta_0L)^3}\left[b_1^2(\text{ln}^2L-\text{ln}L-1)+b_2\right] \\
        &+\frac{1}{(\beta_0L)^4}\left[b_1^3\left(-\text{ln}^3L+\frac{5}{2}\text{ln}^2L+2\text{ln}L-\frac{1}{2}\right)-3b_1b_2\text{ln}L+\frac{b_3}{2}\right]\\
        &+\frac{1}{(\beta_0L)^5}\bigg[3b_2b_1^2(2\text{ln}^2L-\text{ln}L-1)+b_1^4\left(\text{ln}^4L-\frac{13}{3}\text{ln}^3L-\frac{3}{2}\text{ln}^2L+4\text{ln}L+\frac{7}{6}\right)\\
        &-b_3b_1\left(2\text{ln}L+\frac{1}{6}\right)+\left(\frac{5}{3}b_2^2+\frac{1}{3}b_4\right)\bigg] +\mathcal{O}\left(\frac{1}{(\beta_0L)^6}\right),
    \end{split}
\end{equation}
where $b_i=\beta_i/\beta_0$ and $L=\text{ln}(\mu^2/\Lambda_\text{QCD}^2)$. Here, $\Lambda_\text{QCD}^2\approx0.298\text{GeV}$ \cite{Bruno:2017gxd}, is the asymptotic scale of QCD. Then, following the results of \cite{Kajantie:2002wa}, we write the pressure of the quark-gluon plasma as an expansion in the strong coupling constant up to order $g_3^6\ln(1/g_3)$
\begin{equation}\label{hotqcdpressure}
    \begin{split}
        \frac{p_\text{QCD}(T)}{T^4\mu^{-2\epsilon}}=&g_3^0\bigg\{\alpha_{E1}\bigg\}\\
    +&g_3^2\bigg\{\alpha_{E2}\bigg\}\\
        +&\frac{g_3^3}{4\pi}\bigg\{\frac{d_A}{3}\alpha_{E4}^{3/2}\bigg\}\\
        +&\frac{g_3^4}{(4\pi)^2}\bigg\{\alpha_{E3}-d_aC_a\left[\alpha_{E4}\left(\frac{3}{4}+\text{ln}\frac{\mu}{2g_3T\alpha^{1/2}}\right)+\frac{1}{4}\alpha_{E5}\right]\bigg\}\\
        +&\frac{g_3^5}{(4\pi)^3}\bigg\{d_A\alpha_{E4}^{1/2}\left[\frac{1}{2}\alpha_{E6}-C_A^2\left(\frac{89}{24}+\frac{\pi^2}{6}-\frac{11}{6}\text{ln}2\right)\right]\bigg\}\\
        +&g_3^6\frac{d_AC_A}{(4\pi)^4}\bigg\{\left(\alpha_{E6}+\alpha_{E4}\alpha_{E7}\right)\text{ln}\left(g_3\alpha_{E4}^{1/2}\right)\\
       &\hspace{18mm}-8C_A^2\left[\alpha_M\text{ln}\left(g_3\alpha_4^{1/2}\right)+2\alpha_G\text{ln}\left(g_3C_A^{1/2}\right)\right]\bigg\},
    \end{split}
\end{equation}
where 1/$\epsilon$ divergences have been cancelled and details of the group theory factors, as well as each $\alpha_E$, are given below. 

The running of the strong coupling, and thus the $\beta$ functions, will depend on the symmetry breaking pattern of the QCD. Then, the beta functions of \cite{Herzog:2017ohr} are defined in terms of group theory, or colour, factors. Furthermore, the pressure will also depend on these factors, as will be seen. The SM, as well as our chosen composite Higgs model, are both SU($N_c$) gauge theories. Therefore, these group theory factors will be the same for both theories, varying only by differences in $N_c$ and $N_f$. Firstly, we define the Lie algebra of a compact simple Lie group 
\begin{equation}
    T^aT^b-T^bT^a=if^{abc}T^c,
\end{equation}
where $T^a$ are the generators of the representation of the fermions and $f^{abc}$ are structure constants.

We also define the quadratic Casimir operators $C_F$, of the $N_c$-dimensional fundamental representation, $[T^aT^a]_{ik}=C_F\delta_{ik}$ and $C_A$, of the $N_A$-dimensional adjoint representation, $f^{acd}f^{bcd}=C_A^{ab}$. The trace normalisation of the fermion representation is Tr$(T^aT^b)=T_F\delta^{ab}$. We also define the generators of the adjoint representation $[C^a]_{bc}=-if^{abc}$. Higher order group invariants can then be expressed in terms of contractions between the two totally symmetric generator tensors
\begin{equation}
    \begin{split}
        d_F^{abcd}=&\frac{1}{6}\text{Tr}\bigg[T^aT^bT^cT^d+T^aT^bT^dT^c+T^aT^cT^bT^d\\
        &+T^aT^cT^dT^b+T^aT^dT^bT^c+T^aT^dT^cT^b\bigg],\\
        d_A^{abcd}=&\frac{1}{6}\text{Tr}\bigg[C^aC^bC^cC^d+C^aC^bC^dC^c+C^aC^cC^bC^d\\
        &+C^aC^cC^dC^b+C^aC^dC^bC^c+C^aC^dC^cC^b\bigg].
    \end{split}
\end{equation}
The above group theory factors are then given by
\begin{align}
    \begin{split}
        & \quad \quad \quad T_F=\frac{1}{2},\quad  C_A=N_c,\quad  C_F=\frac{N_c^2-1}{2N_c},\quad d_A^{abcd}=N_c^2-1, \quad d_F^{abcd}=N_cN_f,\\
        &\frac{d_A^{abcd} d_A^{abcd}}{N_A}=\frac{Nc^2(N_c^2+36)}{24},\quad
        \frac{d_F^{abcd} d_A^{abcd}}{N_A}=\frac{N_c(N_c^2+6)}{48},\quad \frac{d_F^{abcd} d_F^{abcd}}{N_A}=\frac{N_c^4-6N_c^2+18}{96N_c^2}.   
    \end{split}
\end{align}
Using these factors we are able to describe hot QCD thermodynamics via $\beta$ functions given in \cite{Herzog:2017ohr}, as well as the $\alpha_E$ coefficients appearing in \cref{hotqcdpressure}. These are given as
\begin{equation}
    \begin{split}
        &\alpha_{E1}=\frac{\pi^2}{180}(4d_A+7d_F),\\
        &\alpha_{E2}=-\frac{d_A}{144}(C_A+\frac{5}{2}T_F),\\
        &\begin{split}\alpha_{E3}=&\frac{d_A}{144}\bigg[C_A^2\bigg(\frac{12}{\epsilon}+\frac{194}{3}\ln\frac{\bar{\mu}}{4\pi T}+\frac{116}{5}+4\gamma+\frac{220}{3}\frac{\zeta'(-1)}{\zeta(-1)}-\frac{38}{3}\frac{\zeta'(-3)}{\zeta(-3)}\bigg)\\
        &+C_AT_F\bigg(\frac{12}{\epsilon}+\frac{169}{3}\ln\frac{\bar{\mu}}{4\pi T}+\frac{1121}{60}-\frac{157}{5}\ln2+8\gamma+\frac{146}{3}\frac{\zeta'(-1)}{\zeta(-1)}-\frac{1}{3}\frac{\zeta'(-3)}{\zeta(-3)}\bigg)\\
        &+T_F^2\bigg(\frac{20}{3}\ln\frac{\bar{\mu}}{4\pi T}+\frac{1}{3}-\frac{88}{5}\ln2+4\gamma+\frac{16}{3}\frac{\zeta'(-1)}{\zeta(-1)}-\frac{8}{3}\frac{\zeta'(-3)}{\zeta(-3)}\bigg)\\
        &+C_FT_F\bigg(\frac{105}{4}-24\ln2\bigg)\bigg],\end{split}\\
        &\alpha_{E4}=\frac{1}{3}(C_A+T_F),\\
        &\alpha_{E5}=\frac{2}{3}\bigg[C_A\bigg(\ln\frac{\bar{\mu}}{4\pi T}+\frac{\zeta'(-1)}{\zeta(-1)}\bigg)+T_F\bigg(\ln\frac{\bar{\mu}}{4\pi T}+\frac{1}{2}-\ln2+\frac{\zeta'(-1)}{\zeta(-1)}\bigg)\bigg],\\
        &\begin{split}\alpha_{E6}=&C_A^2\bigg(\frac{22}{9}\ln\frac{\bar{\mu}e^\gamma}{4\pi T}+\frac{5}{9}\bigg)+C_AT_F\bigg(\frac{14}{9}\ln\frac{\bar{\mu}e^\gamma}{4\pi T}-\frac{16}{9}\ln2+1\bigg)\\
        &+T_F^2\bigg(-\frac{8}{9}\ln\frac{\bar{\mu}e^\gamma}{4\pi T}-\frac{16}{9}\ln2+\frac{4}{9}\bigg)-2C_FT_F,\end{split}\\
        &\alpha_{E7}=C_A\bigg(\frac{22}{3}\ln\frac{\bar{\mu}e^\gamma}{4\pi T}+\frac{1}{3}\bigg)-T_F\bigg(\frac{8}{3}\ln\frac{\bar{\mu}e^\gamma}{4\pi T}+\frac{16}{3}\ln2\bigg).
    \end{split}
\end{equation}

\section{Supersymmetric gauge unification}\label{gaugeuniapp}

\begin{figure}[ht!]
    \centering
\includegraphics[width=\linewidth]{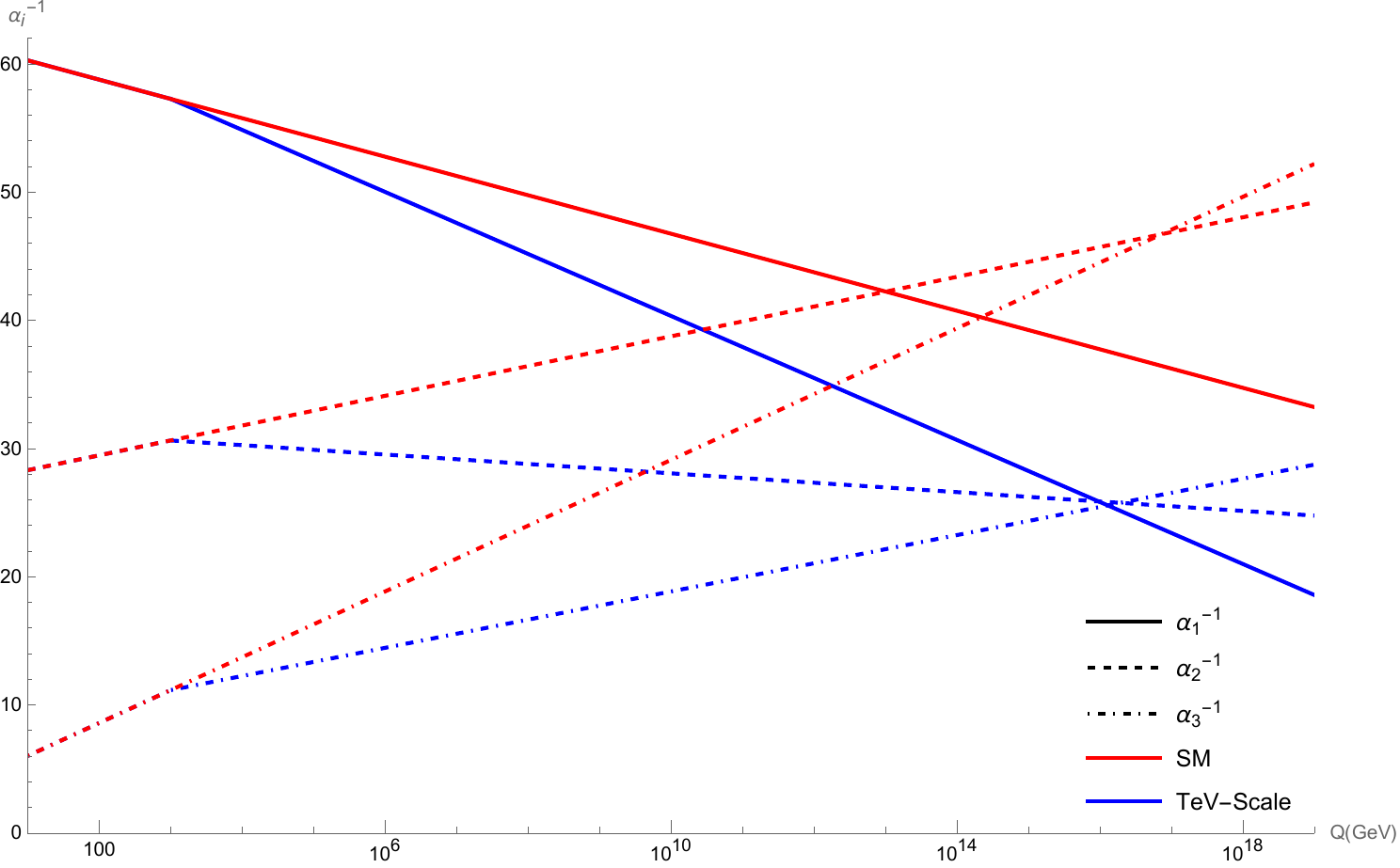}
    \caption{The gauge coupling running as a function of the renormalisation scale, $Q$. We plot the SM case, as well as SUSY breaking at around $1\text{TeV}$, which vary because of the different field content. We see the famous result that, for TeV-scale SUSY, the gauge coupling unify at high energies.}\label{GaugeRunningPlot}
\end{figure}

As previously discussed, an attractive property of SUSY is that all three gauge couplings are able to unify. If this is indeed the case, it would signal that the three different types of interaction within the SM descend, more fundamentally, from one single interaction at much higher energy scales, which is rather aesthetically pleasing. Furthermore, it could shed light on why the charge of the proton and electron are, as far as is known, identical, despite there being no symmetry that relates them. Unification can be seen from solving the 1-loop SUSY RG gauge coupling equations
\begin{equation}\label{SUSYgaugeRGE}
    \frac{\text{d}}{\text{d}t}\alpha_a^{-1}=-\frac{b_a}{2\pi}, 
\end{equation}
where $\alpha_a=g_a^2/4\pi$, $a=(1,2,3)$, $b_a=(33/5,1,-3)$ and $t=\text{ln}(Q/Q_0)$. Here, $Q$ is the RG scale and $Q_0$ is a reference scale. The gauge couplings follow these equations on renormalisation scales larger than the SUSY breaking scale. Thus, we impose boundary conditions at $M_\text{SUSY}$. Below $M_\text{SUSY}$, the gauge couplings follow the usual SM RGEs, with $b_a=(41/10,-19/6,-7)$. We note that these values are clearly smaller than those of the MSSM due to additional degrees of freedom propagating within the loop. We show the running of the gauge couplings in \cref{GaugeRunningPlot}, for both the SM and the MSSM with SUSY breaking at 1 TeV.

The scale at which the gauge couplings unify, i.e. $\alpha^{-1}_1(M_U)=\alpha^{-1}_2(M_U)=\alpha^{-1}_3(M_U)$, is typically referred to as the GUT scale. As we can see from \cref{GaugeRunningPlot}, for TeV scale SUSY breaking the value of $M_\text{GUT}$ is $\mathcal{O}(10^{16})\text{GeV}$. As previously discussed, due to the lack of detection of any sparticles in current collider experiments, there are difficulties motivating such a scenario. A possible reconciliation for this is that SUSY breaking occurs at higher scales and the gauge couplings receive threshold corrections. Threshold corrections are multiplets that gain mass via the same mechanism in which the GUT symmetry is broken. Then, the value of $\alpha^{-1}_a(M_U)$ is given by the UV theory. From the IR perspective, the threshold corrections are given as
\begin{equation}
    \left(\frac{\Delta\lambda_{ij}(M_U)}{12\pi}\right)\equiv\left(\frac{1}{\alpha_i(M_U)}-\frac{1}{\alpha_j(M_U)}\right),
\end{equation}
for $i,j=1,2,3$ and $i\neq j$. 

\begin{figure}[ht!]
    \centering
\includegraphics[width=0.96\linewidth]{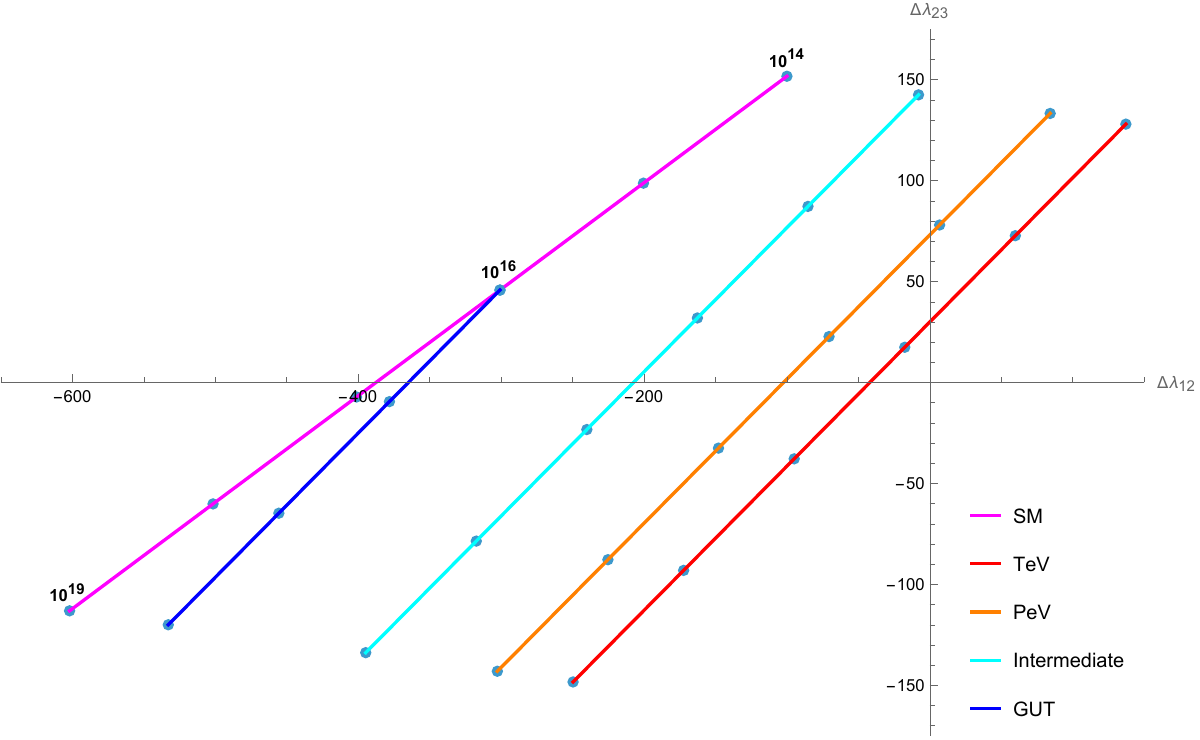}
    \caption{The threshold corrections for each of the SUSY-breaking scenarios we considered, similar to those plotted in \cite{Ellis:2015jwa}. We see that the corrections depend on both the input GUT-scale, as well as the scale of SUSY breaking. However, each of the SUSY breaking scenarios lead to smaller threshold corrections than that of the SM, for a given GUT scale.}\label{ThresholdPlot}
\end{figure}

It should be noted that, if SUSY were to break at larger scales \cite{Hall:2009nd}, the naturalness problem, as well as the hierarchy problem, are no longer solved by the theory. Following \cite{Ellis:2015jwa,Ellis:2017erg}, in \cref{ThresholdPlot} we show $\Delta\lambda_{ij}$ for each considered SUSY breaking scenario. The corrections will also depend on the choice of GUT scale, which we vary from $10^{14}\text{GeV}$ to $10^{19}\text{GeV}$. This provides a visualisation of the IR physics being projected onto the higher scales. In \cite{Ellis:2017erg}, one of the main outputs was the identification of the parameter space in which high-scale SUSY models permit a light Higgs boson, as is observed in colliders. We use these results as boundary conditions for the remaining RGEs, which we then use to determine the SUSY mass spectra later in the section. 

Typically, if the gauge couplings unify at some scale $M_U$, it is assumed that the gaugino masses also unify near that scale. If SUSY were a component of a GUT, this would necessarily be true. Then, we arrive at the following equation
\begin{equation}
    M_1\alpha_1^{-1}=M_2\alpha_2^{-1}=M_3\alpha_3^{-1},
\end{equation}
which is independent of RG scale. Explicitly, the one-loop RGEs of the gaugino mass parameters are given by
\begin{equation}
    \frac{\text{d}}{\text{d}t}M_a=\frac{1}{2\pi}b_a\alpha_aM_a,
\end{equation}
for $a=(1,2,3)$ and the quantities $b_a$ are the same as in \cref{SUSYgaugeRGE}. From these RGEs we can see that, for a given RG scale, the products $\alpha_aM_a$ are each constant. We note that, at tree-level, $M_3(Q)$ defines the running mass of the gluino, discussed in the main text.

\section{Supersymmetry renormalisation group equations}\label{SUSYRGEs}

In this appendix we provide further details of the supersymmetric RGEs solved in \cref{SUSYsection}. For more detail, we advise reading \cite{Martin:1997ns}. Firstly, appearing in the superpotential we have the running of the Yukawa coupling of a scalar $\phi_k$ and 2 fermions $\psi_i\psi_j$ 
\begin{equation}
    \frac{\text{d}}{\text{d}t}y^{ijk}=\gamma_n^iy^{njk}+\gamma_n^jy^{ink}+\gamma_n^ky^{ijn},
\end{equation}
where $\gamma_n^i$ are anomolous dimension matrices associated with the superfields. These can be calculated for the third-family squark and sleptons, as well as the Higgs, which leads to the following running of the individual Yukawa couplings
\begin{equation}
    \begin{split}
        &\frac{\text{d}}{\text{d} t}y_t=\frac{y_t}{16\pi^2}\left[6y_t^*y_t+y_b^*y_b-\frac{64\pi}{3}\alpha_3-12\pi\alpha_2-\frac{52\pi}{15}\alpha_1\right],\\
        &\frac{\text{d}}{\text{d} t}y_b=\frac{y_b}{16\pi^2}\left[6y_b^*y_b+y_t^*y_t+y_\tau^*y_\tau-\frac{64\pi}{3}\alpha_3-12\pi\alpha_2-\frac{28\pi}{15}\alpha_1\right],\\
        &\frac{\text{d}}{\text{d} t}y_\tau=\frac{y_\tau}{16\pi^2}\left[4y_\tau^*y_\tau+3y_b^*y_b-12\pi\alpha_2-\frac{36\pi}{5}\alpha_1\right].
    \end{split}
\end{equation}
Then, we have the final superpotential parameter RGE
\begin{equation}
    \frac{\text{d}}{\text{d} t}\mu=\frac{\mu}{16\pi^2}\left[3y_t^*y_t+3y_b^*y_b+y_\tau^*y_\tau-12\pi\alpha_2-\frac{12\pi}{5}\alpha_1\right].
\end{equation}
It can be useful to approximate the Yukawa couplings: $\textbf{y}_\textbf{u}$, $\textbf{y}_\textbf{d}$ and $\textbf{y}_\textbf{e}$, as their (3,3) family component. In this approximation the holomorphic soft parameters $\textbf{a}_\textbf{u}$, $\textbf{a}_\textbf{d}$ and $\textbf{a}_\textbf{e}$ may be approximated by their (3,3) family components $a_t$, $a_b$ and $a_\tau$ respectively. This leads to following running
\begin{equation*}
 \begin{split}
    &\begin{split}16\pi^2\frac{\text{d}}{\text{d} t}a_t=&a_t\left[18y_t^*y_t+y_b^*y_b-\frac{64\pi}{3}\alpha_3-12\pi\alpha_2-\frac{52\pi}{15}\alpha_1\right]+2a_by_b^*y_t\\
    &+y_t\left[\frac{128\pi}{3}\alpha_3M_3+24\pi\alpha_2M_2+\frac{104\pi}{15}\alpha_1M_1\right],\end{split}\\
    &\begin{split}16\pi^2\frac{\text{d}}{\text{d} t}a_b=&a_b\left[18y_b^*y_b+y_t^*y_t+y_\tau^*y_\tau-\frac{64\pi}{3}\alpha_3-12\pi\alpha_2-\frac{28\pi}{15}\alpha_1\right]+2a_ty_t^*y_b+2a_\tau y_\tau^*y_b\\
    &+y_b\left[\frac{128\pi}{3}\alpha_3M_3+24\pi\alpha_2M_2+\frac{56\pi}{15}\alpha_1M_1\right],\end{split}\\
    &16\pi^2\frac{\text{d}}{\text{d} t}a_\tau=a_\tau\left[12y_\tau^*y_\tau+3y_b^*y_b-12\pi\alpha_2-\frac{36\pi}{5}\alpha_1\right]+6a_by_b^*y_\tau+y_\tau\left[24\pi\alpha_2M_2+\frac{72\pi}{5}\alpha_1M_1\right].
 \end{split}
\end{equation*}
Next, we have the running of the Higgs squared-mass parameters and the third-family squarks and sleptons
\begin{equation}
 \begin{split}
    &16\pi^2\frac{\text{d}}{\text{d}t}m_{H_u}^2=3X_t-24\pi\alpha_2|M_2|^2\\
    &16\pi^2\frac{\text{d}}{\text{d}t}m_{H_d}^2=3X_b+X_\tau-24\pi\alpha_2|M_2|^2-\frac{24\pi}{5}\alpha_1|M_1|^2,\\
        &16\pi^2\frac{\text{d}}{\text{d}t}m_{Q_3}^2=X_t+X_b-\frac{128\pi}{3}\alpha_3|M_3|^2-24\pi\alpha_2|M_2|^2-\frac{8\pi}{15}\alpha_1|M_1|^2,\\
        &16\pi^2\frac{\text{d}}{\text{d}t}m_{\bar{u}_3}^2=2X_t-\frac{128\pi}{3}\alpha_3|M_3|^2-\frac{128\pi}{15}\alpha_1|M_1|^2,\\
        &16\pi^2\frac{\text{d}}{\text{d}t}m_{\bar{d}_3}^2=2X_b-\frac{128\pi}{3}\alpha_3|M_3|^2-\frac{32\pi}{15}\alpha_1|M_1|^2,\\
        &16\pi^2\frac{\text{d}}{\text{d}t}m_{L_3}^2=X_\tau-24\pi\alpha_2|M_2|^2-\frac{24\pi}{5}\alpha_1|M_1|^2,\\
        &16\pi^2\frac{\text{d}}{\text{d}t}m_{\bar{e}_3}^2=2X_\tau-\frac{96\pi}{5}\alpha_1|M_1|^2,
    \end{split}
\end{equation}
where
\begin{equation}
    \begin{split}
        &X_t=2|y_t|^2(m_{H_u}^2+m_{Q_3}^2+m_{\bar{u}_3}^2)+2|a_t|^2,\\
        &X_b=2|y_b|^2(m_{H_d}^2+m_{Q_3}^2+m_{\bar{d}_3}^2)+2|a_b|^2,\\
        &X_\tau=2|y_\tau|^2(m_{H_d}^2+m_{L_3}^2+m_{\bar{e}_3}^2)+2|a_\tau|^2,
    \end{split}
\end{equation}
describe the corrections to the Higgs scalars and third-family squark and slepton RGEs due to the Yukawa terms and holomorphic soft parameters. The other terms in the RGEs are defined by the remaining interactions, namely the first- and second-family squarks and sleptons. The RGEs for each of these scalars are given by 
\begin{equation}
    4\pi\frac{\text{d}}{\text{d}t}m_{\phi_i}^2=-\sum_{a=1,2,3}8C_a(i)\alpha_a|M_a|^2,
\end{equation}
where $C_a(i)$ are Casimir invariants. 

Finally, we describe the boundary conditions used in solving each of the SUSY RGEs. These boundary conditions, obtained from \cite{Ellis:2017erg}, are motivated by matching terms appearing in both the SM and the SUSY Higgs potentials. Specifically, the SM Higgs potential is given as 
\begin{equation}
    V(H)=\frac{\lambda^{SM}_H}{2}\left(|H|^2-v^2\right)^2.
\end{equation}
This is compared to the one-loop matching condition for the SUSY scalar quartic coupling, at the SUSY breaking scale. At tree-level the condition is
\begin{equation}
    \lambda_H^{tree}(M_{SUSY})=\pi\left(\alpha_2(M_{SUSY})+\frac{3}{5}\alpha_1(M_{SUSY})\right)\cos^2(2\beta).
\end{equation}
The one-loop condition is considered due to threshold corrections, which could be large. We show the resulting $\chi^2$-minimised boundary conditions in \cref{SUSYtab}, however please see \cite{Ellis:2017erg} for more details.

\begin{table}[h]
    \centering
    \begin{tabular}{|c|c|c|c|c|c|}
        \hline
        $M_{\text{SUSY}}$ & $m_0$ & $m_{1/2}$ & $\tan\beta$ & $\mu$ \\ \hline
        $5\times10^3$ & $5\times10^3$ & $4.5\times10^4$ & 50 & $2.5\times10^3$ \\ \hline
        $10^6$ & $10^6$ & $10^6$ & 2.5 & $10^6$ \\ \hline
        $10^{10}$ & $10^{10}$ & $10^{10}$ & 1 & $10^{10}$ \\ \hline
        $10^{16}$ & $10^{16}$ & $2\times10^{17}$ & 1 & $10^{13}$ \\ \hline
    \end{tabular}
    \caption{SUSY initial conditions table.   $M_{\text{SUSY}}$, $m_0$, $m_{1/2}$ and $\mu$ are each units of GeV.}
    \label{SUSYtab}
\end{table}

\printbibliography

@article{Agugliaro_2017,
   title={UV-complete composite Higgs models},
   volume={95},
   ISSN={2470-0029},
   url={http://dx.doi.org/10.1103/PhysRevD.95.035019},
   DOI={10.1103/physrevd.95.035019},
   number={3},
   journal={Physical Review D},
   publisher={American Physical Society (APS)},
   author={Agugliaro, Alessandro and Antipin, Oleg and Becciolini, Diego and De Curtis, Stefania and Redi, Michele},
   year={2017},
   month=feb }

@article{Appelquist_2021,
   title={Nearly Conformal Composite Higgs Model},
   volume={126},
   ISSN={1079-7114},
   url={http://dx.doi.org/10.1103/PhysRevLett.126.191804},
   DOI={10.1103/physrevlett.126.191804},
   number={19},
   journal={Physical Review Letters},
   publisher={American Physical Society (APS)},
   author={Appelquist, Thomas and Ingoldby, James and Piai, Maurizio},
   year={2021},
   month=may }

@article{Fayet:1977yc,
    author = "Fayet, Pierre",
    title = "{Spontaneously Broken Supersymmetric Theories of Weak, Electromagnetic and Strong Interactions}",
    reportNumber = "LPTENS 77/11",
    doi = "10.1016/0370-2693(77)90852-8",
    journal = "Phys. Lett. B",
    volume = "69",
    pages = "489",
    year = "1977"
}

@article{Farrar:1978xj,
    author = "Farrar, Glennys R. and Fayet, Pierre",
    title = "{Phenomenology of the Production, Decay, and Detection of New Hadronic States Associated with Supersymmetry}",
    reportNumber = "CALT-68-648",
    doi = "10.1016/0370-2693(78)90858-4",
    journal = "Phys. Lett. B",
    volume = "76",
    pages = "575--579",
    year = "1978"
}

@inproceedings{Haber:1993wf,
    author = "Haber, Howard E.",
    title = "{Introductory low-energy supersymmetry}",
    booktitle = "{Theoretical Advanced Study Institute (TASI 92): From Black Holes and Strings to Particles}",
    eprint = "hep-ph/9306207",
    archivePrefix = "arXiv",
    reportNumber = "SCIPP-92-33",
    pages = "589--686",
    month = "4",
    year = "1993"
}

@article{Csaki:1996ks,
    author = "Csaki, Csaba",
    title = "{The Minimal supersymmetric standard model (MSSM)}",
    eprint = "hep-ph/9606414",
    archivePrefix = "arXiv",
    reportNumber = "MIT-CTP-2542",
    doi = "10.1142/S021773239600062X",
    journal = "Mod. Phys. Lett. A",
    volume = "11",
    pages = "599",
    year = "1996"
}

@article{Martin:1997ns,
    author = "Martin, Stephen P.",
    editor = "Kane, Gordon L.",
    title = "{A Supersymmetry primer}",
    eprint = "hep-ph/9709356",
    archivePrefix = "arXiv",
    reportNumber = "FERMILAB-PUB-97-425-T",
    doi = "10.1142/9789812839657_0001",
    journal = "Adv. Ser. Direct. High Energy Phys.",
    volume = "18",
    pages = "1--98",
    year = "1998"
}

@article{Ellis:2015jwa,
    author = "Ellis, Sebastian A. R. and Wells, James D.",
    title = "{Visualizing gauge unification with high-scale thresholds}",
    eprint = "1502.01362",
    archivePrefix = "arXiv",
    primaryClass = "hep-ph",
    doi = "10.1103/PhysRevD.91.075016",
    journal = "Phys. Rev. D",
    volume = "91",
    number = "7",
    pages = "075016",
    year = "2015"
}

@article{Ellis:2017erg,
    author = "Ellis, Sebastian A. R. and Wells, James D.",
    title = "{High-scale supersymmetry, the Higgs boson mass, and gauge unification}",
    eprint = "1706.00013",
    archivePrefix = "arXiv",
    primaryClass = "hep-ph",
    reportNumber = "MCTP-17-06",
    doi = "10.1103/PhysRevD.96.055024",
    journal = "Phys. Rev. D",
    volume = "96",
    number = "5",
    pages = "055024",
    year = "2017"
}

@article{Martin:2006ub,
    author = "Martin, Stephen P.",
    title = "{Refined gluino and squark pole masses beyond leading order}",
    eprint = "hep-ph/0608026",
    archivePrefix = "arXiv",
    reportNumber = "FERMILAB-PUB-06-595-T",
    doi = "10.1103/PhysRevD.74.075009",
    journal = "Phys. Rev. D",
    volume = "74",
    pages = "075009",
    year = "2006",
    note = "[Erratum: Phys.Rev.D 98, 119901 (2018)]"
}

@article{Pierce:1996zz,
    author = "Pierce, Damien M. and Bagger, Jonathan A. and Matchev, Konstantin T. and Zhang, Ren-jie",
    title = "{Precision corrections in the minimal supersymmetric standard model}",
    eprint = "hep-ph/9606211",
    archivePrefix = "arXiv",
    reportNumber = "SLAC-PUB-7180, JHU-TIPAC-96011",
    doi = "10.1016/S0550-3213(96)00683-9",
    journal = "Nucl. Phys. B",
    volume = "491",
    pages = "3--67",
    year = "1997"
}

@article{Guedens:2002km,
    author = "Guedens, Raf and Clancy, Dominic and Liddle, Andrew R",
    title = "{Primordial black holes in braneworld cosmologies: Formation, cosmological evolution and evaporation}",
    eprint = "astro-ph/0205149",
    archivePrefix = "arXiv",
    doi = "10.1103/PhysRevD.66.043513",
    journal = "Phys. Rev. D",
    volume = "66",
    pages = "043513",
    year = "2002"
}

@article{Gow:2020bzo,
    author = "Gow, Andrew D. and Byrnes, Christian T. and Cole, Philippa S. and Young, Sam",
    title = "{The power spectrum on small scales: Robust constraints and comparing PBH methodologies}",
    eprint = "2008.03289",
    archivePrefix = "arXiv",
    primaryClass = "astro-ph.CO",
    doi = "10.1088/1475-7516/2021/02/002",
    journal = "JCAP",
    volume = "02",
    pages = "002",
    year = "2021"
}

@article{Gorton:2024cdm,
    author = "Gorton, Matthew and Green, Anne M.",
    title = "{How open is the asteroid-mass primordial black hole window?}",
    eprint = "2403.03839",
    archivePrefix = "arXiv",
    primaryClass = "astro-ph.CO",
    doi = "10.21468/SciPostPhys.17.2.032",
    journal = "SciPost Phys.",
    volume = "17",
    number = "2",
    pages = "032",
    year = "2024"
}

@article{Carr:1994ar,
    author = "Carr, Bernard J. and Gilbert, J. H. and Lidsey, James E.",
    title = "{Black hole relics and inflation: Limits on blue perturbation spectra}",
    eprint = "astro-ph/9405027",
    archivePrefix = "arXiv",
    reportNumber = "FERMILAB-PUB-94-109-A",
    doi = "10.1103/PhysRevD.50.4853",
    journal = "Phys. Rev. D",
    volume = "50",
    pages = "4853--4867",
    year = "1994"
}

@article{Chen:2004ft,
    author = "Chen, Pisin",
    editor = "Cline, D. B.",
    title = "{Inflation induced Planck-size black hole remnants as dark matter}",
    eprint = "astro-ph/0406514",
    archivePrefix = "arXiv",
    reportNumber = "SLAC-PUB-10538",
    doi = "10.1016/j.newar.2005.01.015",
    journal = "New Astron. Rev.",
    volume = "49",
    pages = "233--239",
    year = "2005"
}

@article{Aljazaeri:2025ftv,
    author = "Aljazaeri, Amirah and Byrnes, Christian T.",
    title = "{Comprehensively Constraining Ultra-Light Primordial Black Holes Through Relic Formation and Early Mergers}",
    eprint = "2506.16154",
    archivePrefix = "arXiv",
    primaryClass = "astro-ph.CO",
    month = "6",
    year = "2025"
}

@article{Thoss:2024hsr,
    author = "Thoss, Valentin and Burkert, Andreas and Kohri, Kazunori",
    title = "{Breakdown of hawking evaporation opens new mass window for primordial black holes as dark matter candidate}",
    eprint = "2402.17823",
    archivePrefix = "arXiv",
    primaryClass = "astro-ph.CO",
    reportNumber = "KEK-TH-2605;KEK-Cosmo-0339;KEK-QUP-2024-0003",
    doi = "10.1093/mnras/stae1098",
    journal = "Mon. Not. Roy. Astron. Soc.",
    volume = "532",
    number = "1",
    pages = "451--459",
    year = "2024"
}

@article{Dvali:2020wft,
    author = "Dvali, Gia and Eisemann, Lukas and Michel, Marco and Zell, Sebastian",
    title = "{Black hole metamorphosis and stabilization by memory burden}",
    eprint = "2006.00011",
    archivePrefix = "arXiv",
    primaryClass = "hep-th",
    doi = "10.1103/PhysRevD.102.103523",
    journal = "Phys. Rev. D",
    volume = "102",
    number = "10",
    pages = "103523",
    year = "2020"
}

@article{LISA:2017pwj,
    author = "Amaro-Seoane, Pau and others",
    collaboration = "LISA",
    title = "{Laser Interferometer Space Antenna}",
    eprint = "1702.00786",
    archivePrefix = "arXiv",
    primaryClass = "astro-ph.IM",
    month = "2",
    year = "2017"
}

@article{Punturo:2010zz,
    author = "Punturo, M. and others",
    editor = "Ricci, Fulvio",
    title = "{The Einstein Telescope: A third-generation gravitational wave observatory}",
    doi = "10.1088/0264-9381/27/19/194002",
    journal = "Class. Quant. Grav.",
    volume = "27",
    pages = "194002",
    year = "2010"
}

@article{Planck:2018jri,
    author = "Akrami, Y. and others",
    collaboration = "Planck",
    title = "{Planck 2018 results. X. Constraints on inflation}",
    eprint = "1807.06211",
    archivePrefix = "arXiv",
    primaryClass = "astro-ph.CO",
    doi = "10.1051/0004-6361/201833887",
    journal = "Astron. Astrophys.",
    volume = "641",
    pages = "A10",
    year = "2020"
}

@article{Nakama:2016gzw,
    author = "Nakama, Tomohiro and Silk, Joseph and Kamionkowski, Marc",
    title = "{Stochastic gravitational waves associated with the formation of primordial black holes}",
    eprint = "1612.06264",
    archivePrefix = "arXiv",
    primaryClass = "astro-ph.CO",
    doi = "10.1103/PhysRevD.95.043511",
    journal = "Phys. Rev. D",
    volume = "95",
    number = "4",
    pages = "043511",
    year = "2017"
}

@article{Inoue:1982pi,
    author = "Inoue, Kenzo and Kakuto, Akira and Komatsu, Hiromasa and Takeshita, Seiichiro",
    title = "{Aspects of Grand Unified Models with Softly Broken Supersymmetry}",
    reportNumber = "KYUSHU-82-HE-5",
    doi = "10.1143/PTP.68.927",
    journal = "Prog. Theor. Phys.",
    volume = "68",
    pages = "927",
    year = "1982",
    note = "[Erratum: Prog.Theor.Phys. 70, 330 (1983)]"
}

@article{Flores:1982pr,
    author = "Flores, Ricardo A. and Sher, Marc",
    title = "{Higgs Masses in the Standard, Multi-Higgs and Supersymmetric Models}",
    reportNumber = "UCSC-TH-154-82",
    doi = "10.1016/0003-4916(83)90331-7",
    journal = "Annals Phys.",
    volume = "148",
    pages = "95",
    year = "1983"
}

@article{Draper:2016pys,
    author = "Draper, Patrick and Rzehak, Heidi",
    title = "{A Review of Higgs Mass Calculations in Supersymmetric Models}",
    eprint = "1601.01890",
    archivePrefix = "arXiv",
    primaryClass = "hep-ph",
    reportNumber = "DIAS-2016-1, FR-PHENO-2016-001, CP3-ORIGINS-2016-001-DNRF90",
    doi = "10.1016/j.physrep.2016.01.001",
    journal = "Phys. Rept.",
    volume = "619",
    pages = "1--24",
    year = "2016"
}

@article{Byrnes:2018clq,
    author = "Byrnes, Christian T. and Hindmarsh, Mark and Young, Sam and Hawkins, Michael R. S.",
    title = "{Primordial black holes with an accurate QCD equation of state}",
    eprint = "1801.06138",
    archivePrefix = "arXiv",
    primaryClass = "astro-ph.CO",
    doi = "10.1088/1475-7516/2018/08/041",
    journal = "JCAP",
    volume = "08",
    pages = "041",
    year = "2018"
}

@article{Saikawa:2018rcs,
    author = "Saikawa, Ken'ichi and Shirai, Satoshi",
    title = "{Primordial gravitational waves, precisely: The role of thermodynamics in the Standard Model}",
    eprint = "1803.01038",
    archivePrefix = "arXiv",
    primaryClass = "hep-ph",
    reportNumber = "IPMU18-0037, MPP-2018-19",
    doi = "10.1088/1475-7516/2018/05/035",
    journal = "JCAP",
    volume = "05",
    pages = "035",
    year = "2018"
}

@article{Saikawa:2020swg,
    author = "Saikawa, Ken'ichi and Shirai, Satoshi",
    title = "{Precise WIMP Dark Matter Abundance and Standard Model Thermodynamics}",
    eprint = "2005.03544",
    archivePrefix = "arXiv",
    primaryClass = "hep-ph",
    reportNumber = "IPMU20-0047, KANAZAWA-20-03, MPP-2020-62",
    doi = "10.1088/1475-7516/2020/08/011",
    journal = "JCAP",
    volume = "08",
    pages = "011",
    year = "2020"
}

@article{Aoki:2006br,
    author = "Aoki, Y. and Fodor, Z. and Katz, S. D. and Szabo, K. K.",
    title = "{The QCD transition temperature: Results with physical masses in the continuum limit}",
    eprint = "hep-lat/0609068",
    archivePrefix = "arXiv",
    doi = "10.1016/j.physletb.2006.10.021",
    journal = "Phys. Lett. B",
    volume = "643",
    pages = "46--54",
    year = "2006"
}

@article{Borsanyi:2010bp,
    author = "Borsanyi, Szabolcs and Fodor, Zoltan and Hoelbling, Christian and Katz, Sandor D and Krieg, Stefan and Ratti, Claudia and Szabo, Kalman K.",
    collaboration = "Wuppertal-Budapest",
    title = "{Is there still any $T_c$ mystery in lattice QCD? Results with physical masses in the continuum limit III}",
    eprint = "1005.3508",
    archivePrefix = "arXiv",
    primaryClass = "hep-lat",
    reportNumber = "WUB-10-11, MIT-CTP-4152",
    doi = "10.1007/JHEP09(2010)073",
    journal = "JHEP",
    volume = "09",
    pages = "073",
    year = "2010"
}

@article{Borsanyi:2020fev,
    author = "Borsanyi, Szabolcs and Fodor, Zoltan and Guenther, Jana N. and Kara, Ruben and Katz, Sandor D. and Parotto, Paolo and Pasztor, Attila and Ratti, Claudia and Szabo, Kalman K.",
    title = "{QCD Crossover at Finite Chemical Potential from Lattice Simulations}",
    eprint = "2002.02821",
    archivePrefix = "arXiv",
    primaryClass = "hep-lat",
    doi = "10.1103/PhysRevLett.125.052001",
    journal = "Phys. Rev. Lett.",
    volume = "125",
    number = "5",
    pages = "052001",
    year = "2020"
}

@book{Le_Bellac_1996, place={Cambridge}, series={Cambridge Monographs on Mathematical Physics}, title={Thermal Field Theory}, publisher={Cambridge University Press}, author={Le Bellac, Michel}, year={1996}, collection={Cambridge Monographs on Mathematical Physics}
}

@article{Ishii:2002ys,
    author = "Ishii, Noriyoshi and Suganuma, Hideo",
    title = "{A Statistical approach to the QCD phase transition: A Mystery in the critical temperature}",
    eprint = "hep-ph/0210158",
    archivePrefix = "arXiv",
    month = "10",
    year = "2002"
}

@article{PhysRevD.10.2599,
  title = {Baryon structure in the bag theory},
  author = {Chodos, A. and Jaffe, R. L. and Johnson, K. and Thorn, C. B.},
  journal = {Phys. Rev. D},
  volume = {10},
  issue = {8},
  pages = {2599--2604},
  numpages = {0},
  year = {1974},
  month = {Oct},
  publisher = {American Physical Society},
  doi = {10.1103/PhysRevD.10.2599},
  url = {https://link.aps.org/doi/10.1103/PhysRevD.10.2599}
}

@ARTICLE{Abbott2,
       author = {{Abbott}, L.~F. and {Farhi}, Edward and {Wise}, Mark B.},
        title = "{Particle production in the new inflationary cosmology}",
      journal = {Physics Letters B},
         year = 1982,
        month = nov,
       volume = {117},
       number = {1-2},
        pages = {29-33},
          doi = {10.1016/0370-2693(82)90867-X},
       adsurl = {https://ui.adsabs.harvard.edu/abs/1982PhLB..117...29A}
}

@article{Traschen,
  title = {Particle production during out-of-equilibrium phase transitions},
  author = {Traschen, Jennie H. and Brandenberger, Robert H.},
  journal = {Phys. Rev. D},
  volume = {42},
  issue = {8},
  pages = {2491--2504},
  numpages = {0},
  year = {1990},
  month = {Oct},
  publisher = {American Physical Society},
  doi = {10.1103/PhysRevD.42.2491},
  url = {https://link.aps.org/doi/10.1103/PhysRevD.42.2491}
}

@article{Allahverdi,
   title={Reheating in Inflationary Cosmology: Theory and Applications},
   volume={60},
   ISSN={1545-4134},
   url={http://dx.doi.org/10.1146/annurev.nucl.012809.104511},
   DOI={10.1146/annurev.nucl.012809.104511},
   number={1},
   journal={Annual Review of Nuclear and Particle Science},
   publisher={Annual Reviews},
   author={Allahverdi, Rouzbeh and Brandenberger, Robert and Cyr-Racine, Francis-Yan and Mazumdar, Anupam},
   year={2010},
   month=nov, pages={27–51} }

@article{Bassett,
   title={Inflation dynamics and reheating},
   volume={78},
   ISSN={1539-0756},
   url={http://dx.doi.org/10.1103/RevModPhys.78.537},
   DOI={10.1103/revmodphys.78.537},
   number={2},
   journal={Reviews of Modern Physics},
   publisher={American Physical Society (APS)},
   author={Bassett, Bruce A. and Tsujikawa, Shinji and Wands, David},
   year={2006},
   month=may, pages={537–589} }

@article{Jedamzik:2024wtq,
    author = "Jedamzik, Karsten",
    title = "{Primordial black hole formation during cosmic phase transitions}",
    eprint = "2406.11417",
    archivePrefix = "arXiv",
    primaryClass = "astro-ph.CO",
    month = "6",
    year = "2024"
}

@article{Husdal:2016haj,
    author = "Husdal, Lars",
    title = "{On Effective Degrees of Freedom in the Early Universe}",
    eprint = "1609.04979",
    archivePrefix = "arXiv",
    primaryClass = "astro-ph.CO",
    doi = "10.3390/galaxies4040078",
    journal = "Galaxies",
    volume = "4",
    number = "4",
    pages = "78",
    year = "2016"
}

@article{Kajantie:2002wa,
    author = "Kajantie, K. and Laine, M. and Rummukainen, K. and Schroder, Y.",
    title = "{The Pressure of hot QCD up to g6 ln(1/g)}",
    eprint = "hep-ph/0211321",
    archivePrefix = "arXiv",
    reportNumber = "CERN-TH-2002-334, HIP-2002-62-TH, MIT-CTP-3325",
    doi = "10.1103/PhysRevD.67.105008",
    journal = "Phys. Rev. D",
    volume = "67",
    pages = "105008",
    year = "2003"
}

@article{Herzog:2017ohr,
    author = "Herzog, F. and Ruijl, B. and Ueda, T. and Vermaseren, J. A. M. and Vogt, A.",
    title = "{The five-loop beta function of Yang-Mills theory with fermions}",
    eprint = "1701.01404",
    archivePrefix = "arXiv",
    primaryClass = "hep-ph",
    reportNumber = "NIKHEF-2017-001, LTH-1117",
    doi = "10.1007/JHEP02(2017)090",
    journal = "JHEP",
    volume = "02",
    pages = "090",
    year = "2017"
}

@article{Wu:2019mky,
    author = "Wu, Xing-Gang and Shen, Jian-Ming and Du, Bo-Lun and Huang, Xu-Dong and Wang, Sheng-Quan and Brodsky, Stanley J.",
    title = "{The QCD renormalization group equation and the elimination of fixed-order scheme-and-scale ambiguities using the principle of maximum conformality}",
    eprint = "1903.12177",
    archivePrefix = "arXiv",
    primaryClass = "hep-ph",
    reportNumber = "SLAC-PUB-17403",
    doi = "10.1016/j.ppnp.2019.05.003",
    journal = "Prog. Part. Nucl. Phys.",
    volume = "108",
    pages = "103706",
    year = "2019"
}

@article{Kniehl:2006bg,
    author = "Kniehl, B. A. and Kotikov, A. V. and Onishchenko, A. I. and Veretin, O. L.",
    title = "{Strong-coupling constant with flavor thresholds at five loops in the anti-MS scheme}",
    eprint = "hep-ph/0607202",
    archivePrefix = "arXiv",
    reportNumber = "DESY-06-074",
    doi = "10.1103/PhysRevLett.97.042001",
    journal = "Phys. Rev. Lett.",
    volume = "97",
    pages = "042001",
    year = "2006"
}

@article{Bruno:2017gxd,
    author = "Bruno, Mattia and Dalla Brida, Mattia and Fritzsch, Patrick and Korzec, Tomasz and Ramos, Alberto and Schaefer, Stefan and Simma, Hubert and Sint, Stefan and Sommer, Rainer",
    collaboration = "ALPHA",
    title = "{QCD Coupling from a Nonperturbative Determination of the Three-Flavor $\Lambda$ Parameter}",
    eprint = "1706.03821",
    archivePrefix = "arXiv",
    primaryClass = "hep-lat",
    reportNumber = "CERN-TH-2017-129, DESY-17-088, WUB-17-03",
    doi = "10.1103/PhysRevLett.119.102001",
    journal = "Phys. Rev. Lett.",
    volume = "119",
    number = "10",
    pages = "102001",
    year = "2017"
}

@ARTICLE{Tachikawa:2013kta,
       author = {{Tachikawa}, Yuji},
        title = "{N=2 supersymmetric dynamics for pedestrians}",
      journal = {arXiv e-prints},
         year = 2013,
        month = dec,
          eid = {arXiv:1312.2684},
        pages = {arXiv:1312.2684},
          doi = {10.48550/arXiv.1312.2684},
archivePrefix = {arXiv},
       eprint = {1312.2684},
 primaryClass = {hep-th},
       adsurl = {https://ui.adsabs.harvard.edu/abs/2013arXiv1312.2684T}
}

@article{Borsanyi:2016ksw,
    author = "Borsanyi, Sz. and others",
    title = "{Calculation of the axion mass based on high-temperature lattice quantum chromodynamics}",
    eprint = "1606.07494",
    archivePrefix = "arXiv",
    primaryClass = "hep-lat",
    reportNumber = "DESY-16-105",
    doi = "10.1038/nature20115",
    journal = "Nature",
    volume = "539",
    number = "7627",
    pages = "69--71",
    year = "2016"
}

@article{Borsanyi:2013bia,
    author = "Borsanyi, Szabocls and Fodor, Zoltan and Hoelbling, Christian and Katz, Sandor D. and Krieg, Stefan and Szabo, Kalman K.",
    title = "{Full result for the QCD equation of state with 2+1 flavors}",
    eprint = "1309.5258",
    archivePrefix = "arXiv",
    primaryClass = "hep-lat",
    doi = "10.1016/j.physletb.2014.01.007",
    journal = "Phys. Lett. B",
    volume = "730",
    pages = "99--104",
    year = "2014"
}

@article{Bazavov:2017dsy,
    author = "Bazavov, A. and Petreczky, P. and Weber, J. H.",
    title = "{Equation of State in 2+1 Flavor QCD at High Temperatures}",
    eprint = "1710.05024",
    archivePrefix = "arXiv",
    primaryClass = "hep-lat",
    doi = "10.1103/PhysRevD.97.014510",
    journal = "Phys. Rev. D",
    volume = "97",
    number = "1",
    pages = "014510",
    year = "2018"
}

@article{Bresciani:2025vxw,
    author = "Bresciani, Matteo and Brida, Mattia Dalla and Giusti, Leonardo and Pepe, Michele",
    title = "{QCD Equation of State with Nf=3 Flavors up to the Electroweak Scale}",
    eprint = "2501.11603",
    archivePrefix = "arXiv",
    primaryClass = "hep-lat",
    doi = "10.1103/PhysRevLett.134.201904",
    journal = "Phys. Rev. Lett.",
    volume = "134",
    number = "20",
    pages = "201904",
    year = "2025"
}

@article{Navarrete:2024ruu,
    author = {Navarrete, Pablo and Schr{\"o}der, York},
    title = "{The g$^{6}$ pressure of hot Yang-Mills theory: canonical form of the integrand}",
    eprint = "2408.15830",
    archivePrefix = "arXiv",
    primaryClass = "hep-ph",
    doi = "10.1007/JHEP11(2024)037",
    journal = "JHEP",
    volume = "11",
    pages = "037",
    year = "2024"
}

@article{Braaten:1995jr,
    author = "Braaten, Eric and Nieto, Agustin",
    title = "{Free energy of QCD at high temperature}",
    eprint = "hep-ph/9510408",
    archivePrefix = "arXiv",
    reportNumber = "OHSTPY-HEP-T-95-020",
    doi = "10.1103/PhysRevD.53.3421",
    journal = "Phys. Rev. D",
    volume = "53",
    pages = "3421--3437",
    year = "1996"
}

@article{Bellwied:2015lba,
    author = "Bellwied, R. and Borsanyi, S. and Fodor, Z. and Katz, S. D. and Pasztor, A. and Ratti, C. and Szabo, K. K.",
    title = "{Fluctuations and correlations in high temperature QCD}",
    eprint = "1507.04627",
    archivePrefix = "arXiv",
    primaryClass = "hep-lat",
    doi = "10.1103/PhysRevD.92.114505",
    journal = "Phys. Rev. D",
    volume = "92",
    number = "11",
    pages = "114505",
    year = "2015"
}

@article{Zeldovich,
       author = {{Zeldovich}, Ya. B. and {Novikov}, I.~D.},
        title = "{The Hypothesis of Cores Retarded during Expansion and the Hot Cosmological Model}",
      journal = "Soviet Astronomy",
         year = 1967,
        month = feb,
       volume = {10},
        pages = {602},
       adsurl = {https://ui.adsabs.harvard.edu/abs/1967SvA....10..602Z}
}

@article{Carr1,
    author = {Carr, B. J. and Hawking, S. W.},
    title = "{Black Holes in the Early Universe}",
    journal = {Monthly Notices of the Royal Astronomical Society},
    volume = {168},
    number = {2},
    pages = {399-415},
    year = {1974},
    month = {08},
    doi = {10.1093/mnras/168.2.399},
    url = {https://doi.org/10.1093/mnras/168.2.399},
    eprint = {https://academic.oup.com/mnras/article-pdf/168/2/399/8079885/mnras168-0399.pdf},
}

@article{Hawking,
    author = {Hawking, Stephen},
    title = "{Gravitationally Collapsed Objects of Very Low Mass}",
    journal = {Monthly Notices of the Royal Astronomical Society},
    volume = {152},
    number = {1},
    pages = {75-78},
    year = {1971},
    month = {04},
    issn = {0035-8711},
    doi = {10.1093/mnras/152.1.75},
    url = {https://doi.org/10.1093/mnras/152.1.75},
    eprint = {https://academic.oup.com/mnras/article-pdf/152/1/75/9360899/mnras152-0075.pdf},
}

@ARTICLE{Chapline,
       author = {{Chapline}, G.~F.},
        title = "{Cosmological effects of primordial black holes}",
      journal = "Nature",
         year = 1975,
        month = jan,
       volume = {253},
       number = {5489},
        pages = {251-252},
          doi = {10.1038/253251a0},
       adsurl = {https://ui.adsabs.harvard.edu/abs/1975Natur.253..251C}
}

@article{Green,
   title={Primordial black holes as a dark matter candidate},
   volume={48},
   ISSN={1361-6471},
   url={http://dx.doi.org/10.1088/1361-6471/abc534},
   DOI={10.1088/1361-6471/abc534},
   number={4},
   journal={Journal of Physics G: Nuclear and Particle Physics},
   publisher={IOP Publishing},
   author={Green, Anne M and Kavanagh, Bradley J},
   year={2021},
   month=feb, pages={043001} }

@article{Abbott,
   title={Binary Black Hole Mergers in the First Advanced LIGO Observing Run},
   volume={6},
   ISSN={2160-3308},
   url={http://dx.doi.org/10.1103/PhysRevX.6.041015},
   DOI={10.1103/physrevx.6.041015},
   number={4},
   journal={Physical Review X},
   publisher={American Physical Society (APS)},
   author={Abbott, B.P. et al.},
   year={2016},
   month=oct }

@article{Abbott1,
   title={GWTC-3: Compact Binary Coalescences Observed by LIGO and Virgo during the Second Part of the Third Observing Run},
   volume={13},
   ISSN={2160-3308},
   url={http://dx.doi.org/10.1103/PhysRevX.13.041039},
   DOI={10.1103/physrevx.13.041039},
   number={4},
   journal={Physical Review X},
   publisher={American Physical Society (APS)},
   author={Abbott, R. et al.},
   year={2023},
   month=dec }

@article{Bird,
   title={Did LIGO Detect Dark Matter?},
   volume={116},
   ISSN={1079-7114},
   url={http://dx.doi.org/10.1103/PhysRevLett.116.201301},
   DOI={10.1103/physrevlett.116.201301},
   number={20},
   journal={Physical Review Letters},
   publisher={American Physical Society (APS)},
   author={Bird, Simeon and Cholis, Ilias and Muñoz, Julian B. and Ali-Ha{\"i}moud, Yacine and Kamionkowski, Marc and Kovetz, Ely D. and Raccanelli, Alvise and Riess, Adam G.},
   year={2016},
   month=may }

@article{Carr2,
   title={Primordial black holes as dark matter candidates},
   ISSN={2590-1990},
   url={http://dx.doi.org/10.21468/SciPostPhysLectNotes.48},
   DOI={10.21468/scipostphyslectnotes.48},
   journal={SciPost Physics Lecture Notes},
   publisher={Stichting SciPost},
   author={Carr, Bernard and K{\"u}hnel, Florian},
   year={2022},
   month=may }

@article{KAGRA:2013rdx,
    author = "Abbott, B. P. and others",
    collaboration = "KAGRA, LIGO Scientific, Virgo",
    title = "{Prospects for observing and localizing gravitational-wave transients with Advanced LIGO, Advanced Virgo and KAGRA}",
    eprint = "1304.0670",
    archivePrefix = "arXiv",
    primaryClass = "gr-qc",
    reportNumber = "LIGO-P1200087, VIR-0288A-12, JGW-P1808427",
    doi = "10.1007/s41114-020-00026-9",
    journal = "Living Rev. Rel.",
    volume = "19",
    pages = "1",
    year = "2016"
}

@article{Franciolini,
   title={From inflation to black hole mergers and back again: Gravitational-wave data-driven constraints on inflationary scenarios with a first-principle model of primordial black holes across the QCD epoch},
   volume={106},
   ISSN={2470-0029},
   url={http://dx.doi.org/10.1103/PhysRevD.106.123526},
   DOI={10.1103/physrevd.106.123526},
   number={12},
   journal={Physical Review D},
   publisher={American Physical Society (APS)},
   author={Franciolini, Gabriele and Musco, Ilia and Pani, Paolo and Urbano, Alfredo},
   year={2022},
   month=dec }

@article{Escriva1,
   title={Simulations of PBH formation at the QCD epoch and comparison with the GWTC-3 catalog},
   volume={2023},
   ISSN={1475-7516},
   url={http://dx.doi.org/10.1088/1475-7516/2023/05/004},
   DOI={10.1088/1475-7516/2023/05/004},
   number={05},
   journal={Journal of Cosmology and Astroparticle Physics},
   publisher={IOP Publishing},
   author={Escriv{\`a}, Albert and Bagui, Eleni and Clesse, Sebastien},
   year={2023},
   month=may, pages={004} }

@article{NANOGrav,
doi = {10.3847/2041-8213/acdac6},
url = {https://dx.doi.org/10.3847/2041-8213/acdac6},
year = {2023},
month = {jun},
publisher = {The American Astronomical Society},
volume = {951},
number = {1},
pages = {L8},
author = {Gabriella Agazie et. al},
title = {The NANOGrav 15 yr Data Set: Evidence for a Gravitational-wave Background},
journal = {The Astrophysical Journal Letters},
}

@article{EPTA,
   title={The second data release from the European Pulsar Timing Array: III. Search for gravitational wave signals},
   volume={678},
   ISSN={1432-0746},
   url={http://dx.doi.org/10.1051/0004-6361/202346844},
   DOI={10.1051/0004-6361/202346844},
   journal={Astronomy \& Astrophysics},
   publisher={EDP Sciences},
   author={Antoniadis, J. et al.},
   year={2023},
   month=oct, pages={A50} }

@article{PPTA,
   title={Search for an Isotropic Gravitational-wave Background with the Parkes Pulsar Timing Array},
   volume={951},
   ISSN={2041-8213},
   url={http://dx.doi.org/10.3847/2041-8213/acdd02},
   DOI={10.3847/2041-8213/acdd02},
   number={1},
   journal={The Astrophysical Journal Letters},
   publisher={American Astronomical Society},
   author={Reardon, Daniel J. et al.},
   year={2023},
   month=jun, pages={L6} }

@article{CPTA,
   title={Searching for the Nano-Hertz Stochastic Gravitational Wave Background with the Chinese Pulsar Timing Array Data Release I},
   volume={23},
   ISSN={1674-4527},
   url={http://dx.doi.org/10.1088/1674-4527/acdfa5},
   DOI={10.1088/1674-4527/acdfa5},
   number={7},
   journal={Research in Astronomy and Astrophysics},
   publisher={IOP Publishing},
   author={Xu, Heng et al.},
   year={2023},
   month=jun, pages={075024} 
}

@article{Pritchard:2024vix,
    author = "Pritchard, Xavier and Byrnes, Christian T.",
    title = "{Constraining the impact of standard model phase transitions on primordial black holes}",
    eprint = "2407.16563",
    archivePrefix = "arXiv",
    primaryClass = "astro-ph.CO",
    doi = "10.1088/1475-7516/2025/01/076",
    journal = "JCAP",
    volume = "01",
    pages = "076",
    year = "2025"
}

@article{Iovino:2024tyg,
    author = {Iovino, A. J. and Perna, G. and Riotto, A. and Veerm{\"a}e, H.},
    title = "{Curbing PBHs with PTAs}",
    eprint = "2406.20089",
    archivePrefix = "arXiv",
    primaryClass = "astro-ph.CO",
    doi = "10.1088/1475-7516/2024/10/050",
    journal = "JCAP",
    volume = "10",
    pages = "050",
    year = "2024"
}

@article{Esser:2025pnt,
    author = "Esser, Nicolas and Filion, Carrie and De Rijcke, Sven and Kallivayalil, Nitya and Richstein, Hannah and Tinyakov, Peter and Wyse, Rosemary F. G.",
    title = "{Constraints on asteroid-mass primordial black holes in dwarf galaxies using Hubble Space Telescope photometry}",
    eprint = "2503.03352",
    archivePrefix = "arXiv",
    primaryClass = "astro-ph.GA",
    doi = "10.1051/0004-6361/202554687",
    journal = "Astron. Astrophys.",
    volume = "698",
    pages = "A290",
    year = "2025"
}

@article{Ando,
    author = "Ando, Kenta and Inomata, Keisuke and Kawasaki, Masahiro",
    title = "{Primordial black holes and uncertainties in the choice of the window function}",
    eprint = "1802.06393",
    archivePrefix = "arXiv",
    primaryClass = "astro-ph.CO",
    reportNumber = "IPMU-18-0033",
    doi = "10.1103/PhysRevD.97.103528",
    journal = "Phys. Rev. D",
    volume = "97",
    number = "10",
    pages = "103528",
    year = "2018"
}

@article{Young,
   title={The primordial black hole formation criterion re-examined: Parametrisation, timing and the choice of window function},
   volume={29},
   ISSN={1793-6594},
   url={http://dx.doi.org/10.1142/S0218271820300025},
   DOI={10.1142/s0218271820300025},
   number={02},
   journal={International Journal of Modern Physics D},
   publisher={World Scientific Pub Co Pte Lt},
   author={Young, Sam},
   year={2019},
   month=nov, pages={2030002} }

@ARTICLE{Press,
       author = {{Press}, William H. and {Schechter}, Paul},
        title = "{Formation of Galaxies and Clusters of Galaxies by Self-Similar Gravitational Condensation}",
      journal = "Astrophys J.",
         year = 1974,
        month = feb,
       volume = {187},
        pages = {425-438},
          doi = {10.1086/152650},
       adsurl = {https://ui.adsabs.harvard.edu/abs/1974ApJ...187..425P}
}

@ARTICLE{Bardeen,
       author = {{Bardeen}, J.~M. and {Bond}, J.~R. and {Kaiser}, N. and {Szalay}, A.~S.},
        title = "{The Statistics of Peaks of Gaussian Random Fields}",
      journal = "Astrophys J.",
         year = 1986,
        month = may,
       volume = {304},
        pages = {15},
          doi = {10.1086/164143},
       adsurl = {https://ui.adsabs.harvard.edu/abs/1986ApJ...304...15B}
}

@article{aghanim,
   title={Planck2018 results: VI. Cosmological parameters},
   volume={641},
   ISSN={1432-0746},
   url={http://dx.doi.org/10.1051/0004-6361/201833910},
   DOI={10.1051/0004-6361/201833910},
   journal={Astronomy \& Astrophysics},
   publisher={EDP Sciences},
   author={Aghanim, N. et al.},
   year={2020},
   month=sep, pages={A6} }

@article{Kalaja:2019uju,
    author = "Kalaja, Alba and Bellomo, Nicola and Bartolo, Nicola and Bertacca, Daniele and Matarrese, Sabino and Musco, Ilia and Raccanelli, Alvise and Verde, Licia",
    title = "{From Primordial Black Holes Abundance to Primordial Curvature Power Spectrum (and back)}",
    eprint = "1908.03596",
    archivePrefix = "arXiv",
    primaryClass = "astro-ph.CO",
    doi = "10.1088/1475-7516/2019/10/031",
    journal = "JCAP",
    volume = "10",
    pages = "031",
    year = "2019"
}

@article{Musco2,
   title={Primordial black hole formation in the early universe: critical behaviour and self-similarity},
   volume={30},
   ISSN={1361-6382},
   url={http://dx.doi.org/10.1088/0264-9381/30/14/145009},
   DOI={10.1088/0264-9381/30/14/145009},
   number={14},
   journal={Classical and Quantum Gravity},
   publisher={IOP Publishing},
   author={Musco, Ilia and Miller, John C},
   year={2013},
   month=jun, pages={145009} }

@article{Stamou,
    author = "Stamou, Ioanna",
    title = "{Mechanisms for producing Primordial Black Holes from Inflationary Models Beyond Fine-Tuning}",
    eprint = "2404.14321",
    archivePrefix = "arXiv",
    primaryClass = "astro-ph.CO",
    month = "4",
    year = "2024"
}

@article{37,
    author = "Cole, Philippa S. and Gow, Andrew D. and Byrnes, Christian T. and Patil, Subodh P.",
    title = "{Primordial black holes from single-field inflation: a fine-tuning audit}",
    eprint = "2304.01997",
    archivePrefix = "arXiv",
    primaryClass = "astro-ph.CO",
    doi = "10.1088/1475-7516/2023/08/031",
    journal = "JCAP",
    volume = "08",
    pages = "031",
    year = "2023"
}

@article{Qin,
    author = "Qin, Wenzer and Geller, Sarah R. and Balaji, Shyam and McDonough, Evan and Kaiser, David I.",
    title = "{Planck constraints and gravitational wave forecasts for primordial black hole dark matter seeded by multifield inflation}",
    eprint = "2303.02168",
    archivePrefix = "arXiv",
    primaryClass = "astro-ph.CO",
    reportNumber = "MIT-CTP/5525",
    doi = "10.1103/PhysRevD.108.043508",
    journal = "Phys. Rev. D",
    volume = "108",
    number = "4",
    pages = "043508",
    year = "2023"
}

@article{Domenech:2021ztg,
    author = "Dom{\`e}nech, Guillem",
    title = "{Scalar Induced Gravitational Waves Review}",
    eprint = "2109.01398",
    archivePrefix = "arXiv",
    primaryClass = "gr-qc",
    doi = "10.3390/universe7110398",
    journal = "Universe",
    volume = "7",
    number = "11",
    pages = "398",
    year = "2021"
}

@article{Aldecoa-Tamayo:2025dqe,
    author = "Aldecoa-Tamayo, Itzi and Byrnes, Christian T. and Seery, David",
    title = "{Primordial black holes in Randall-Sundrum: Cosmological signatures}",
    eprint = "2509.13409",
    archivePrefix = "arXiv",
    primaryClass = "astro-ph.CO",
    month = "9",
    year = "2025"
}

@article{Braaten:1995ju,
    author = "Braaten, Eric and Nieto, Agustin",
    title = "{On the convergence of perturbative QCD at high temperature}",
    eprint = "hep-ph/9508406",
    archivePrefix = "arXiv",
    reportNumber = "NUHEP-TH-95-10",
    doi = "10.1103/PhysRevLett.76.1417",
    journal = "Phys. Rev. Lett.",
    volume = "76",
    pages = "1417--1420",
    year = "1996"
}

@article{Andersen:2010wu,
    author = "Andersen, Jens O. and Leganger, Lars E. and Strickland, Michael and Su, Nan",
    title = "{NNLO hard-thermal-loop thermodynamics for QCD}",
    eprint = "1009.4644",
    archivePrefix = "arXiv",
    primaryClass = "hep-ph",
    doi = "10.1016/j.physletb.2010.12.070",
    journal = "Phys. Lett. B",
    volume = "696",
    pages = "468--472",
    year = "2011"
}

@article{Kneur:2015moa,
    author = "Kneur, J. -L. and Pinto, M. B.",
    title = "{Renormalization Group Optimized Perturbation Theory at Finite Temperatures}",
    eprint = "1508.02610",
    archivePrefix = "arXiv",
    primaryClass = "hep-ph",
    doi = "10.1103/PhysRevD.92.116008",
    journal = "Phys. Rev. D",
    volume = "92",
    number = "11",
    pages = "116008",
    year = "2015"
}

@article{Kneur:2021feo,
    author = {Kneur, Jean-Lo{\"i}c and Pinto, Marcus Benghi and Restrepo, Tulio E.},
    title = "{QCD pressure: Renormalization group optimized perturbation theory confronts lattice}",
    eprint = "2101.02124",
    archivePrefix = "arXiv",
    primaryClass = "hep-ph",
    doi = "10.1103/PhysRevD.104.L031502",
    journal = "Phys. Rev. D",
    volume = "104",
    number = "3",
    pages = "L031502",
    year = "2021"
}

@article{Liu_2016,
   title={Patterns of strong coupling for LHC searches},
   volume={2016},
   ISSN={1029-8479},
   url={http://dx.doi.org/10.1007/JHEP11(2016)141},
   DOI={10.1007/jhep11(2016)141},
   number={11},
   journal={Journal of High Energy Physics},
   publisher={Springer Science and Business Media LLC},
   author={Liu, Da and Pomarol, Alex and Rattazzi, Riccardo and Riva, Francesco},
   year={2016},
   month=nov }

@article{Glioti_2025,
   title={Exploring the flavor symmetry landscape},
   volume={18},
   ISSN={2542-4653},
   url={http://dx.doi.org/10.21468/SciPostPhys.18.6.201},
   DOI={10.21468/scipostphys.18.6.201},
   number={6},
   journal={SciPost Physics},
   publisher={Stichting SciPost},
   author={Glioti, Alfredo and Rattazzi, Riccardo and Ricci, Lorenzo and Vecchi, Luca},
   year={2025},
   month=jun }

@book{Panico_2016,
   title={The Composite Nambu-Goldstone Higgs},
   ISBN={9783319226170},
   ISSN={1616-6361},
   url={http://dx.doi.org/10.1007/978-3-319-22617-0},
   DOI={10.1007/978-3-319-22617-0},
   journal={Lecture Notes in Physics},
   publisher={Springer International Publishing},
   author={Panico, Giuliano and Wulzer, Andrea},
   year={2016} }

@article{Panico_2013,
   title={On the tuning and the mass of the composite Higgs},
   volume={2013},
   ISSN={1029-8479},
   url={http://dx.doi.org/10.1007/JHEP03(2013)051},
   DOI={10.1007/jhep03(2013)051},
   number={3},
   journal={Journal of High Energy Physics},
   publisher={Springer Science and Business Media LLC},
   author={Panico, Giuliano and Redi, Michele and Tesi, Andrea and Wulzer, Andrea},
   year={2013},
   month=mar }

@article{Panero,
  title = {Thermodynamics of the QCD Plasma and the Large-$N$ Limit},
  author = {Panero, Marco},
  journal = {Phys. Rev. Lett.},
  volume = {103},
  issue = {23},
  pages = {232001},
  numpages = {4},
  year = {2009},
  month = {Dec},
  publisher = {American Physical Society},
  doi = {10.1103/PhysRevLett.103.232001},
  url = {https://link.aps.org/doi/10.1103/PhysRevLett.103.232001}
}

@article{DeGrand:2021zjw,
    author = "DeGrand, Thomas",
    title = "{Finite temperature properties of QCD with two flavors and three, four and five colors}",
    eprint = "2102.01150",
    archivePrefix = "arXiv",
    primaryClass = "hep-lat",
    doi = "10.1103/PhysRevD.103.094513",
    journal = "Phys. Rev. D",
    volume = "103",
    number = "9",
    pages = "094513",
    year = "2021"
}

@article{Engels_1997,
   title={Thermodynamics of four-flavour QCD with improved staggered fermions},
   volume={396},
   ISSN={0370-2693},
   url={http://dx.doi.org/10.1016/S0370-2693(97)00114-7},
   DOI={10.1016/s0370-2693(97)00114-7},
   number={1–4},
   journal={Physics Letters B},
   publisher={Elsevier BV},
   author={Engels, J. and Joswig, R. and Karsch, F. and Laermann, E. and L{\"u}tgemeier, M. and Petersson, B.},
   year={1997},
   month=mar, pages={210–216} }

@article{Ihssen:2024miv,
    author = "Ihssen, Friederike and Pawlowski, Jan M. and Sattler, Franz R. and Wink, Nicolas",
    title = "{Towards quantitative precision in functional QCD I}",
    eprint = "2408.08413",
    archivePrefix = "arXiv",
    primaryClass = "hep-ph",
    month = "8",
    year = "2024"
}

@article{Morgante_2023,
   title={Gravitational waves from dark SU(3) Yang-Mills theory},
   volume={107},
   ISSN={2470-0029},
   url={http://dx.doi.org/10.1103/PhysRevD.107.036010},
   DOI={10.1103/physrevd.107.036010},
   number={3},
   journal={Physical Review D},
   publisher={American Physical Society (APS)},
   author={Morgante, Enrico and Ramberg, Nicklas and Schwaller, Pedro},
   year={2023},
   month=feb }

@article{Cuteri:2018wci,
    author = "Cuteri, Francesca and Philipsen, Owe and Sciarra, Alessandro",
    title = "{Progress on the nature of the QCD thermal transition as a function of quark flavors and masses}",
    eprint = "1811.03840",
    archivePrefix = "arXiv",
    primaryClass = "hep-lat",
    doi = "10.22323/1.334.0170",
    journal = "PoS",
    volume = "LATTICE2018",
    pages = "170",
    year = "2018"
}

@article{Cuteri:2021ikv,
    author = "Cuteri, Francesca and Philipsen, Owe and Sciarra, Alessandro",
    title = "{On the order of the QCD chiral phase transition for different numbers of quark flavours}",
    eprint = "2107.12739",
    archivePrefix = "arXiv",
    primaryClass = "hep-lat",
    doi = "10.1007/JHEP11(2021)141",
    journal = "JHEP",
    volume = "11",
    pages = "141",
    year = "2021"
}

@article{Klinger:2025xxb,
    author = "Klinger, Jan Philipp and Kaiser, Reinhold and Philipsen, Owe",
    title = "{The order of the chiral phase transition in massless many-flavour lattice QCD}",
    eprint = "2501.19251",
    archivePrefix = "arXiv",
    primaryClass = "hep-lat",
    doi = "10.22323/1.466.0172",
    journal = "PoS",
    volume = "LATTICE2024",
    pages = "172",
    year = "2025"
}

@article{Dvali:2025sog,
    author = "Dvali, Gia",
    title = "{Swift Memory Burden in Merging Black Holes: how information load affects black hole's classical dynamics}",
    eprint = "2509.22540",
    archivePrefix = "arXiv",
    primaryClass = "hep-th",
    month = "9",
    year = "2025"
}

@article{tHooft:1973alw,
    author = "'t Hooft, Gerard",
    editor = "Taylor, J. C.",
    title = "{A Planar Diagram Theory for Strong Interactions}",
    reportNumber = "CERN-TH-1786",
    doi = "10.1016/0550-3213(74)90154-0",
    journal = "Nucl. Phys. B",
    volume = "72",
    pages = "461",
    year = "1974"
}

@article{Witten:1979kh,
    author = "Witten, Edward",
    title = "{Baryons in the 1/n Expansion}",
    reportNumber = "HUTP-79-A007",
    doi = "10.1016/0550-3213(79)90232-3",
    journal = "Nucl. Phys. B",
    volume = "160",
    pages = "57--115",
    year = "1979"
}

@article{Escriva:2023nzn,
    author = "Escriv{\`a}, Albert and Tada, Yuichiro and Yoo, Chul-Moon",
    title = "{Primordial black holes and induced gravitational waves from a smooth crossover beyond standard model theories}",
    eprint = "2311.17760",
    archivePrefix = "arXiv",
    primaryClass = "astro-ph.CO",
    doi = "10.1103/PhysRevD.110.063521",
    journal = "Phys. Rev. D",
    volume = "110",
    number = "6",
    pages = "063521",
    year = "2024"
}

@article{Barnard:2014tla,
    author = "Barnard, James and Gherghetta, Tony and Ray, Tirtha Sankar and Spray, Andrew",
    title = "{The Unnatural Composite Higgs}",
    eprint = "1409.7391",
    archivePrefix = "arXiv",
    primaryClass = "hep-ph",
    doi = "10.1007/JHEP01(2015)067",
    journal = "JHEP",
    volume = "01",
    pages = "067",
    year = "2015"
}

@article{Coleman:1967ad,
    author = "Coleman, Sidney R. and Mandula, J.",
    editor = "Zichichi, A.",
    title = "{All Possible Symmetries of the S Matrix}",
    doi = "10.1103/PhysRev.159.1251",
    journal = "Phys. Rev.",
    volume = "159",
    pages = "1251--1256",
    year = "1967"
}

@article{Kushwaha:2025zpz,
    author = "Kushwaha, Ashu and Suyama, Teruaki",
    title = "{Excursion Set Approach to Primordial Black Holes: Cloud-in-Cloud and Mass Function Revisited}",
    eprint = "2509.25871",
    archivePrefix = "arXiv",
    primaryClass = "astro-ph.CO",
    month = "9",
    year = "2025"
}

@article{Montefalcone:2025akm,
    author = "Montefalcone, Gabriele and Hooper, Dan and Freese, Katherine and Kelso, Chris and Kuhnel, Florian and Sandick, Pearl",
    title = "{Does Memory Burden Open a New Mass Window for Primordial Black Holes as Dark Matter?}",
    eprint = "2503.21005",
    archivePrefix = "arXiv",
    primaryClass = "astro-ph.CO",
    reportNumber = "UTWI-09-2025, NORDITA-2025-015",
    month = "3",
    year = "2025"
}

@article{ATLAS:2015yey,
    author = "Aad, Georges and others",
    collaboration = "ATLAS, CMS",
    title = "{Combined Measurement of the Higgs Boson Mass in $pp$ Collisions at $\sqrt{s}=7$ and 8 TeV with the ATLAS and CMS Experiments}",
    eprint = "1503.07589",
    archivePrefix = "arXiv",
    primaryClass = "hep-ex",
    reportNumber = "ATLAS-HIGG-2014-14, CMS-HIG-14-042, CERN-PH-EP-2015-075",
    doi = "10.1103/PhysRevLett.114.191803",
    journal = "Phys. Rev. Lett.",
    volume = "114",
    pages = "191803",
    year = "2015"
}

@article{Kajantie:1996mn,
    author = "Kajantie, K. and Laine, M. and Rummukainen, K. and Shaposhnikov, Mikhail E.",
    title = "{Is there a~ hot electroweak phase transition at $m_H \gtrsim m_W$?}",
    eprint = "hep-ph/9605288",
    archivePrefix = "arXiv",
    reportNumber = "CERN-TH-96-126, HD-THEP-96-15, IUHET-333",
    doi = "10.1103/PhysRevLett.77.2887",
    journal = "Phys. Rev. Lett.",
    volume = "77",
    pages = "2887--2890",
    year = "1996"
}

@article{Cline:1998hy,
    author = "Cline, James M. and Moore, Guy D.",
    title = "{Supersymmetric electroweak phase transition: Baryogenesis versus experimental constraints}",
    eprint = "hep-ph/9806354",
    archivePrefix = "arXiv",
    reportNumber = "MCGILL-98-11",
    doi = "10.1103/PhysRevLett.81.3315",
    journal = "Phys. Rev. Lett.",
    volume = "81",
    pages = "3315--3318",
    year = "1998"
}

@article{Bruggisser:2018mrt,
    author = "Bruggisser, Sebastian and Von Harling, Benedict and Matsedonskyi, Oleksii and Servant, G{\'e}raldine",
    title = "{Electroweak Phase Transition and Baryogenesis in Composite Higgs Models}",
    eprint = "1804.07314",
    archivePrefix = "arXiv",
    primaryClass = "hep-ph",
    reportNumber = "DESY-17-229",
    doi = "10.1007/JHEP12(2018)099",
    journal = "JHEP",
    volume = "12",
    pages = "099",
    year = "2018"
}

@article{Hindmarsh:2013xza,
    author = "Hindmarsh, Mark and Huber, Stephan J. and Rummukainen, Kari and Weir, David J.",
    title = "{Gravitational waves from the sound of a first order phase transition}",
    eprint = "1304.2433",
    archivePrefix = "arXiv",
    primaryClass = "hep-ph",
    reportNumber = "HIP-2013-07-TH",
    doi = "10.1103/PhysRevLett.112.041301",
    journal = "Phys. Rev. Lett.",
    volume = "112",
    pages = "041301",
    year = "2014"
}

@Inbook{Hald2007,
title="De Moivre's Normal Approximation to the Binomial, 1733, and Its Generalization",
bookTitle="A History of Parametric Statistical Inference from Bernoulli to Fisher, 1713--1935",
year="2007",
publisher="Springer New York",
address="New York, NY",
pages="17--24",
isbn="978-0-387-46409-1",
doi="10.1007/978-0-387-46409-1_3",
url="https://doi.org/10.1007/978-0-387-46409-1_3"
}

@incollection{STIGLER2005329,
title = {Chapter 24 - P.S. Laplace, Théorie analytique des probabilités, first edition (1812); Essai philosophique sur les probabilités, first edition (1814)},
editor = {I. Grattan-Guinness and Roger Cooke and Leo Corry and Pierre Crépel and Niccolo Guicciardini},
booktitle = {Landmark Writings in Western Mathematics 1640-1940},
publisher = {Elsevier Science},
address = {Amsterdam},
pages = {329-340},
year = {2005},
isbn = {978-0-444-50871-3},
doi = {https://doi.org/10.1016/B978-044450871-3/50105-4},
url = {https://www.sciencedirect.com/science/article/pii/B9780444508713501054},
author = {Stephen M. Stigler}
}

@article{Bardeen:1985tr,
    author = "Bardeen, James M. and Bond, J. R. and Kaiser, Nick and Szalay, A. S.",
    title = "{The Statistics of Peaks of Gaussian Random Fields}",
    reportNumber = "FERMILAB-PUB-85-148-A, NSF-ITP-85-93",
    doi = "10.1086/164143",
    journal = "Astrophys. J.",
    volume = "304",
    pages = "15--61",
    year = "1986"
}

@article{Germani:2025fkh,
    author = "Germani, Cristiano and Gorji, Mohammad Ali and Uwabo-Niibo, Michiru and Yamaguchi, Masahide",
    title = "{Peaks sphericity of non-Gaussian random fields}",
    eprint = "2503.05434",
    archivePrefix = "arXiv",
    primaryClass = "astro-ph.CO",
    doi = "10.1088/1475-7516/2025/09/052",
    journal = "JCAP",
    volume = "09",
    pages = "052",
    year = "2025"
}

@article{Ebrahimian:2025syf,
    author = "Ebrahimian, Ehsan and Abolhasani, Ali Akbar and Mirbabayi, Mehrdad",
    title = "{Primordial black hole formation in matter domination}",
    eprint = "2507.18312",
    archivePrefix = "arXiv",
    primaryClass = "gr-qc",
    month = "7",
    year = "2025"
}

@article{Ye:2025wif,
    author = "Ye, Weitao and Gong, Yungui and Harada, Tomohiro and Kang, Zhaofeng and Kohri, Kazunori and Saito, Daiki and Yoo, Chul-Moon",
    title = "{Primordial Black Hole Formation and Spin in Matter Domination Revisited}",
    eprint = "2508.10070",
    archivePrefix = "arXiv",
    primaryClass = "gr-qc",
    reportNumber = "NU-QG-9, RUP-25-19, KUNS-3070, KEK-TH-2753, KEK-Cosmo-0391",
    month = "8",
    year = "2025"
}

@article{Aldabergenov:2022rfc,
    author = "Aldabergenov, Yermek and Addazi, Andrea and Ketov, Sergei V.",
    title = "{Inflation, SUSY breaking, and primordial black holes in modified supergravity coupled to chiral matter}",
    eprint = "2206.02601",
    archivePrefix = "arXiv",
    primaryClass = "astro-ph.CO",
    reportNumber = "IPMU22-0033",
    doi = "10.1140/epjc/s10052-022-10654-w",
    journal = "Eur. Phys. J. C",
    volume = "82",
    number = "8",
    pages = "681",
    year = "2022"
}

@article{Ferretti_2014,
   title={Fermionic UV completions of composite Higgs models},
   volume={2014},
   ISSN={1029-8479},
   url={http://dx.doi.org/10.1007/JHEP03(2014)077},
   DOI={10.1007/jhep03(2014)077},
   number={3},
   journal={Journal of High Energy Physics},
   publisher={Springer Science and Business Media LLC},
   author={Ferretti, Gabriele and Karateev, Denis},
   year={2014},
   month=mar }

@article{Cacciapaglia_2014,
   title={Fundamental composite (Goldstone) Higgs dynamics},
   volume={2014},
   ISSN={1029-8479},
   url={http://dx.doi.org/10.1007/JHEP04(2014)111},
   DOI={10.1007/jhep04(2014)111},
   number={4},
   journal={Journal of High Energy Physics},
   publisher={Springer Science and Business Media LLC},
   author={Cacciapaglia, Giacomo and Sannino, Francesco},
   year={2014},
   month=apr }

@article{Carrasco_2014,
   title={Up, down, strange and charm quark masses with Nf= 2+1+1 twisted mass lattice QCD},
   volume={887},
   ISSN={0550-3213},
   url={http://dx.doi.org/10.1016/j.nuclphysb.2014.07.025},
   DOI={10.1016/j.nuclphysb.2014.07.025},
   journal={Nuclear Physics B},
   publisher={Elsevier BV},
   author={Carrasco, N. and Deuzeman, A. and Dimopoulos, P. and Frezzotti, R. and Giménez, V. and Herdoiza, G. and Lami, P. and Lubicz, V. and Palao, D. and Picca, E. and Reker, S. and Riggio, L. and Rossi, G.C. and Sanfilippo, F. and Scorzato, L. and Simula, S. and Tarantino, C. and Urbach, C. and Wenger, U.},
   year={2014},
   month=oct, pages={19–68} }

@book{Weinberg_1996, place={Cambridge}, title={The Quantum Theory of Fields}, publisher={Cambridge University Press}, author={Weinberg, Steven}, year={1996}}

@article{Dong_2021,
   title={UV completed composite Higgs model with heavy composite partners},
   volume={104},
   ISSN={2470-0029},
   url={http://dx.doi.org/10.1103/PhysRevD.104.035013},
   DOI={10.1103/physrevd.104.035013},
   number={3},
   journal={Physical Review D},
   publisher={American Physical Society (APS)},
   author={Dong, Zi-Yu and Guan, Cong-Sen and Ma, Teng and Shu, Jing and Xue, Xiao},
   year={2021},
   month=aug }

@article{Banerjee_2017,
   title={Improving fine-tuning in composite Higgs models},
   volume={96},
   ISSN={2470-0029},
   url={http://dx.doi.org/10.1103/PhysRevD.96.035040},
   DOI={10.1103/physrevd.96.035040},
   number={3},
   journal={Physical Review D},
   publisher={American Physical Society (APS)},
   author={Banerjee, Avik and Bhattacharyya, Gautam and Ray, Tirtha Sankar},
   year={2017},
   month=aug }

@article{Murnane_2019,
   title={Exploring fine-tuning of the Next-to-Minimal Composite Higgs Model},
   volume={2019},
   ISSN={1029-8479},
   url={http://dx.doi.org/10.1007/JHEP04(2019)076},
   DOI={10.1007/jhep04(2019)076},
   number={4},
   journal={Journal of High Energy Physics},
   publisher={Springer Science and Business Media LLC},
   author={Murnane, Daniel and White, Martin and Williams, Anthony G.},
   year={2019},
   month=apr }

@misc{fujikura2023,
      title={Cosmological Phase Transitions in Composite Higgs Models}, 
      author={Kohei Fujikura and Yuichiro Nakai and Ryosuke Sato and Yaoduo Wang},
      year={2023},
      eprint={2306.01305},
      archivePrefix={arXiv},
      primaryClass={hep-ph},
      url={https://arxiv.org/abs/2306.01305}, 
}

@article{Germani:2017bcs,
    author = "Germani, Cristiano and Prokopec, Tomislav",
    title = "{On primordial black holes from an inflection point}",
    eprint = "1706.04226",
    archivePrefix = "arXiv",
    primaryClass = "astro-ph.CO",
    reportNumber = "ICCUB-17-012",
    doi = "10.1016/j.dark.2017.09.001",
    journal = "Phys. Dark Univ.",
    volume = "18",
    pages = "6--10",
    year = "2017"
}

@article{Khlopov:1985fch,
    author = "Khlopov, M. Yu. and Malomed, B. A. and Zeldovich, Ia. B. and Zeldovich, Ya. B.",
    title = "{Gravitational instability of scalar fields and formation of primordial black holes}",
    doi = "10.1093/mnras/215.4.575",
    journal = "Mon. Not. Roy. Astron. Soc.",
    volume = "215",
    number = "4",
    pages = "575--589",
    year = "1985"
}

@article{Ketov:2021fww,
    author = "Ketov, Sergei V.",
    editor = "Isenberg, James A. and Cleaver, Gerald B. and Shao, Lijing and Olmo, Gonzalo J. and Tommei, Giacomo",
    title = "{Multi-Field versus Single-Field in the Supergravity Models of Inflation and Primordial Black Holes}",
    doi = "10.3390/universe7050115",
    journal = "Universe",
    volume = "7",
    number = "5",
    pages = "115",
    year = "2021"
}

@article{Jedamzik:1996mr,
    author = "Jedamzik, Karsten",
    title = "{Primordial black hole formation during the QCD epoch}",
    eprint = "astro-ph/9605152",
    archivePrefix = "arXiv",
    doi = "10.1103/PhysRevD.55.R5871",
    journal = "Phys. Rev. D",
    volume = "55",
    pages = "5871--5875",
    year = "1997"
}

@article{Musco:2023dak,
    author = "Musco, Ilia and Jedamzik, Karsten and Young, Sam",
    title = "{Primordial black hole formation during the QCD phase transition: Threshold, mass distribution, and abundance}",
    eprint = "2303.07980",
    archivePrefix = "arXiv",
    primaryClass = "astro-ph.CO",
    doi = "10.1103/PhysRevD.109.083506",
    journal = "Phys. Rev. D",
    volume = "109",
    number = "8",
    pages = "083506",
    year = "2024"
}

@article{Bhaumik:2025vlb,
    author = "Bhaumik, Nilanjandev and Guo, Huai-Ke and Liu, Si-Jiang",
    title = "{Extended mass distribution of PBHs during QCD phase transition: SGWB and mini-EMRIs}",
    eprint = "2509.25083",
    archivePrefix = "arXiv",
    primaryClass = "astro-ph.CO",
    month = "9",
    year = "2025"
}

@article{Franciolini:2023wjm,
    author = "Franciolini, Gabriele and Racco, Davide and Rompineve, Fabrizio",
    title = "{Footprints of the QCD Crossover on Cosmological Gravitational Waves at Pulsar Timing Arrays}",
    eprint = "2306.17136",
    archivePrefix = "arXiv",
    primaryClass = "astro-ph.CO",
    reportNumber = "CERN-TH-2023-080",
    doi = "10.1103/PhysRevLett.132.081001",
    journal = "Phys. Rev. Lett.",
    volume = "132",
    number = "8",
    pages = "081001",
    year = "2024",
    note = "[Erratum: Phys.Rev.Lett. 133, 189901 (2024)]"
}

@article{Carr:2019kxo,
    author = {Carr, Bernard and Clesse, Sebastien and Garc{\'\i}a-Bellido, Juan and K{\"u}hnel, Florian},
    title = "{Cosmic conundra explained by thermal history and primordial black holes}",
    eprint = "1906.08217",
    archivePrefix = "arXiv",
    primaryClass = "astro-ph.CO",
    doi = "10.1016/j.dark.2020.100755",
    journal = "Phys. Dark Univ.",
    volume = "31",
    pages = "100755",
    year = "2021"
}

@article{Alonso-Monsalve:2023brx,
    author = "Alonso-Monsalve, Elba and Kaiser, David I.",
    title = "{Primordial Black Holes with QCD Color Charge}",
    eprint = "2310.16877",
    archivePrefix = "arXiv",
    primaryClass = "hep-ph",
    reportNumber = "MIT-CTP/5624",
    doi = "10.1103/PhysRevLett.132.231402",
    journal = "Phys. Rev. Lett.",
    volume = "132",
    number = "23",
    pages = "231402",
    year = "2024"
}

@misc{LEP,
      title={A Combination of Preliminary Electroweak Measurements and Constraints on the Standard Model}, 
      author={The LEP Collaboration and ALEPH Collaboration and DELPHI Collaboration and L3 Collaboration and OPAL Collaboration and the LEP Electroweak Working Group},
      year={2007},
      eprint={hep-ex/0612034},
      archivePrefix={arXiv},
      primaryClass={hep-ex},
      url={https://arxiv.org/abs/hep-ex/0612034}, 
}

@article{20251,
title = {The stairway to heaven},
journal = {Physics Reports},
volume = {1115},
pages = {1-2},
year = {2025},
note = {CMS physics results from the first decade of LHC data},
issn = {0370-1573},
doi = {https://doi.org/10.1016/j.physrep.2025.01.004},
url = {https://www.sciencedirect.com/science/article/pii/S0370157325000249}
}

@article{AtlasDrel,
   title={Measurements of differential Z boson production cross sections in proton-proton collisions at {$ \sqrt{s} $} = 13TeV},
   volume={2019},
   ISSN={1029-8479},
   url={http://dx.doi.org/10.1007/JHEP12(2019)061},
   DOI={10.1007/jhep12(2019)061},
   number={12},
   journal={Journal of High Energy Physics},
   publisher={Springer Science and Business Media LLC},
   author={Sirunyan, A. M. et al},
   year={2019},
   month=dec }

@article{Isidori_2024,
   title={The standard model effective field theory at work},
   volume={96},
   ISSN={1539-0756},
   url={http://dx.doi.org/10.1103/RevModPhys.96.015006},
   DOI={10.1103/revmodphys.96.015006},
   number={1},
   journal={Reviews of Modern Physics},
   publisher={American Physical Society (APS)},
   author={Isidori, Gino and Wilsch, Felix and Wyler, Daniel},
   year={2024},
   month=mar }

@misc{gouttenoire2022,
      title={Beyond the Standard Model Cocktail}, 
      author={Yann Gouttenoire},
      year={2022},
      eprint={2207.01633},
      archivePrefix={arXiv},
      primaryClass={hep-ph},
      url={https://arxiv.org/abs/2207.01633}, 
}

@misc{craig2022,
      title={Naturalness: A Snowmass White Paper}, 
      author={Nathaniel Craig},
      year={2022},
      eprint={2205.05708},
      archivePrefix={arXiv},
      primaryClass={hep-ph},
      url={https://arxiv.org/abs/2205.05708}, 
}

@misc{peskin2025,
      title={What is the Hierarchy Problem?}, 
      author={Michael E. Peskin},
      year={2025},
      eprint={2505.00694},
      archivePrefix={arXiv},
      primaryClass={hep-ph},
      url={https://arxiv.org/abs/2505.00694}, 
}

@article{PhysRevD.14.1667,
  title = {Gauge-symmetry hierarchies},
  author = {Gildener, Eldad},
  journal = {Phys. Rev. D},
  volume = {14},
  issue = {6},
  pages = {1667--1672},
  numpages = {0},
  year = {1976},
  month = {Sep},
  publisher = {American Physical Society},
  doi = {10.1103/PhysRevD.14.1667},
  url = {https://link.aps.org/doi/10.1103/PhysRevD.14.1667}
}

@article{PhysRevD.20.2619,
  title = {Dynamics of spontaneous symmetry breaking in the Weinberg-Salam theory},
  author = {Susskind, Leonard},
  journal = {Phys. Rev. D},
  volume = {20},
  issue = {10},
  pages = {2619--2625},
  numpages = {0},
  year = {1979},
  month = {Nov},
  publisher = {American Physical Society},
  doi = {10.1103/PhysRevD.20.2619},
  url = {https://link.aps.org/doi/10.1103/PhysRevD.20.2619}
}

@article{Belfatto_2020,
   title={The CKM unitarity problem: a trace of new physics at the TeV scale?},
   volume={80},
   ISSN={1434-6052},
   url={http://dx.doi.org/10.1140/epjc/s10052-020-7691-6},
   DOI={10.1140/epjc/s10052-020-7691-6},
   number={2},
   journal={The European Physical Journal C},
   publisher={Springer Science and Business Media LLC},
   author={Belfatto, Benedetta and Beradze, Revaz and Berezhiani, Zurab},
   year={2020},
   month=feb }

@book{Schwartz_2013, place={Cambridge}, title={Quantum Field Theory and the Standard Model}, publisher={Cambridge University Press}, author={Schwartz, Matthew D.}, year={2013}}

@article{WITTEN1981513,
title = {Dynamical breaking of supersymmetry},
journal = {Nuclear Physics B},
volume = {188},
number = {3},
pages = {513-554},
year = {1981},
issn = {0550-3213},
doi = {https://doi.org/10.1016/0550-3213(81)90006-7},
url = {https://www.sciencedirect.com/science/article/pii/0550321381900067},
author = {Edward Witten}
}

@article{DIMOPOULOS1979237,
title = {Mass without scalars},
journal = {Nuclear Physics B},
volume = {155},
number = {1},
pages = {237-252},
year = {1979},
issn = {0550-3213},
doi = {https://doi.org/10.1016/0550-3213(79)90364-X},
url = {https://www.sciencedirect.com/science/article/pii/055032137990364X},
author = {Savas Dimopoulos and Leonard Susskind}
}

@misc{contino2010tasi,
      title={Tasi 2009 lectures: The Higgs as a Composite Nambu-Goldstone Boson}, 
      author={Roberto Contino},
      year={2010},
      eprint={1005.4269},
      archivePrefix={arXiv},
      primaryClass={hep-ph},
      url={https://arxiv.org/abs/1005.4269}, 
}

@misc{krippendorf2010,
      title={Cambridge Lectures on Supersymmetry and Extra Dimensions}, 
      author={Sven Krippendorf and Fernando Quevedo and Oliver Schlotterer},
      year={2010},
      eprint={1011.1491},
      archivePrefix={arXiv},
      primaryClass={hep-th},
      url={https://arxiv.org/abs/1011.1491}, 
}

@article{CS_KI_1996,
   title={THE MINIMAL SUPERSYMMETRIC STANDARD MODEL},
   volume={11},
   ISSN={1793-6632},
   url={http://dx.doi.org/10.1142/S021773239600062X},
   DOI={10.1142/s021773239600062x},
   number={08},
   journal={Modern Physics Letters A},
   publisher={World Scientific Pub Co Pte Lt},
   author={CSÁKI, CSABA},
   year={1996},
   month=mar, pages={599–613} }

@article{Chatrchyan_2012,
   title={Observation of a new boson at a mass of 125 GeV with the CMS experiment at the LHC},
   volume={716},
   ISSN={0370-2693},
   url={http://dx.doi.org/10.1016/j.physletb.2012.08.021},
   DOI={10.1016/j.physletb.2012.08.021},
   number={1},
   journal={Physics Letters B},
   publisher={Elsevier BV},
   author={Chatrchyan, S et al},
   year={2012},
   month=sep, pages={30–61} }

@article{Aad_2012,
   title={Observation of a new particle in the search for the Standard Model Higgs boson with the ATLAS detector at the LHC},
   volume={716},
   ISSN={0370-2693},
   url={http://dx.doi.org/10.1016/j.physletb.2012.08.020},
   DOI={10.1016/j.physletb.2012.08.020},
   number={1},
   journal={Physics Letters B},
   publisher={Elsevier BV},
   author={Aad, G et al},
   year={2012},
   month=sep, pages={1–29} }

@article{Giudice_2007,
   title={The strongly-interacting light Higgs},
   volume={2007},
   ISSN={1029-8479},
   url={http://dx.doi.org/10.1088/1126-6708/2007/06/045},
   DOI={10.1088/1126-6708/2007/06/045},
   number={06},
   journal={Journal of High Energy Physics},
   publisher={Springer Science and Business Media LLC},
   author={Giudice, Gian Francesco and Grojean, Christophe and Pomarol, Alex and Rattazzi, Riccardo},
   year={2007},
   month=jun, pages={045–045} }

@article{Kumar:2025jfi,
    author = "Kumar, Utkarsh",
    title = "{Primordial gravitational wave background as a probe of primordial black holes}",
    eprint = "2507.10033",
    archivePrefix = "arXiv",
    primaryClass = "gr-qc",
    doi = "10.1103/ymhr-9711",
    journal = "Phys. Rev. D",
    volume = "112",
    number = "8",
    pages = "084027",
    year = "2025"
}

@article{Escriva:2024ivo,
    author = "Escriv{\`a}, Albert and Inui, Ryoto and Tada, Yuichiro and Yoo, Chul-Moon",
    title = "{LISA forecast on a smooth crossover beyond the standard model through the scalar-induced gravitational waves}",
    eprint = "2404.12591",
    archivePrefix = "arXiv",
    primaryClass = "astro-ph.CO",
    doi = "10.1103/PhysRevD.111.023528",
    journal = "Phys. Rev. D",
    volume = "111",
    number = "2",
    pages = "023528",
    year = "2025"
}

@article{Hall:2009nd,
    author = "Hall, Lawrence J. and Nomura, Yasunori",
    title = "{A Finely-Predicted Higgs Boson Mass from A Finely-Tuned Weak Scale}",
    eprint = "0910.2235",
    archivePrefix = "arXiv",
    primaryClass = "hep-ph",
    reportNumber = "UCB-PTH-09-31",
    doi = "10.1007/JHEP03(2010)076",
    journal = "JHEP",
    volume = "03",
    pages = "076",
    year = "2010"
}

@article{Maki:1962mu,
    author = "Maki, Ziro and Nakagawa, Masami and Sakata, Shoichi",
    title = "{Remarks on the unified model of elementary particles}",
    doi = "10.1143/PTP.28.870",
    journal = "Prog. Theor. Phys.",
    volume = "28",
    pages = "870--880",
    year = "1962"
}

@article{Super-Kamiokande:1998kpq,
    author = "Fukuda, Y. and others",
    collaboration = "Super-Kamiokande",
    title = "{Evidence for oscillation of atmospheric neutrinos}",
    eprint = "hep-ex/9807003",
    archivePrefix = "arXiv",
    reportNumber = "BU-98-17, ICRR-REPORT-422-98-18, UCI-98-8, KEK-PREPRINT-98-95, LSU-HEPA-5-98, UMD-98-003, SBHEP-98-5, TKU-PAP-98-06, TIT-HPE-98-09",
    doi = "10.1103/PhysRevLett.81.1562",
    journal = "Phys. Rev. Lett.",
    volume = "81",
    pages = "1562--1567",
    year = "1998"
}

@article{Peccei:1977hh,
    author = "Peccei, R. D. and Quinn, Helen R.",
    title = "{CP Conservation in the Presence of Instantons}",
    reportNumber = "ITP-568-STANFORD",
    doi = "10.1103/PhysRevLett.38.1440",
    journal = "Phys. Rev. Lett.",
    volume = "38",
    pages = "1440--1443",
    year = "1977"
}

@article{Weinberg:1977ma,
    author = "Weinberg, Steven",
    title = "{A New Light Boson?}",
    reportNumber = "HUTP-77/A074",
    doi = "10.1103/PhysRevLett.40.223",
    journal = "Phys. Rev. Lett.",
    volume = "40",
    pages = "223--226",
    year = "1978"
}

@article{Wilczek:1977pj,
    author = "Wilczek, Frank",
    title = "{Problem of Strong  $P$  and  $T$  Invariance in the Presence of Instantons}",
    reportNumber = "Print-77-0939 (COLUMBIA)",
    doi = "10.1103/PhysRevLett.40.279",
    journal = "Phys. Rev. Lett.",
    volume = "40",
    pages = "279--282",
    year = "1978"
}

@article{Kobayashi:1973fv,
    author = "Kobayashi, Makoto and Maskawa, Toshihide",
    title = "{CP Violation in the Renormalizable Theory of Weak Interaction}",
    reportNumber = "KUNS-242",
    doi = "10.1143/PTP.49.652",
    journal = "Prog. Theor. Phys.",
    volume = "49",
    pages = "652--657",
    year = "1973"
}

@article{Froggatt:1978nt,
    author = "Froggatt, C. D. and Nielsen, Holger Bech",
    title = "{Hierarchy of Quark Masses, Cabibbo Angles and CP Violation}",
    reportNumber = "CERN-TH-2519",
    doi = "10.1016/0550-3213(79)90316-X",
    journal = "Nucl. Phys. B",
    volume = "147",
    pages = "277--298",
    year = "1979"
}

@article{Englert:1964et,
    author = "Englert, F. and Brout, R.",
    editor = "Taylor, J. C.",
    title = "{Broken Symmetry and the Mass of Gauge Vector Mesons}",
    doi = "10.1103/PhysRevLett.13.321",
    journal = "Phys. Rev. Lett.",
    volume = "13",
    pages = "321--323",
    year = "1964"
}

@article{Higgs:1964pj,
    author = "Higgs, Peter W.",
    editor = "Taylor, J. C.",
    title = "{Broken Symmetries and the Masses of Gauge Bosons}",
    doi = "10.1103/PhysRevLett.13.508",
    journal = "Phys. Rev. Lett.",
    volume = "13",
    pages = "508--509",
    year = "1964"
}

@article{Weinberg:1967tq,
    author = "Weinberg, Steven",
    title = "{A Model of Leptons}",
    doi = "10.1103/PhysRevLett.19.1264",
    journal = "Phys. Rev. Lett.",
    volume = "19",
    pages = "1264--1266",
    year = "1967"
}

@article{Kirzhnits:1972iw,
    author = "Kirzhnits, D. A.",
    title = "{Weinberg model in the hot universe}",
    journal = "JETP Lett.",
    volume = "15",
    pages = "529--531",
    year = "1972"
}

@article{Weinberg:1974hy,
    author = "Weinberg, Steven",
    title = "{Gauge and Global Symmetries at High Temperature}",
    reportNumber = "PRINT-74-0689 (HARVARD)",
    doi = "10.1103/PhysRevD.9.3357",
    journal = "Phys. Rev. D",
    volume = "9",
    pages = "3357--3378",
    year = "1974"
}

@article{Linde:1978px,
    author = "Linde, Andrei D.",
    title = "{Phase Transitions in Gauge Theories and Cosmology}",
    reportNumber = "LEBEDEV-78-166",
    doi = "10.1088/0034-4885/42/3/001",
    journal = "Rept. Prog. Phys.",
    volume = "42",
    pages = "389",
    year = "1979"
}

@ARTICLE{1981MNRAS.195..467S,
       author = "Sato, K.",
        title = "{First-order phase transition of a vacuum and the expansion of the Universe}",
      journal = {Mon. Not. R. Astron. Soc.},
         year = 1981,
        month = may,
       volume = {195},
        pages = {467-479},
          doi = {10.1093/mnras/195.3.467},
       adsurl = {https://ui.adsabs.harvard.edu/abs/1981MNRAS.195..467S}
}

@article{Sasaki:2018dmp,
    author = "Sasaki, Misao and Suyama, Teruaki and Tanaka, Takahiro and Yokoyama, Shuichiro",
    title = "{Primordial black holes{\textemdash}perspectives in gravitational wave astronomy}",
    eprint = "1801.05235",
    archivePrefix = "arXiv",
    primaryClass = "astro-ph.CO",
    doi = "10.1088/1361-6382/aaa7b4",
    journal = "Class. Quant. Grav.",
    volume = "35",
    number = "6",
    pages = "063001",
    year = "2018"
}

@article{Merchand:2025bzt,
    author = "Merchand, Marco",
    title = "{Exploring Ultra-Slow-Roll Inflation in Composite Pseudo-Nambu-Goldstone Boson Models: Implications for Primordial Black Holes and Gravitational Waves}",
    eprint = "2510.15460",
    archivePrefix = "arXiv",
    primaryClass = "astro-ph.CO",
    month = "10",
    year = "2025"
}

@article{Motohashi:2017kbs,
    author = "Motohashi, Hayato and Hu, Wayne",
    title = "{Primordial Black Holes and Slow-Roll Violation}",
    eprint = "1706.06784",
    archivePrefix = "arXiv",
    primaryClass = "astro-ph.CO",
    doi = "10.1103/PhysRevD.96.063503",
    journal = "Phys. Rev. D",
    volume = "96",
    number = "6",
    pages = "063503",
    year = "2017"
}

@article{Garcia-Bellido:2017mdw,
    author = "Garcia-Bellido, Juan and Ruiz Morales, Ester",
    title = "{Primordial black holes from single field models of inflation}",
    eprint = "1702.03901",
    archivePrefix = "arXiv",
    primaryClass = "astro-ph.CO",
    reportNumber = "IFT-UAM-CSIC-17-007, CERN-TH-2017-196",
    doi = "10.1016/j.dark.2017.09.007",
    journal = "Phys. Dark Univ.",
    volume = "18",
    pages = "47--54",
    year = "2017"
}

@article{Lu:2022yuc,
    author = "Lu, Philip and Takhistov, Volodymyr and Fuller, George M.",
    title = "{Signatures of a High Temperature QCD Transition in the Early Universe}",
    eprint = "2212.00156",
    archivePrefix = "arXiv",
    primaryClass = "astro-ph.CO",
    reportNumber = "IPMU22-0064, KEK-QUP-2022-0017, KEK-TH-2476, KEK-Cosmo-0303",
    doi = "10.1103/PhysRevLett.130.221002",
    journal = "Phys. Rev. Lett.",
    volume = "130",
    number = "22",
    pages = "221002",
    year = "2023"
}

@article{Agashe_2005,
   title={The minimal composite Higgs model},
   volume={719},
   ISSN={0550-3213},
   url={http://dx.doi.org/10.1016/j.nuclphysb.2005.04.035},
   DOI={10.1016/j.nuclphysb.2005.04.035},
   number={1–2},
   journal={Nuclear Physics B},
   publisher={Elsevier BV},
   author={Agashe, Kaustubh and Contino, Roberto and Pomarol, Alex},
   year={2005},
   month=jul, pages={165–187} }

@article{KAPLAN1984187,
title = {Composite Higgs scalars},
journal = {Physics Letters B},
volume = {136},
number = {3},
pages = {187-190},
year = {1984},
issn = {0370-2693},
doi = {https://doi.org/10.1016/0370-2693(84)91178-X},
url = {https://www.sciencedirect.com/science/article/pii/037026938491178X},
author = {David B. Kaplan and Howard Georgi and Savas Dimopoulos}
}

@article{KAPLAN1984183,
title = {SU(2) × U(1) breaking by vacuum misalignment},
journal = {Physics Letters B},
volume = {136},
number = {3},
pages = {183-186},
year = {1984},
issn = {0370-2693},
doi = {https://doi.org/10.1016/0370-2693(84)91177-8},
url = {https://www.sciencedirect.com/science/article/pii/0370269384911778},
author = {David B. Kaplan and Howard Georgi}
}

@article{Gripaios_2009,
   title={Beyond the minimal composite Higgs model},
   volume={2009},
   ISSN={1029-8479},
   url={http://dx.doi.org/10.1088/1126-6708/2009/04/070},
   DOI={10.1088/1126-6708/2009/04/070},
   number={04},
   journal={Journal of High Energy Physics},
   publisher={Springer Science and Business Media LLC},
   author={Gripaios, Ben and Pomarol, Alex and Riva, Francesco and Serra, Javi},
   year={2009},
   month=apr, pages={070–070} }

@article{10.1093/ptep/ptaa104,
    author = {Particle Data Group},
    title = {Review of Particle Physics},
    journal = {Progress of Theoretical and Experimental Physics},
    volume = {2020},
    number = {8},
    pages = {083C01},
    year = {2020},
    month = {08},
    issn = {2050-3911},
    doi = {10.1093/ptep/ptaa104},
    url = {https://doi.org/10.1093/ptep/ptaa104},
    eprint = {https://academic.oup.com/ptep/article-pdf/2020/8/083C01/34673722/ptaa104.pdf},
}
\end{document}